\documentclass{article}

\usepackage[includeheadfoot,top=10 mm, bottom=15 mm, left=25 mm, right=25 mm]{geometry}
\usepackage{algpseudocode}
\usepackage{alltt}
\usepackage{amsbsy}
\usepackage{amscd}
\usepackage{amsfonts}
\usepackage{amsmath}
\usepackage{amssymb}
\usepackage{amsthm}
\usepackage{boldtensors}
\usepackage{booktabs}
\usepackage{enumitem}
\usepackage{fancyhdr}
\usepackage{float}
\usepackage{geometry}
\usepackage{graphicx}
\usepackage{multirow}
\usepackage{siunitx}
\usepackage{supertabular}
\usepackage[dvipsnames]{xcolor}
\usepackage{color}
\usepackage{tikz}
\usepackage{tikz-3dplot}
\usepackage{url}
\usepackage{verbatim}
\usepackage{xspace}
\usepackage{cases}
\usepackage{graphics}

\usepackage[nomessages]{fp}
\usepackage[english]{babel}
\usepackage[boxed]{algorithm}
\usepackage[bf,tight]{subfigure}
\usepackage[utf8]{inputenc}
\usepackage[T1]{fontenc}
\usepackage{csquotes}
\usepackage{ae,aecompl}
\usepackage[abs]{overpic}
\usepackage[normalem]{ulem}
\usepackage[font=footnotesize]{caption}

\usepackage[colorlinks]{hyperref}
\usepackage{cleveref}

\usepackage[maxcitenames=2, giveninits=true, isbn=false, doi=false]{biblatex}
\renewbibmacro*{url+urldate}{}
\DeclareFieldFormat{labelnumberwidth}{#1\adddot}

\usepackage{microtype}

\usetikzlibrary{arrows,calc,decorations.markings,math,arrows.meta}
\usetikzlibrary{shapes.geometric,patterns}

\newcommand{\op}[1]{\operatorname{\text{#1}}}

\newcommand{\Div}{\op{Div}}
\newcommand{\Grad}{\op{Grad}}
\newcommand{\cC}{\set{C}}
\newcommand{\cS}{\set{S}}

\newcommand{\cV}{\set{V}}
\newcommand{\set}[1]{\mathcal{#1}}
\newcommand*\diff{\mathop{}\!\mathrm{d}}

\providecommand{\keywords}[1]{\textbf{Keywords:} #1}

\newtheorem{remark}{Remark}

\begin{document}

\pagestyle{fancyplain}

\lhead [\fancyplain{}{\emph{A. Mota, D. Koliesnikova, I. Tezaur, J. Hoy}}]
{\fancyplain{}{\emph{A. Mota, D. Koliesnikova, I. Tezaur, J. Hoy}}}

\rhead
[\fancyplain{}{\emph{Contact via the Schwarz Alternating Method}}]
{\fancyplain{}{\emph{Contact via the Schwarz Alternating Method}}}

\title{A Fundamentally New Coupled Approach to Contact Mechanics via the Dirichlet-Neumann Schwarz Alternating Method}

\author{\Large Alejandro Mota$^1$\thanks{Email: amota@sandia.gov},
  Daria Koliesnikova$^1$, Irina Tezaur$^1$, Jonathan Hoy$^2$
  \\
  \\
  $^1$Sandia National Laboratories\\
  Livermore, CA 94550, USA\\
  \\
  $^2$Air Force Research Laboratory\\
  Edwards, CA 93523, USA
}

\date{\today}

\maketitle

\begin{abstract}
  Contact phenomena are essential in understanding the behavior of mechanical systems. Existing computational approaches for simulating mechanical contact often encounter numerical issues, such as inaccurate physical predictions, energy conservation errors, and unwanted oscillations. We introduce an alternative technique, rooted in the non-overlapping Schwarz alternating method, originally developed for domain decomposition. In multi-body contact scenarios, this method treats each body as a separate, non-overlapping domain and prevents interpenetration using an alternating Dirichlet-Neumann iterative process. This approach has a strong theoretical foundation, eliminates the need for contact constraints, and offers flexibility, making it well-suited for multiscale and multiphysics applications.

  We conducted a numerical comparison between the Schwarz method and traditional methods like Lagrange multiplier and penalty methods, focusing on a benchmark impact problem. Our results indicate that the Schwarz alternating method surpasses traditional methods in several key areas: it provides more accurate predictions for various measurable quantities and demonstrates exceptional energy conservation capabilities. To address the issue of unwanted oscillations in contact velocities and forces, we explored various algorithms and stabilization techniques, ultimately opting for the naïve-stabilized Newmark scheme for its simplicity and effectiveness. Furthermore, we validated the efficiency of the Schwarz method in a three-dimensional impact problem, highlighting its innate capacity to accommodate different mesh topologies, time integration schemes, and time steps for each interacting body.

\end{abstract}


\section{Introduction} \label{sec:intro}

Understanding the behavior of mechanical systems during contact is crucial, whether those systems are engineered or naturally occurring. This understanding is particularly important for various environmental conditions, such as when surfaces touch, slide, or experience impacts. While the methods for simulating the general behavior of mechanical systems are well-established, the simulation of contact mechanics still poses considerable challenges due to the complex nature of contact phenomena, including nonlinearities and lack of smoothness in calculations \cite{Wriggers:2006, Acary:2016, Doyen:2011, Krause:2012}.

In the existing literature, contact problems are typically divided into two main categories \cite{DiStasio:2019}: persistent contact problems that exhibit consistent velocities, forces, and accelerations, and impact problems characterized by abrupt changes in velocity, leading to locally undefined forces and accelerations. Another way to categorize contact problems is by the nature of the surfaces involved: smooth contact involves surfaces without any sharp edges, kinks, or corners, while non-smooth contact pertains to surfaces with such irregularities \cite{Kane.etal:1999}. This work concentrates on the dynamics of impact problems involving smooth surfaces. Nevertheless, we posit that our approach could be extended to both persistent and non-smooth contact scenarios, which we aim to investigate in future work.

Traditional numerical simulations of contact mechanics involve two main steps. The first step is the contact detection phase, which relies on proximity search algorithms to identify when contact between domains occurs. These algorithms have garnered attention, primarily due to their importance in other fields like video game development and robotics \cite{Jimenez.etal:2001, Kockare.etal:2007, Haddadin.etal:2017}. The second step is enforcing contact constraints to prevent interpenetration between bodies. Current methods typically require solving for all bodies involved in contact as a monolithic system, limiting the flexibility to adapt individual domain-specific numerical features, such as time steps or time integrators. Additionally, these methods often necessitate problem-specific parameters that can influence their accuracy and stability. Issues such as poor energy conservation and artificial oscillations remain persistent concerns \cite{Doyen:2011, Krause:2012}. Another challenge is the discretization of the contact interface, where special consideration is needed for different types of discretizations \cite{Francavilla:1975, Zavarise:2009, Puso:2004, Puso.Laursen:2004, Yang:2008}.

Thus, despite some improvements and development of new techniques over the years, designing a numerical method that is accurate, efficient, stable, and conserves energy remains a complex and ongoing research topic in the contact dynamics community.

This paper presents a fundamentally new approach for simulating mechanical contact, based on the Schwarz alternating method \cite{Schwarz:1870}. Originally designed for domain decomposition, this method is well-suited for addressing mechanical contact. It allows for treating each body as a separate, non-overlapping domain and employs an alternating Dirichlet-Neumann iterative process for contact enforcement. The algorithm leverages existing well-established finite element methods, utilizing standard Dirichlet and Neumann boundary conditions and classic time integration schemes like the Newmark-$\beta$ time integrator. The primary goal is to develop a robust, precise, and efficient method for contact mechanics, which can be integrated seamlessly into existing production-level simulation codes where intrusive modifications are not feasible. Before delving into the new methodology and the paper's main contributions (Section \ref{sec:contrib}), a brief review of popular traditional methods for contact enforcement is provided (Section \ref{sec:conventional}). The computational hurdles associated with energy conservation and the emergence of undesired oscillations in numerical contact enforcement are subsequently addressed in Section \ref{sec:stability_and_energy_conservation}.

\subsection{Overview of several conventional techniques for contact enforcement}
\label{sec:conventional}

Traditional approaches to contact mechanics integrate contact constraints into the energy variational framework of the problem as elaborated in Section \ref{sec:conventional_contact}. These constraints can be enforced either strongly or weakly using a variety of methods. Prominent among these are the penalty method \cite{Hunek:1993, Faltus:2020}, the Lagrange multiplier method \cite{Carpenter:1991, Belytschko.Xiao:2003}, the augmented Lagrangian method \cite{Belytschko.Xiao:2003, Simo:1992, Zienkiewicz.etal:2005}, and the Nitsche method \cite{Wriggers:2008, Mlika:2018, Chouly:2019, Chouly:2015}. A brief overview of the pros and cons of these techniques is presented here; for in-depth technical details, readers are referred to seminal works \cite{Simo:1985, Oden:1984, Bathe:1985, Gallego:1989, Wriggers:1985, Simo:1992, Carpenter:1991, Laursen:2003, Wriggers:2006, Gilardi:2002} and references cited therein.

The penalty approach \cite{Hunek:1993, Mota.etal:2003, Wriggers.Rust.Reddy:2016, Faltus:2020} serves as a straightforward technique for handling mechanical contact problems. It imposes a contact force proportional to the penetration distance via a penalty coefficient $\tau$. Due to its simplicity, the method is readily integrable into pre-existing computational frameworks. However, the method's precision and robustness are highly contingent on the selection of the penalty coefficient, which is inherently problem-specific. Employing overly small values for $\tau$ can lead to significant interpenetration and consequently to unreliable solutions. Conversely, choosing excessively high values for $\tau$ can undermine numerical stability and introduce fluctuations or errors in the resulting contact forces, both issues linked to ill-conditioning \cite{Wriggers.Zavarise:2004}.

In the Lagrange multiplier method \cite{Carpenter:1991, Belytschko.Xiao:2003, Popp.etal:2012, Wriggers.Rust.Reddy:2016}, contact constraints are weakly enforced via Lagrange multipliers within a mixed formulation. As a result, in contrast to the penalty method, the method guarantees exact satisfaction of contact conditions, thus ensuring consistency. Moreover, the approach is free from the requirement of tuning empirical parameters. Nevertheless, the Lagrange multiplier technique comes with its own set of challenges. Specifically, the design of the Lagrange multiplier finite element space must fulfill the \emph{inf-sup} condition \cite{Boffi.Brezzi.Fortin:2013}. Also, the integration of this approach into existing high-performance computing (HPC) platforms, like Sierra's Solid Mechanics (Sierra/SM) software \cite{Sierra:2011}, can be a complex endeavor. The formulation further results in an indefinite saddle point problem at the discrete level, often complicating numerical resolution and potentially necessitating specialized preconditioning techniques.

The augmented Lagrangian approach is suggested in literature as a composite strategy, encapsulating the merits of both the Lagrange multiplier and penalty methods \cite{Belytschko.Xiao:2003, Simo:1992, Zienkiewicz.etal:2005, Alart:1991, Glowinski:1989, Solberg.Jones.Papadopoulos:2007, Puso.etal:2008}. Functioning as a synthesis of these techniques, the method iteratively refines contact forces—essentially Lagrange multipliers—utilizing a penalty-based algorithm. Advantages of the augmented Lagrangian strategy over both its parent methods are numerous: it minimizes the need for parameter calibration, effectively mitigates issues of ill-conditioning that may afflict a penalty method \cite{Wriggers.Zavarise:2004}, and imposes contact constraints exactly as the Lagrange multiplier method, albeit without requiring the solution of a larger system. While the \emph{inf-sup} condition does inherently factor into the augmented Lagrangian formulation, it can be circumvented via a successive, independent update of both primal and dual degrees of freedom, facilitated by the application of Uzawa's algorithm \cite{Mlika:2018}. Nevertheless, the method is not without drawbacks: it exhibits a linear rate of convergence for the pressure field, undermining its overall quadratic convergence \cite{Boffi.Brezzi.Fortin:2013}, and presents challenges when integrating it into existing computational codes.

Emerging in recent literature, the Nitsche method \cite{Wriggers:2008, Mlika:2018, Chouly:2019, Chouly:2015} offers an alternative contact mechanism. This method originates from a unique variational formulation by \textcite{Nitsche:1971}. It results in a matrix equation for the primary variables that closely resembles the augmented Lagrangian formulation. Much like the augmented Lagrangian approach, the Nitsche method modifies the energy functional under optimization by incorporating both a contact constraint term and a stabilization term. The stabilization term is specifically added to mitigate the ill-conditioning in the global equation system formed by the Nitsche method. The interrelationship between the augmented Lagrangian and Nitsche methods is extensively discussed in \textcite{Wriggers:2008, Mlika:2018, Chouly:2019, Chouly:2015}. In essence, the Nitsche formulation can be viewed as a variant of the augmented Lagrangian method, where contact stresses are introduced as a weighted average \cite{Wriggers:2008}. Consequently, both methods share a majority of their respective strengths and weaknesses.

\subsection{Contributions and differentiating features}
\label{sec:contrib}

The new approach proposed in this paper extends our prior contributions in Schwarz multiscale coupling \cite{Mota:2017, Mota:2022, Barnett:2022} to address various complexities in numerical simulations of contact phenomena. Unlike traditional methods that integrate contact constraints directly into the variational formulation, as discussed in Section \ref{sec:conventional}, the Schwarz alternating method considers each body as an independent, non-overlapping domain. Interpenetration is prevented through an iterative Dirichlet-Neumann procedure. This technique is grounded on a rigorous theoretical framework and allows for different meshes, material models, solvers, and time-integration schemes for each participating body, as demonstrated in our earlier work \cite{Mota:2017, Mota:2022}. These advantages are particularly valuable in scenarios involving multiscale and multiphysics contact situations. Moreover, the method's foundation on a straightforward Dirichlet-Neumann iteration allows for easy integration into existing production codes without necessitating significant modifications to their infrastructure. While our newly proposed technique bears similarities to various existing domain decomposition algorithms focused on mechanical contact \cite{Dostal:2009, Dostal:2019, Blanze:1996, Oumaziz:2018}, it distinguishes itself in several key aspects detailed below.

The Total Finite Element Tear and Interconnect (TFETI) method by \textcite{Dostal:2009, Dostal:2019} also relies on a non-overlapping domain decomposition around a contact boundary, much like our approach. Nevertheless, our method employs a straightforward Dirichlet-Neumann iterative process between the contacting subdomains, facilitating a minimally invasive implementation. In contrast, the TFETI method requires formulating and solving a Quadratic Programming (QP) problem with inequality constraints, while enforcing prescribed displacements through a Lagrange multiplier framework.

Another approach for contact problems that also utilizes domain decomposition is the LArge Time INcrement (LATIN) method \cite{Blanze:1996, Oumaziz:2018}. This method employs a modular sub-structured strategy, alternating between independent and parallel global linear simulations over non-overlapping subdomains and local nonlinear problems at the subdomain interfaces. In \textcite{Blanze:1996}, the computation of contact forces is achieved through a mismatch in displacements at the boundaries between subdomains. Meanwhile, \textcite{Oumaziz:2018} introduce a technique to circumvent the use of Robin boundary conditions, commonly found in LATIN methods, by incorporating an additional layer of elements around the subdomain boundaries. In contrast to these methods, our approach leverages alternating Neumann and Dirichlet boundary conditions on the contact interface. Consequently, the sole similarity between our method and \textcite{Blanze:1996, Oumaziz:2018} lies in the overarching concept of domain decomposition.

Our methodology also exhibits similarities with the work by \textcite{Eck:2002}, who introduce a contact-Neumann Schwarz-type iterative algorithm designed for frictional contact problems. The primary distinction between our technique and that of \textcite{Eck:2002} is that the latter requires solving a standard contact problem in one of the subdomains using conventional contact constraints. On the other hand, our approach utilizes a straightforward Dirichlet-Neumann iteration, allowing us to sidestep the need for traditional contact constraints and their conventional enforcement mechanisms.

Lastly, although the core iterative procedure in our methodology bears similarity to the strategy employed by \textcite{Bochev:2017}, it is important to note that the focus of their work in is on interface coupling, not on contact problems. Both methods utilize a non-overlapping alternating Dirichlet-Neumann Schwarz iteration, but \textcite{Bochev:2017} target different application areas.

\subsection{Stability and energy conservation}
\label{sec:stability_and_energy_conservation}

As previously noted, solving contact dynamics problems is particularly challenging due to their inherently nonlinear nature. Conventional contact algorithms coupled with standard time integration methods—such as Newmark-$\beta$, generalized $\alpha$, or Hilber–Hughes–Taylor (HHT) schemes—frequently lead to computational issues. These can range from undesirable oscillations in specific quantities of interest at best, to severe problems like poor energy conservation or even energy blow-ups at worst \cite{Krause:2012, Wriggers:2006, Acary:2016, Doyen:2011}.

\subsubsection{Energy conservation properties}
\label{sec:energy_conserv}

As evidenced in \textcite{Krause:2012}, the incorporation of contact constraints via conventional methods can adversely affect the system's conservation properties, such as energy and momentum conservation. The authors' findings make it clear that for energy conservation, the numerical scheme must ensure zero relative velocities in the normal direction at the contact interface. Achieving this condition in a fully discrete setting with traditional contact methods is problematic. Numerous algorithms have been developed to address these challenges \cite{Jean:1999, Moreau:1999, Laursen:1997, Laursen:2002, Deuflhard:2008, Khenous:2008, Hughes:1978, Simo:1992}, though these often require complex and invasive modifications to existing methods.

In contrast, the Schwarz contact method introduced in this paper naturally enforces this crucial condition via its specific contact boundary conditions. As further detailed in Section \ref{sec:schwarz}, the velocities at the contact boundaries are inherently equal due to the imposition of Dirichlet boundary conditions. Our numerical experiments, presented in Sections~\ref{sec:results}-\ref{sec:results_3D}, confirm that our Schwarz contact method excels at conserving energy without requiring any supplemental algorithms—a feat often elusive for traditional methods.

\subsubsection{Techniques for reducing artificial oscillations}
\label{sec:chatter_intro}

One pervasive issue that appears in simulations of contact or impact problems is the emergence of undesirable oscillations in both contact forces and velocities. This stability issue is not exclusive to traditional methods; it is also present in our Schwarz contact method. To mitigate these spurious oscillations, various strategies have been developed. These can be categorized into several groups: contact enforcement techniques \cite{Carpenter:1991, Jean:1999, Moreau:1999, Paoli:2001, Paoli:2002, Paoli:2002a}, mass
redistribution methods \cite{Khenous:2008, Doyen:2009, Hager:2008, Monjaraz:2022}, stabilization methods \cite{Doyen:2011, Kane.etal:1999, Deuflhard:2008}, and adapted time integration schemes \cite{Fung:2003, Chaudhary.Bathe:1986, Chung.Lee:1994, Tchamwa:1999}.

Contact enforcement techniques, commonly referred to as non-smooth contact dynamics methods, were initially developed by \textcite{Jean:1999, Moreau:1999} within the context of rigid body dynamics. The objective is to implement a velocity-based contact law to completely describe the impact problem. Used in conjunction with the $\theta$-method for time integration, this approach generally ensures stability in both displacement and velocity, as well as acceptable energy behavior. Building on the work of \textcite{Jean:1999, Moreau:1999}, \textcite{Paoli:2001, Paoli:2002, Paoli:2002a} proposed a scheme that employs a similar contact law, but formulated in terms of position. When specific parameters are chosen, this approach resembles the central difference (explicit) scheme by \textcite{Carpenter:1991}. The integration of additional impact laws into the Schwarz contact method appears to be non-trivial and fairly invasive, which is why we have opted not to consider these techniques in our paper.

Mass redistribution methods have been introduced by \textcite{Khenous:2008, Doyen:2009, Hager:2008}. These strategies exploit the observation that numerical instabilities and oscillations often originate from the inertia of the contact boundary. To address this, the authors suggest formulating a new mass matrix where the mass is re-allocated in a way that removes mass from the contact boundary nodes while preserving the original mass matrix's invariants (e.g., total mass, center of gravity, and moments of inertia). Various formulations for this new mass matrix exist. In \textcite{Khenous:2008, Doyen:2009}, the authors recommend nullifying the coefficients of the mass matrix associated to displacements of the contact boundary. A subsequent modified mass matrix is then generated by solving an optimization problem that constrains the total mass, center of gravity, and moments of inertia. An alternative approach by \textcite{Hager:2008} focuses on reconstructing the mass matrix using quadrature formulas. This method is shown to be a stable interpolation-operator and standard-quadrature hybrid that maintains the original mass matrix's properties. Despite proofs confirming the well-posedness of problems with the new mass matrix, these techniques introduce new complexities, such as the need to solve additional optimization problems, or potential alteration of original mass matrix properties like sparsity patterns. Consequently, these methods are not considered here.

Stabilization methods \cite{Kane.etal:1999, Deuflhard:2008, Doyen:2011} share similarities with mass redistribution techniques in that they aim to mitigate instabilities originating from the contact boundary's inertia. Nevertheless, these methods achieve stabilization without altering the mass matrix. Various procedures are employed to eliminate the non-physical components of the boundary forces. For instance, \textcite{Kane.etal:1999} treat the contact accelerations in a fully implicit manner, \textcite{Deuflhard:2008} modify the predictor step, and \textcite{Doyen:2011} set the contact accelerations to zero. Given their generic nature, these techniques present promising avenues for incorporation into our Schwarz contact method to address the issue of spurious oscillations, as discussed in Section \ref{sec:chatter}.

An alternative set of approaches focuses on specialized time integration methods designed to eliminate unwanted high-frequency oscillations. Commonly, these methods employ algorithmic damping or numerical dissipation. Examples include the Wilson $\theta$-method, Houbolt method, generalized $\alpha$ method, collocation method, and modified variants of the Newmark-$\beta$ scheme \cite{Fung:2003, Chung.Hulbert:1993, Wriggers:2006}. In our work, we aim to utilize the well-established Newmark-$\beta$ integrator; thus, only schemes built on this integrator are examined. For more details, see Section \ref{sec:chatter}. Due to their more invasive implementation requirements and their unavailability in most production codes of interest to us, other time integration techniques are not considered in our investigation.

A detailed discussion on some of these time integration methods, their algorithmic characteristics, and a numerical comparison to identify the best fit for the Schwarz contact method are presented in Section \ref{sec:chatter}.

\subsection{Organization of the paper} \label{sec:organization}

This paper is organized as follows:
Section~\ref{sec:intro} introduces the problem of mechanical contact, highlights its significance in modeling and simulation, and offers an overview of existing work. The section also outlines how we intend to approach contact modeling using the Schwarz method.
Section~\ref{sec:mechanics} establishes the foundational variational framework for the general solid mechanics problem, incorporating both spatial and temporal discretizations.
Section~\ref{sec:conventional_contact} provides the fundamental framework of conventional contact methods, setting up a comparative stage for our Schwarz-based approach.
Section~\ref{sec:schwarz} delves into our innovative Schwarz alternating contact formulation, which leverages a decomposition into subdomains and alternating Dirichlet-Neumann (position-traction) boundary conditions.
Section~\ref{sec:results} presents numerical results for a well-known one-dimensional (1D) impact benchmark with an analytical solution. This 1D model serves as a simplified but effective test bed for evaluating the merits of the Schwarz algorithm, eliminating complexities inherent to multi-dimensional analyses.
Section~\ref{sec:chatter} addresses the method's limitations, presenting a specific strategy designed to minimize spurious oscillations in certain key metrics without sacrificing its accuracy or energy conservation features.
Section~\ref{sec:results_3D} extends the validation to three-dimensional impact scenarios, corroborating the method's robustness across various performance indicators, including its adaptability to diverse mesh topologies and time-stepping schemes.
Section~\ref{sec:conc} concludes by summarizing the key findings and pointing to potential avenues for future research.
\section{Solid mechanics problem formulation} \label{sec:mechanics}

To establish our notation, we begin with the standard variational formulation for finite deformation mechanics without incorporating contact forces. Consider a time interval $I := \{ t \in [t_0, t_N] \}$, where $t_0 < t_N$ and $t_0, t_N \in \mathbb{R}$. Let $\Omega \subset \mathbb{R}^3$ be a regular open set representing the body, and let its motion be governed by the mapping $~x = ~\varphi(X, t): \Omega \times I \rightarrow \mathbb{R}^3$. Here, $~X \in \Omega$ represents the material points and $t \in I$ is the time variable.

The boundary of the body $\partial \Omega$ is partitioned into a position-prescribed boundary $\partial_{~\varphi} \Omega$ and a traction-prescribed boundary $\partial_{~T} \Omega$, both with mutually exclusive domains: $\partial_{~\varphi} \Omega \cap \partial_{~T} \Omega = \emptyset$. The unit normal to the boundary is denoted as $~N$.

The Dirichlet boundary conditions, or the prescribed positions, are given by $~\chi: \partial_{~\varphi} \Omega \times I \rightarrow \mathbb{R}^3$. The Neumann boundary conditions, or the prescribed tractions, are given by $~T: \partial_{~T} \Omega \times I \rightarrow \mathbb{R}^3$.

The deformation gradient $~F$ is defined as $~F := \nabla ~\varphi$. Let $~x_{t_0} \equiv ~X: \Omega \rightarrow \mathbb{R}^3$ and $~v_{t_0}: \Omega \rightarrow \mathbb{R}^3$ represent the initial position and velocity at time $t_0$, respectively. Additionally, let $\rho ~B: \Omega \rightarrow \mathbb{R}^3$ denote the body force, where $\rho$ is the mass density in the reference configuration.

We introduce the kinetic energy of the body as
\begin{equation}\label{eq:kinetic-energy}
  T(\dot{~\varphi})  :=  \frac{1}{2} \int_{\Omega} \rho \dot{~\varphi} \cdot \dot{~\varphi} \ \diff V,
\end{equation}
and its potential energy as
\begin{equation}\label{eq:potential-energy}
  V(~\varphi)  :=  \int_{\Omega} A(~F, ~Z) \ \diff V  -  \int_{\Omega} \rho ~B \cdot ~\varphi \ \diff V  -
  \int_{\partial_{~T} \Omega} ~T \cdot ~\varphi \ \diff S,
\end{equation}
where $A(~F, ~Z)$ is the Helmholtz free-energy density and $~Z$ represents a set of internal variables. The Lagrangian of the body is then defined as
\begin{equation}\label{eq:Lagrangian}
  L(~\varphi, \dot{~\varphi}) := T(\dot{~\varphi}) - V(~\varphi),
\end{equation}
which leads to the action functional
\begin{equation}\label{eq:action-functional}
  S[~\varphi] := \int_{I} L(~\varphi, \dot{~\varphi}) \ \diff t.
\end{equation}

According to Hamilton's Variational Principle, the equation of motion is derived by identifying the critical point of the action functional $S[~\varphi]$ within the Sobolev space $W_2^1(\Omega \times I)$. This space consists of all functions that are both square-integrable and possess square-integrable first derivatives. The endpoints of the deformation mapping are fixed at $t_0$ and $t_N$. Define
\begin{equation} \label{eq:S}
    \cS :=  \left\{~\varphi \in W_2^1(\Omega \times I)  :   ~\varphi = ~\chi \; \mathrm{on} \; \partial_{~\varphi} \Omega \times I;
    ~\varphi = ~x_{t_0} \; \mathrm{on} \; \Omega \times t_0;
    ~\varphi = ~x_{t_N} \; \mathrm{on} \; \Omega \times t_N \right\}
\end{equation}
and
\begin{equation} \label{eq:V}
    \cV :=  \left\{~\xi \in W_2^1(\Omega \times I) :  ~\xi = ~0 \; \mathrm{on} \; \partial_{~\varphi} \Omega \times I   \cup
    \Omega \times t_0   \cup  \Omega \times t_N \right\}.
\end{equation}
Here, $~\xi$ serves as a test function. This formulation leads to
\begin{equation} \label{eq:var}
  \begin{split}
    \begin{aligned}
      \delta S := D S[~\varphi] (~\xi)
      & =
      \int_{I} \left(
        \frac{\partial L}{\partial ~\varphi} \cdot ~\xi
        +
        \frac{\partial L}{\partial \dot{~\varphi}} \cdot \dot{~\xi}
      \right) \ \diff t
      =
      \int_{I} \left(
        \frac{\partial L}{\partial ~\varphi}
        -
        \frac{\diff}{\diff t}
        \frac{\partial L}{\partial \dot{~\varphi}}
      \right) \cdot ~\xi \ \diff t
      \\
      & =
      \int_{I} \left[
        \int_{\Omega} \left(
        \rho ~B \cdot ~\xi - ~P : \Grad ~\xi
        +
        \rho \dot{~\varphi} \cdot \dot{~\xi}
        \right) \ \diff V
        +
        \int_{\partial_{~T} \Omega} ~T \cdot ~\xi \ \diff S
      \right] \ \diff t
      \\
      & =
      \int_{I} \left[
        \int_{\Omega} \left(
          \Div ~P + \rho ~B - \rho \ddot{~\varphi}
        \right) \cdot ~\xi \ \diff V
        +
        \int_{\partial_{~T} \Omega} ~T \cdot ~\xi \ \diff S
      \right] \ \diff t
      =
      0,
    \end{aligned}
  \end{split}
\end{equation}
where $~P = \partial A / \partial ~F$ denotes the first Piola-Kirchhoff stress.
The Euler-Lagrange equation corresponding to \eqref{eq:action-functional} is
then
\begin{equation} \label{eq:euler-lagrange}
  \Div ~P + \rho ~B
  =
  \rho \ddot{~\varphi}
  \quad \mathrm{in} \quad
  \Omega \times I,
\end{equation}
with initial conditions
\begin{equation} \label{eq:initial-conditions}
  \begin{split}
    \begin{aligned}
      ~\varphi(~X, t_0)  & =  ~x_{t_0}   & \mathrm{in} & \quad   \Omega,      \\
      \dot{~\varphi}(~X, t_0)    & =  ~v_{t_0}
      & \mathrm{in} & \quad
      \Omega,
    \end{aligned}
  \end{split}
\end{equation}
and boundary conditions
\begin{equation} \label{eq:boundary-conditions}
  \begin{split}
    \begin{aligned}
      ~\varphi(~X, t)
      & =
      ~\chi
      & \mathrm{on} & \quad
      \partial_{~\varphi} \Omega  \times I,
      \\
      ~P ~N
      & =
      ~T
      & \mathrm{on} & \quad
      \partial_{~T} \Omega \times I.
    \end{aligned}
  \end{split}
\end{equation}
Upon discretizing the variational form \eqref{eq:var} in space using the classical Galerkin finite element method (FEM) \cite{Hughes:2000}, we arrive at the semi-discrete matrix equation
\begin{equation} \label{eq:semidiscrete}
    ~M \ddot{~u} + ~f^{\mathrm{int}}= ~f^{\mathrm{ext}}.
\end{equation}
In \eqref{eq:semidiscrete}, $~M$ is the mass matrix, $~u := ~\varphi(~X, t) - ~X$ represents the displacement vector, $\ddot{~u}$ is the acceleration vector, $~f^{\mathrm{ext}}$ denotes applied external forces, and $~f^{\mathrm{int}}$ accounts for internal forces arising from mechanical and other effects within the material. In problems involving mechanical contact, $~f^{\mathrm{ext}}$ will also include a contact force contribution.

To derive a fully discrete problem, a time-integration scheme is applied to \eqref{eq:semidiscrete}. A commonly used scheme for solid mechanics problems is the Newmark-$\beta$ method \cite{Rao:2017}. Further details on both implicit and explicit Newmark-$\beta$ algorithms are discussed in Section \ref{sec:chatter}.

\section{Conventional contact methods} \label{sec:conventional_contact}

The aim of this section is to outline the fundamental framework within which traditional contact methods operate. This serves to highlight the distinctions between such conventional methods and the Schwarz contact approach. Specifically, we focus on the contrast between formulating contact as a constraint in conventional methods and treating it as a coupled problem in the Schwarz method.

\subsection{Contact constraints} \label{sec:constraint}

Traditional contact methods, as outlined in Section \ref{sec:conventional}, often rely on constraint-based formulations tied to the potential energy \eqref{eq:potential-energy}. To elaborate, let $\cC$ represent the set of permissible configurations $~\varphi$ where no interpenetration occurs \cite{Kane.etal:1999}. The set's \emph{indicator function} is defined as
\begin{equation} \label{eq:indicator}
  \begin{aligned}
    I_{\cC}(~\varphi) :=
    \begin{cases}
      0, & \text{if } ~\varphi \in \cC,
      \\
      \infty, & \text{if } ~\varphi \notin \cC.
    \end{cases}
    \end{aligned}
\end{equation}
This indicator function can be incorporated into the potential energy expression \eqref{eq:potential-energy} as follows
\begin{equation}\label{eq:potential-energy-indicator}
  V(~\varphi)
  :=
  \int_{\Omega} A(~F, ~Z) \ \diff V
  -
  \int_{\Omega} \rho ~B \cdot ~\varphi \ \diff V
  +
  \int_{\Omega} I_{\cC}(~\varphi) \ \diff V
  -
  \int_{\partial_{~T} \Omega} ~T \cdot ~\varphi \ \diff S.
\end{equation}
Conventional methods aim to satisfy the constraints, usually zero gap and/or zero gap rate, set by the indicator function via various approaches (see Section \ref{sec:conventional}). It should be noted, however, that these methods can lead to over-constraint, resulting in surface locking \cite{Puso:2004, Jones:2001}. Conversely, under-constraining may produce configurations where interpenetration occurs.

\subsection{Formulation and implementation}

While a comprehensive discussion on the formulation and implementation of traditional contact methods is beyond the scope of this paper (see Section \ref{sec:results}), we direct the interested reader to relevant literature mentioned in Section \ref{sec:conventional}. Our implementation of the conventional methods is built upon enforcing a zero-gap constraint between interacting bodies. Below is a brief description of the baseline conventional methods used in comparison to our proposed Schwarz alternating method.

\paragraph{Implicit Lagrange multiplier method.}
Our implementation relies on equation (6) from \textcite{Carpenter:1991}. As detailed in Section~\ref{sec:comparison}, we employ the Backward-Euler integration scheme for its reduced susceptibility to numerical instability.

\paragraph{Explicit Lagrange multiplier method.}
This method is based on equation (10) from \textcite{Carpenter:1991} and further expands on the work of \textcite{Katona:1985}.

\paragraph{Implicit penalty method.}
In this approach, the contact force $\lambda$ is proportional to the gap $g$, which measures the extent of overlap between the two interacting domains. The proportionality constant is the penalty parameter, $\tau$. The Newmark-$\beta$ method \cite{Newmark:1959} serves as the underlying integration scheme. Because the gap at the next time step is not known a priori, an iterative procedure is required to resolve it, continuing until convergence.

\paragraph{Explicit penalty method.}
This method mirrors the explicit Lagrange multiplier approach as laid out in \textcite{Carpenter:1991}. However, the contact force $\lambda$ is estimated as $\lambda \approx \lambda_n = \tau g$, where $\tau$ is the penalty parameter and $g$ is the overlapping distance between the domains, circumventing the need to solve for $\lambda$ as an independent variable.

\section{The Schwarz alternating method for contact} \label{sec:schwarz}

The focus of this paper is to present and assess a \emph{novel} strategy for simulating mechanical contact using the Schwarz alternating method. Originally introduced by \textcite{Schwarz:1870} in 1870, this method serves as an iterative domain decomposition technique. Our previous adaptations of the Schwarz method for continuum-to-continuum coupling in both quasistatic and dynamic solid mechanics are found in \cite{Mota:2017} and \cite{Mota:2022}.

In these prior works, the overarching concept is to partition a given physical domain $\Omega$ into $d$ overlapping subdomains, represented as $\Omega = \cup_{i=1}^d \Omega^i$. The governing partial differential equations (PDEs) are then solved within each of these subdomains iteratively, with information exchange facilitated via Dirichlet boundary conditions (BCs). It is important to note that the superscript $i$ designates quantities that are specific to the subdomain $\Omega^i$.

The Schwarz alternating method offers several advantages over traditional monolithic discretization methods. Among these are the ability to easily couple different mesh resolutions, different element types, and even diverse time-integration schemes without introducing any inaccuracies or artifacts. Additionally, the method comes with provable guarantees of convergence \cite{Mota:2017, Mota:2022}. These advantageous features make the Schwarz method particularly promising for applications in contact dynamics, a topic that will be explored next.

\subsection{Overlapping and non-overlapping Schwarz methods}

\begin{figure}
    \centering
      \subfigure[Overlapping Schwarz]{ \includegraphics[page=1,width=0.4\textwidth]
                      {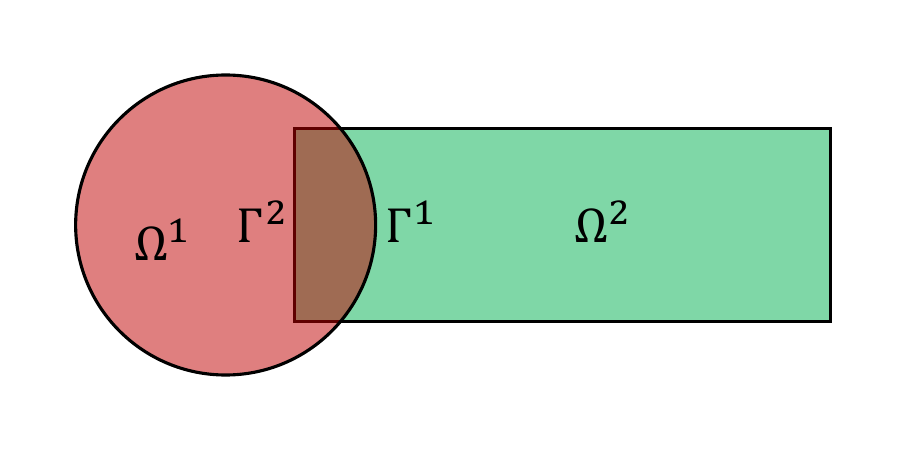} \label{fig:dd_overlapping} }
      \subfigure[Non-overlapping Schwarz]{ \includegraphics[page=2,width=0.4\textwidth]
                      {plots/schwarz_illustration.pdf} \label{fig:dd_nonoverlapping}}
    \caption{(a) Overlapping Schwarz method. Information is transferred between $\Omega^1$ and $\Omega^2$ through the Schwarz boundaries $\Gamma^1 = \partial \Omega^1 \cap \Omega^2$ and $\Gamma^2 = \partial \Omega^2 \cap \Omega^1$. (b) Non-overlapping Schwarz method. Information is relayed via a defined \emph{contact boundary}, $\Gamma$, which is the intersection of the boundaries of all domains in contact.}
    \label{fig:dd}
\end{figure}

Without loss of generality, consider a decomposition of $\Omega$ into two specific subdomains $\Omega^1$ and $\Omega^2$, as depicted in Figure \ref{fig:dd_overlapping}. Information is transferred between these subdomains through the Schwarz boundaries $\Gamma^1 = \partial \Omega^1 \cap \Omega^2$ and $\Gamma^2 = \partial \Omega^2 \cap \Omega^1$, which are also shown in Figure \ref{fig:dd_overlapping}.

Our primary goal is to develop an alternative contact method that avoids body interpenetration upon impact, circumventing the need for traditional contact constraints discussed in Section \ref{sec:conventional}. A key insight is that contact among $d$ domains can be formulated as a coupled problem involving $d$ non-overlapping subdomains. This concept is illustrated for a $d=2$ case in Figure \ref{fig:dd_nonoverlapping}.

In such scenarios, the non-overlapping version of the Schwarz alternating method—as described by \textcite{LionsNonOverlap:1988, Zanolli:1987}—can be effectively utilized to manage contact between domains. Interpenetration is precluded through an iterative Schwarz-based procedure, where information is relayed via a defined \emph{contact boundary}, $\Gamma$, which is the intersection of the boundaries of all domains in contact—mathematically, $\Gamma = \{ \partial \Omega^i \}_{i=1}^d \cap_{i \neq j}  \{ \partial \Omega^j \}_{j=1}^d$, as illustrated in Figure \ref{fig:dd_nonoverlapping} for $d=2$. For each individual domain $\Omega^i$, its respective contact boundary is designated as $\Gamma^i = \partial \Omega^i \cap \Gamma$. It is worth noting that the concept of a contact boundary becomes relevant only when actual contact occurs between the domains.

\begin{figure}
    \centering
      \includegraphics[trim={10mm 35mm 50mm 30mm},clip,width=0.8\textwidth]{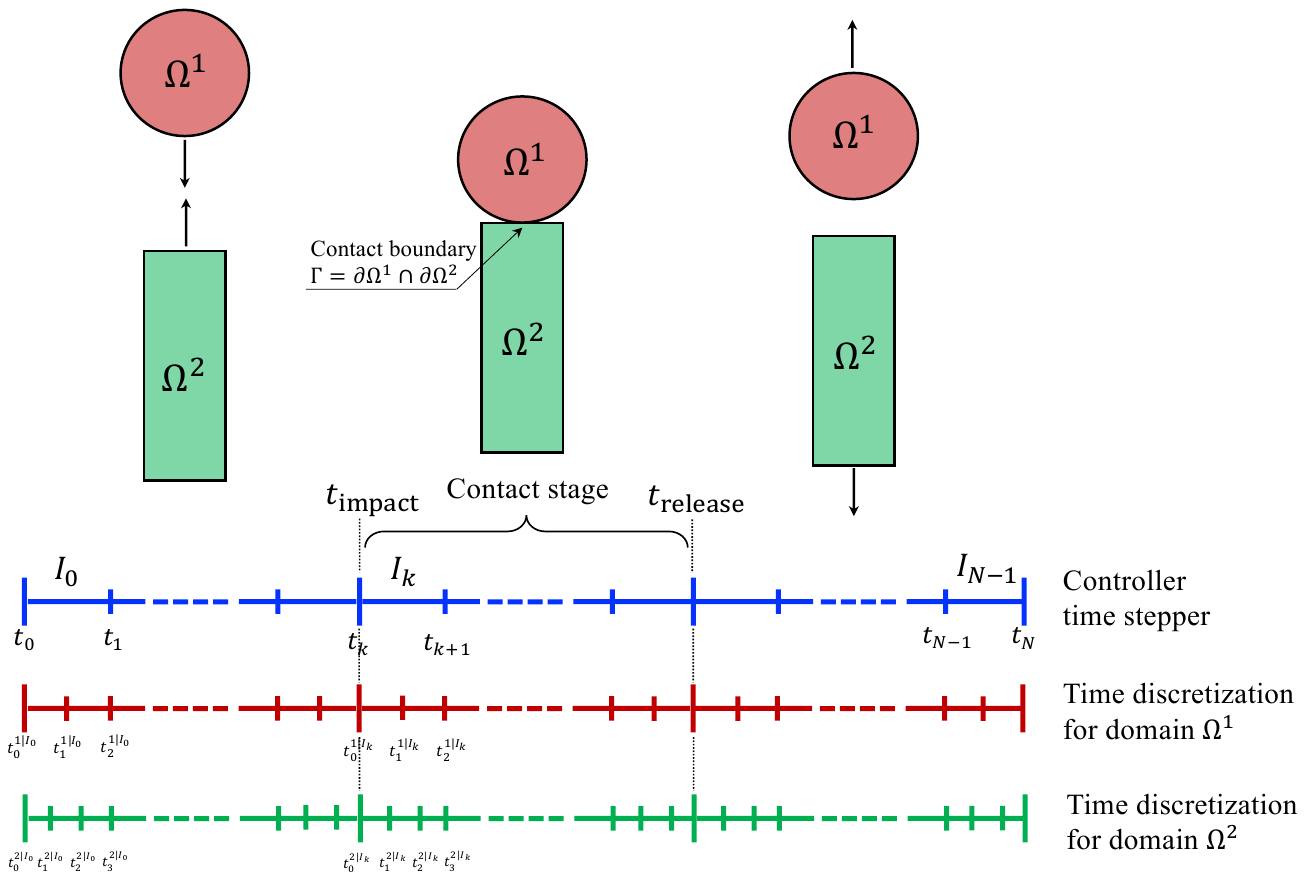}
         \caption{Contact for two subdomains $\Omega^1$ and $\Omega^2$. A global controller time-stepper partitions the simulation time into $N$ global intervals, denoted as $I_k$. The scheme also defines $N+1$ global time stops, $t_0 < t_1 < \ldots < t_{N} \in \mathbb{R}$. These global time stops serve as reference points for significant events, such as impact or release times in our context. In parallel, we also define local time integrators and time steps for each domain.}
        \label{fig:domains}
\end{figure}

In the following sections, we delve into both the theoretical framework and algorithmic implementation of the Schwarz alternating method as applied to contact mechanics. Specifically, we focus on a general scenario involving an arbitrary number $d$ of open, bounded domains—denoted as $\Omega^i$ and $\Omega^j$—that are disjoint, i.e., $\Omega^i \cap \Omega^j = \emptyset$ for $1 \leq i, j \leq d$ and $i \neq j$. These domains are in motion towards one another, as depicted in the left-most panel of Figure \ref{fig:domains}.

Adopting the framework from \textcite{Mota:2022}, we introduce a global controller time-stepper that partitions the simulation time into $N$ global intervals, denoted as $I_k$. The scheme also defines $N+1$ global time \emph{stops}, $t_0 < t_1 < \ldots < t_{N} \in \mathbb{R}$. Each interval $I_k$ is given by $I_k := \{t \in [t_{k}, t_{k+1}]\}$, where $k \in \{0, \ldots, N-1\}$ and $N \in \mathbb{N}$, as illustrated in Figure \ref{fig:domains}.  These global time stops serve as reference points for significant events, such as impact or release times in our context. In parallel, we also define local time integrators and time steps for each domain, as depicted in Figure \ref{fig:domains}. Importantly, our approach allows for different domains to employ distinct time integrators and step sizes.

For instance, in Figure \ref{fig:domains}, domain $\Omega^1$ utilizes time steps $\Delta t^1 := \frac{I_k}{l_\mathrm{max}}$ with $l_\mathrm{max} = 2$, and defines local time stops as $\{t^{1|I_{k}}_l\}_{l=0}^{l_\mathrm{max}} \in I_k$, i.e., $t^{1|I_{k}}_0$, $t^{1|I_{k}}_1$, and $t^{1|I_{k}}_2$ in this specific example. Conversely, domain $\Omega^2$ uses time steps $\Delta t^2 := \frac{I_k}{l_\mathrm{max}}$ where $l_\mathrm{max} = 3$, and sets 4 local stops $\{t^{2|I_{k}}_l\}_{l=0}^{l_\mathrm{max}} \in I_k$, namely $t^{2|I_{k}}_0$, $t^{2|I_{k}}_1$, $t^{2|I_{k}}_2$, and $t^{2|I_{k}}_3$.

\subsection{Impact and release detection criteria} \label{sec:contact_criteria}

A critical aspect of any contact algorithm is to dynamically identify the timing and domains involved in impact or release events, as well as their associated contact boundaries. These elements are not known \emph{a priori} and must be determined during the simulation.

In this work, we employ a Boolean variable called \emph{active contact} to indicate the contact state of the system for each global time interval $I_{k} := \{t \in [t_{k}, t_{k+1}]\}$. The determination of this variable relies on a triad of criteria:
\begin{itemize}
\item \emph{Persistence} indicates whether contact was active in the preceding time interval $I_{k-1}$. If \emph{persistence} is true then check for the \emph{compression} criterion; otherwise check for the \emph{overlap} criterion.
\item \emph{Compression} evaluates the contact tractions $~P ~N \cdot ~N$ at the contact boundary $\Gamma$. The criterion holds if these tractions are compressive (positive). Contact is released when these tractions are either zero or negative, thus making \emph{compression} false.
\item \emph{Overlap} checks for any interpenetration between domains. It is considered true when bodies begin to overlap or penetrate each other, that is, there is a positive interpenetrating volume $\mathrm{vol}\cap_{i=1}^d\Omega_i$.
\end{itemize}

In the context of this paper's impact/release detection framework, the \emph{overlap} criterion is chiefly responsible for detecting impacts, while the \emph{compression} criterion serves as a key indicator for release events. This approach is further elaborated in Algorithm \ref{alg:contact_criteria}.

\begin{algorithm}
  \caption{Impact and release conditions. Their determination relies on a triad of criteria: \emph{persistence} indicates whether contact was active in the preceding time interval; \emph{compression} evaluates the contact tractions at the contact boundaries to determine if they are compressive; and \emph{overlap} checks for any interpenetration between domains.}
  \footnotesize
  \begin{algorithmic}[1]
  \State \emph{persistence} at time interval $I_k$ $\gets$ \emph{active contact} at time interval $I_{k-1}$
   \If{\emph{persistence} == true}
    \Comment{contact status \emph{active} at the previous time interval}
   \State \emph{compression} $\gets ~P ~N \cdot ~N > 0$ anywhere on $\Gamma$
    \If{\emph{compression} == true}
    \State \emph{active contact} $\gets$ true
      \Comment{contact sustained}
     \Else{}
     \State \emph{active contact} $\gets$ false
     \Comment{release detected}
    \EndIf{}
    \Else{}
    \Comment{contact status \emph{not active} at the previous time interval}
     \State \emph{overlap} $\gets \mathrm{vol}\cap_{i=1}^d\Omega_i > 0$
     \If{\emph{overlap} == true}
      \State \emph{active contact} $\gets$ true
       \Comment{impact detected}
       \Else{}
     \State \emph{active  contact} $\gets$ false
     \Comment{no contact}
    \EndIf{}
   \EndIf{}
  \end{algorithmic}
  \label{alg:contact_criteria}
\end{algorithm}

Consequently, when \emph{active contact} is set to \emph{true}—indicating detected contact—the Schwarz contact algorithm is initiated, as discussed in Section \ref{sec:contact_algo}. It is important to recognize that the state of the system can change, and events like impact or release can occur between controller time stops within a given controller interval $I_k$. To accurately capture such dynamic changes, the contact criteria are evaluated at the conclusion of each controller time interval, specifically at stop $t_{k+1}$. If the \emph{active contact} status changes, computations for the relevant global time interval are re-executed. This procedure is encapsulated in Algorithm \ref{alg:contact}.

\begin{algorithm}
    \footnotesize
    \begin{algorithmic}[1]
        \State $k \gets 0$
        \State Initialize for $I_0$: \emph{active contact} $\gets$ false
        \For{$k \in [0, N-1]$ }   \Comment{controller time interval $I_k$}
            \If{\emph{active contact} == true}
                \Comment{contact}
                \State apply Schwarz contact method to compute solutions for domains  $\{\Omega^i\}_{i=1}^d$ in $I_k$ using Algorithm \ref{alg:schwarz}
            \Else{}
                \Comment{no contact}
                \State apply standard procedure to compute solutions for  \eqref{eq:euler-lagrange}-\eqref{eq:boundary-conditions} for domains $\{\Omega^i\}_{i=1}^d$ in $I_k$
            \EndIf{}
            \State  \emph{persistence} $\gets$ \emph{active contact}
            \Comment{save contact status from the previous time interval}
            \State update \emph{active contact} using Algorithm  \ref{alg:contact_criteria}
            \Comment{check contact criteria}
            \If{\emph{persistence} $\neq$ \emph{active contact}}
                \State repeat current time interval $I_k$: go to step 4
            \EndIf{}
            \State $k \gets k+1$
        \EndFor
    \end{algorithmic}
    \caption{Simulation workflow for Schwarz contact. Computation of the solution in each subdomain can be carried out by whatever means available. Before contact is detected, each body is treated independently. During the contact stage, the Schwarz alternating method is applied in order to handle the contact and prevent bodies from interpenetrating. When the contact criteria are no longer fulfilled, the dynamics are again resolved separately in each body without the Schwarz iterative process, and without Schwarz (contact) boundary conditions.}
    \label{alg:contact}
\end{algorithm}

The criteria presented in this paper bear both similarities and differences to the classic Signorini conditions, which are commonly used in contact mechanics. These conditions were introduced by \textcite{Signorini:1959} and further elaborated by \textcite{Fichera:1972}. Both sets of conditions aim to prevent interpenetration between contacting bodies and to ensure appropriate contact behavior. Like the Signorini conditions, our \emph{compression} and \emph{overlap} criteria deal with establishing a contact pressure and avoiding penetration. The \emph{compression} criterion, which ensures that the contact tractions are compressive, parallels the Signorini condition that requires the contact pressure to be non-negative. On the other hand, the \emph{overlap} criterion, which detects positive interpenetrating volume, can be viewed as an extension of Signorini's gap condition that requires the gap between bodies to be non-negative.

Our method, however, introduces a new dimension in the form of the \emph{persistence} criterion, which depends on the contact state of the previous time interval. This adds a temporal aspect to the contact model that is not directly accounted for in the Signorini conditions. Furthermore, our framework is tailored to dynamically identify impact or release events, making it more adaptive in comparison. While the Signorini conditions are often static and applied at each time step without considering the previous state, our criteria collectively contribute to a more nuanced and dynamic contact algorithm. This is especially pertinent for simulations involving complex geometries and varying contact conditions.

\begin{remark}
The Signorini conditions in contact mechanics can be considered as a specialized instance of the Karush-Kuhn-Tucker (KKT) conditions, which are more broadly employed in optimization to handle both inequality and equality constraints. In this context, the Lagrange multipliers in the KKT formulation can be interpreted as reaction forces at the points of contact. Therefore, the KKT conditions offer a more comprehensive mathematical framework that includes the Signorini conditions as a subset, specifically when addressing frictionless contact scenarios in the field of solid mechanics.
\end{remark}

\subsection{Simulation workflow} \label{sec:contact_algo}

Algorithm \ref{alg:contact} and Figure \ref{fig:domains} present a generic approach used to handle contact problems with the Schwarz alternating method.

Suppose that the bodies are not in contact at the initial time $t_0$ and are moving toward each other, as shown in the top left panel of Figure \ref{fig:domains}. Before contact is detected, each body is treated independently, that is, the governing PDEs are solved separately in each domain $\{\Omega^i\}_{i=1}^d$ with their own regular Dirichlet and/or Neumann boundary conditions (if applicable), see \eqref{eq:euler-lagrange}-\eqref{eq:boundary-conditions}.

The second step, referred to as contact stage, begins when \emph{active contact} is set to \emph{true} (contact between bodies is detected), see Algorithm \ref{alg:contact_criteria}. As shown in the top middle panel of Figure \ref{fig:domains}, we mark the beginning of the current time interval as $t_{\mathrm{impact}}$, and define the contact boundary between bodies which are involved in contact. During the contact stage, the Schwarz alternating method is applied in order to handle the contact and prevent bodies from interpenetrating. As stated earlier, computation of the solution in each subdomain can be carried out by whatever means available, in a close analog to the Schwarz alternating method for multiscale coupling \cite{Mota:2017,  Mota:2022, Barnett:2022}. Thus, each subdomain can advance its own solution within a given global time interval $I_k$ using its own time integrator, and its own time step, see Figure \ref{fig:domains}. The detailed presentation of the Schwarz contact iterative algorithm is given in Section \ref{sec:schwarz_formulation}.

The last phase starts when bodies separate, at $t_{\mathrm{release}}$, when the contact criteria are no longer fulfilled, as shown in the top right panel of Figure~\ref{fig:domains}. The dynamics are again resolved separately in each body without the Schwarz iterative process, and without Schwarz (contact) boundary conditions (the contact boundary is no longer defined).

\subsection{Formulation of the Schwarz contact method}
\label{sec:schwarz_formulation}

Initially shown by \textcite{LionsNonOverlap:1988, Zanolli:1987}, achieving a convergent Schwarz method for non-overlapping domains requires specialized transmission conditions. While Dirichlet-Dirichlet boundary conditions guarantee convergence for overlapping domains \cite{Mota:2017, Mota:2022} as seen in Figure \ref{fig:dd_overlapping}, convergence in non-overlapping domains can be attained through Robin-Robin \cite{LionsNonOverlap:1988, Gerardo-Giorda:2013, Gander:2008, Deng:2003, Lui:2001} or alternating Dirichlet-Neumann \cite{Zanolli:1987, Funaro:1988, Cote:2005, Kwok:2014} boundary conditions (Figure \ref{fig:dd_nonoverlapping}).

In our current work, we adopt the latter strategy, implementing alternating Dirichlet-Neumann boundary conditions for the mechanical problem defined by \eqref{eq:var}. Throughout each time interval $I_k$ within the contact phase, the Schwarz iterative method proceeds by solving Euler-Lagrange equation sequences, as per \eqref{eq:schwarz1}--\eqref{eq:schwarz_ics}, augmented by additional contact boundary conditions. For illustrative purposes, we focus on two bodies, $\Omega^1$ and $\Omega^2$.

Let $n =1, 2, \ldots$ represent the Schwarz iteration count, with $~\varphi^{(n)}$, $\dot{~\varphi}^{(n)}$, and $\ddot{~\varphi}^{(n)}$ denoting the position, velocity, and acceleration at the $n^{\mathrm{th}}$ iteration, respectively. The first Piola-Kirchhoff stress at the $n^{\mathrm{th}}$ iteration is given by $~P^{(n)}$.

Spatial transfer operators, denoted $\mathcal{P}^{~\varphi}_{(\Gamma^i \to \Gamma^j) \times I_k}$ and $\mathcal{P}^{~T}_{(\Gamma^i \to \Gamma^j) \times I_k}$, are used to project position and traction data from body $\Omega^i$ to the contact boundary $\Gamma^j$ within $\Omega^j$, $i \neq j$. These operators play a crucial role in the prescription of Schwarz contact boundary conditions. Additionally, the boundary conditions selected inherently fulfill the requirements for energy conservation, as detailed in Section \ref{sec:energy_conserv}.

In most scenarios, both spatial and temporal information transfers are necessary to specify appropriate contact boundary conditions. This enables the use of different time steps across different domains. Hence, Dirichlet and Neumann values undergo temporal interpolation before the spatial operators are applied, a procedure elaborated in Section \ref{sec:transfer}.

It is worth noting that all interacting bodies may have distinct constitutive models, material properties like mass density $\rho$, body forces, etc. These attributes should be individually defined for each body with a superscript $i$, although for ease of presentation, this specific notation is left out here.

To initiate the Schwarz iterations, data required for contact boundary conditions are transferred from the corresponding preceding time steps, further discussed in Section \ref{sec:transfer}. For clarity and simplicity in presentation, we focus on a two-body system in our equations. Although extensions to systems with $n$ bodies are conceptually direct, they inherently introduce greater complexity in mathematical representation. The equations for a two-body system are
\begin{align}
    \mathrm{Domain} \; \Omega^1:
    &
    \begin{cases}
        ~\Div ~P^{(n)} + \rho ~B = \rho \ddot{~\varphi}^{(n)}
        &
        \mathrm{in}  \; \Omega^1 \times I_k,
        \\
        ~\varphi^{(n)}(~X, t) = ~\chi
        &
        \mathrm{on} \; \partial_{\varphi}\Omega^1 \times I_k,
        \\
        ~P^{(n)}~N = ~T
        &
        \mathrm{on} \; \partial_{~T}\Omega^1 \times I_k,
        \\
        ~\varphi^{(n)}(~X, t) = \mathcal{P}^{~\varphi}_{\Gamma^2 \to \Gamma^1}[~\varphi^{(n-1)}(\Omega^2, t)]
        &
        \mathrm{on} \; \Gamma^1 \times I_k,
        \\
        \dot{~\varphi}^{(n)}(~X, t) = \mathcal{P}^{~\varphi}_{\Gamma^2 \to \Gamma^1}[\dot{~\varphi}^{(n-1)}(\Omega^2, t)]
        &
        \mathrm{on} \; \Gamma^1 \times I_k,
        \\
        \ddot{~\varphi}^{(n)}(~X, t) = \mathcal{P}^{~\varphi}_{\Gamma^2 \to \Gamma^1}[\ddot{~\varphi}^{(n-1)}(\Omega^2, t)]
        &
        \mathrm{on} \; \Gamma^1 \times I_k,
        \label{eq:schwarz1}
    \end{cases}
    \\
    \mathrm{Domain} \; \Omega^2:
    &
    \begin{cases}
        ~\Div ~P^{(n)} + \rho ~B = \rho \ddot{~\varphi}^{(n)}
        &
        \qquad \; \mathrm{in} \; \Omega^2 \times I_k,
        \\
        ~\varphi^{(n)}(~X, t) = ~\chi
        &
        \qquad \; \mathrm{on} \; \partial_{\varphi}\Omega^2 \times I_k,
        \\
        ~P^{(n)}~N = ~T
        &
        \qquad \; \mathrm{on} \; \partial_{~T}\Omega^2 \times I_k,
        \\
        ~P^{(n)}~N = \mathcal{P}^{~T}_{\Gamma^1 \to \Gamma^2}[~T^{(n)}(\Omega^1, t)]
        &
        \qquad \; \mathrm{on} \; \Gamma^2 \times I_k,
        \label{eq:schwarz2}
    \end{cases}
\end{align}
with initial conditions prescribed in terms of position and velocity, $~x_{t_k}^{i}$ and $~v_{t_k}^{i}$, respectively, for $\Omega^i$, $i=1,2$
\begin{equation} \label{eq:schwarz_ics}
    \begin{aligned}
    ~\varphi^{(n)}(~X, t_k) &= ~x_{t_k}^{i} &\mathrm{in} & \quad \Omega^i
    \\
    \dot{~\varphi}^{(n)}(~X, t_k) &= ~v_{t_k}^{i}  &\mathrm{in} & \quad \Omega^i.
    \end{aligned}
\end{equation}
The Schwarz contact algorithm for a two-body scenario is concisely outlined in Algorithm \ref{alg:schwarz}.
The iterative process of the Schwarz method continues until pre-determined convergence criteria are met.
In this context, both relative and absolute convergence metrics are employed. Error evaluation is performed using the Euclidean norm to measure the position differences between consecutive Schwarz iterations as
\begin{equation} \label{eq:convergence-errors}
    \begin{aligned}
        \epsilon_{\mathrm{abs}}^{(n)}
        & :=
        \sqrt{
            ||\triangle ~x^1||^2
            +
            ||\triangle ~x^2||^2
        },
        \\
        \epsilon_{\mathrm{rel}}^{(n)}
        & :=
        \sqrt{
            \frac{||\triangle ~x^1||^2}{||~x^1||^2}
            +
            \frac{||\triangle ~x^2||^2}{||~x^2||^2}
        },
    \end{aligned}
\end{equation}
where $~x^1$ and $~x^2$ are the positions for subdomains 1 and 2 for the $n^{\mathrm{th}}$ Schwarz iteration, and $\triangle ~x^1$ and $\triangle ~x^2$ are the differences in positions for the $n^{\mathrm{th}}$ and $(n-1)^{\mathrm{th}}$ Schwarz iterations.

\begin{algorithm}
    \footnotesize
    \begin{algorithmic}[1]
        \For{time interval $I_k$, $t \in [t_k, t_{k+1}]$}
            \State Schwarz iteration count: $n \gets 1$
            \Repeat
                \Comment{Schwarz iterative process}
                \State Domain $\Omega^1$
                \For{$t^1_{j|I_k} \in I_k$}
                \Comment{Time stops for $\Omega^1$}
                    \State  \quad $~\varphi^{(n)}(~X, t^1_{j|I_k})  \gets ~\chi \; \mathrm{on} \; \partial_{\varphi}\Omega^1 \times t^1_{j|I_k}$
                    \Comment{Regular Dirichlet BC}
                    \State  \quad $~P^{(n)}~N \gets ~T \; \mathrm{on} \; \partial_{~T}\Omega^1 \times t^1_{j|I_k}$
                    \Comment{Regular Neumann BC}
                    \State \quad $~\varphi^{(n)}(~X, t^1_{j|I_k}) \gets \mathcal{P}^{~\varphi}_{\Gamma^2 \to \Gamma^1}[~\varphi^{(n-1)}(\Omega^2, t^1_{j|I_k})] \; \mathrm{on} \; \Gamma^1 \times t^1_{j|I_k}$
                    \Comment{Contact Dirichlet BC}
                    \State \quad $\dot{~\varphi}^{(n)}(~X, t^1_{j|I_k}) \gets \mathcal{P}^{~\varphi}_{\Gamma^2 \to \Gamma^1}[\dot{~\varphi}^{(n-1)}(\Omega^2, t^1_{j|I_k})] \; \mathrm{on} \; \Gamma^1 \times t^1_{j|I_k}$
                    \State \quad $\ddot{~\varphi}^{(n)}(~X, t^1_{j|I_k})  \gets  \mathcal{P}^{~\varphi}_{\Gamma^2 \to \Gamma^1}[\ddot{~\varphi}^{(n-1)}(\Omega^2, t^1_{j|I_k})] \; \mathrm{on} \; \Gamma^1 \times t^1_{j|I_k}$
                    \State \quad Solve  $~\Div ~P^{(n)} + \rho ~B = \rho \ddot{~\varphi}^{(n)} \; \mathrm{in} \; \Omega^1 \times t^1_{j|I_k} $
                    \Comment{Dynamic problem}
                \EndFor
                \State Domain $\Omega^2$
                \For{$t^2_{j|I_k} \in I_k$}
                \Comment{Time stops for $\Omega^2$}
                    \State \quad $~\varphi^{(n)}(~X, t^2_{j|I_k}) \gets ~\chi \; \mathrm{on} \; \partial_{\varphi}\Omega^2 \times t^2_{j|I_k}$
                    \Comment{Regular Dirichlet BC}
                    \State \quad $~P^{(n)}~N \gets ~T \; \mathrm{on} \; \partial_{~T}\Omega^2 \times t^2_{j|I_k} $
                    \Comment{Regular Neumann BC}
                    \State \quad $~P^{(n)}~N \gets \mathcal{P}^{~T}_{\Gamma^1 \to \Gamma^2 }[~T^{(n)}(\Omega^1, t^2_{j|I_k})] \; \mathrm{on} \; \Gamma^2 \times t^2_{j|I_k} $
                    \Comment{Contact Neumann BC}
                    \State \quad Solve $~\Div ~P^{(n)} + \rho ~B = \rho \ddot{~\varphi}^{(n)} \; \mathrm{in} \; \Omega^2 \times t^2_{j|I_k}$
                    \Comment{Dynamic problem}
                \EndFor
                \State $n \gets n + 1$
            \Until $\epsilon_{\mathrm{abs}}^{(n)} \leq \mathrm{tol}_{\mathrm{abs}}$ or $\epsilon_{\mathrm{rel}}^{(n)} \leq \mathrm{tol}_{\mathrm{rel}}$ \eqref{eq:convergence-errors} or maximum number of iterations is reached
        \EndFor
        \State $k \gets k + 1$ \Comment{advance to the next controller time interval}
    \end{algorithmic}
    \caption{The Schwarz contact method for a controller time interval $I_k$ for the specific case of two bodies. It proceeds by solving Euler-Lagrange equation sequences, as per \eqref{eq:schwarz1}--\eqref{eq:schwarz_ics}, augmented by alternating Dirichlet-Neumann contact boundary conditions.}
    \label{alg:schwarz}
\end{algorithm}

As shown by \textcite{Krause:2012}, traditional contact techniques can compromise the conservation properties of a system, such as energy or momentum conservation. The analysis by \citeauthor{Krause:2012} highlights that to ensure energy conservation, the numerical scheme must obtain zero relative velocities in the normal direction at the contact boundary. Achieving this condition in a fully discrete setting is challenging using standard constraint-based contact approaches. Importantly, the Schwarz method naturally satisfies this condition due to its unique Dirichlet boundary conditions, which equate velocities at contact boundaries.

\begin{remark}
For illustrative purposes, in \eqref{eq:schwarz1}--\eqref{eq:schwarz2}, we use Dirichlet boundary conditions for the contact boundary of domain $\Omega^1$ and Neumann conditions for domain $\Omega^2$. Our numerical investigations reveal that the sequence in which these Dirichlet and Neumann conditions are alternated does not influence the convergence rate of the Schwarz method. Remarkably, one can even switch the order of these conditions at each controller time step without impacting the outcome. The crucial element is to maintain the alternation of boundary conditions.
\end{remark}

\begin{remark}
Upon the initial detection of contact at $t_{\mathrm{impact}}$, some overlap between $\Omega^1$ and $\Omega^2$ is likely. The Schwarz procedure outlined in Algorithm \ref{alg:schwarz} naturally handles this by eliminating the overlap, resulting in a single, distinct contact boundary $\Gamma$ such that $\Omega^1 \cap \Omega^2 = \Gamma$.
\end{remark}

\begin{remark}
In this work, we prefer the alternating Dirichlet-Neumann formulation of the Schwarz method over the Robin-Robin formulation. This is because Dirichlet and Neumann conditions are more easily integrated into existing solid mechanics codes like Sandia's Sierra/SM \cite{Sierra:2011}, thus minimizing the disruption to legacy software. Furthermore, it is worth mentioning that our proposed Schwarz contact scheme does not incorporate a damping or relaxation parameter, a feature common in many non-overlapping Schwarz algorithms \cite{Zanolli:1987, Funaro:1988, Cote:2005, Kwok:2014, Bochev:2017}. Exploration of different boundary condition sets and additional parameters is left for future investigations.
\end{remark}

\subsection{Information transfer in the Schwarz contact algorithm} \label{sec:transfer}

\begin{figure}
     \centering
            \subfigure[Red domain as source, green domain as destination]
            {\includegraphics[page=1,width=0.48\textwidth]{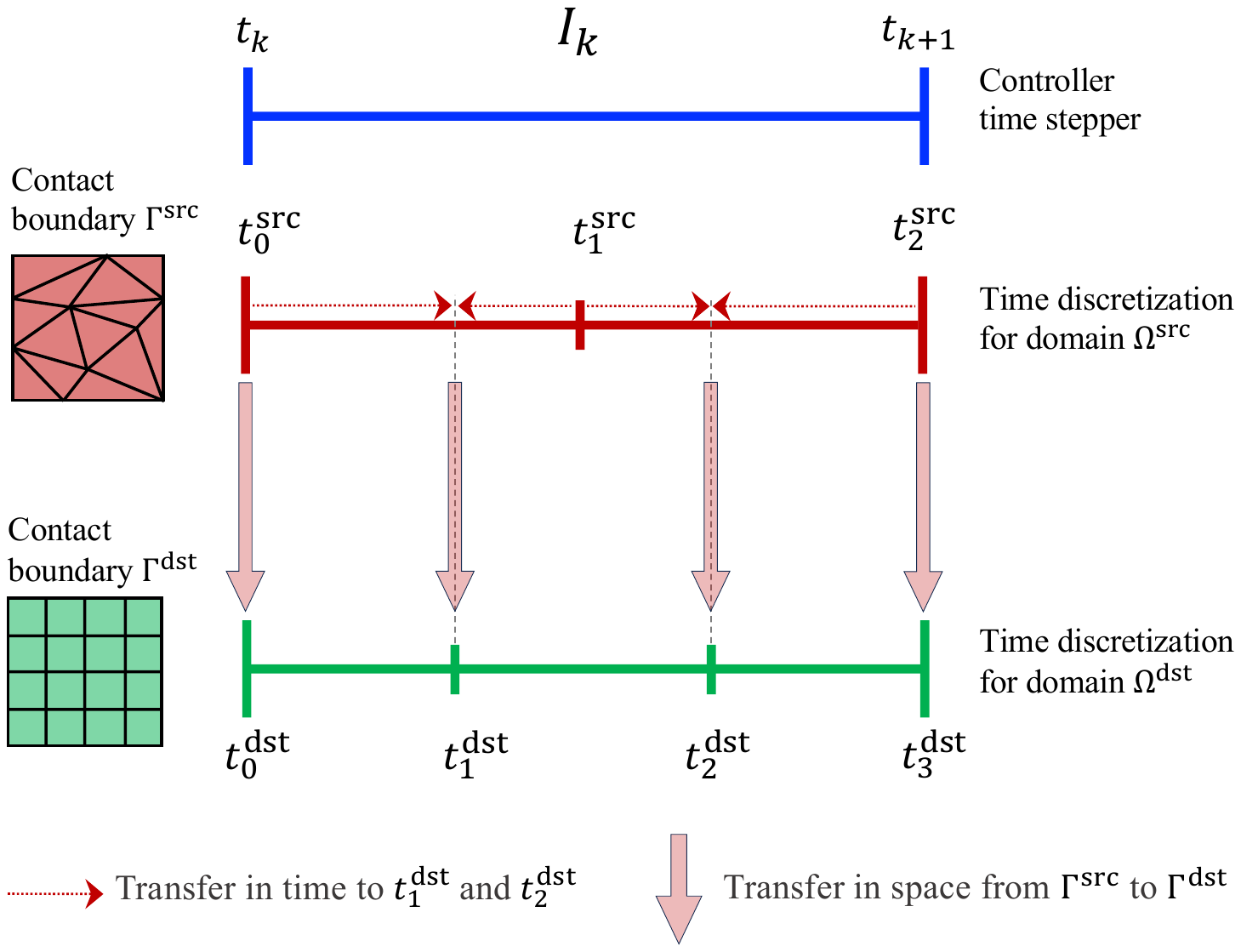} \label{fig:projections_domains_a}}
            \subfigure[Green domain as source, red domain as destination]
            {\includegraphics[page=2,width=0.48\textwidth]{plots/transfer_Schwarz_new_1.pdf} \label{fig:projections_domains_b}}
        \caption{Field transfer in the Schwarz contact algorithm within one controller time step during the contact stage. (a) The red domain serves as the source domain $\Omega^{\mathrm{src}}$, while the green domain functions as the destination domain $\Omega^{\mathrm{dst}}$. These domains have respective time stops $\{t^{\mathrm{src}}_l\}_{l=0}^{2}$ and $\{t^{\mathrm{dst}}_l\}_{l=0}^{3}$, as well as corresponding contact boundaries $\Gamma^{\mathrm{src}}$ and $\Gamma^{\mathrm{dst}}$. (b) The roles of the domains are reversed.}
        \label{fig:projections_domains}
\end{figure}

As previously mentioned, the Schwarz contact algorithm offers considerable flexibility, allowing each body to use its own time steps and integrators. This means that within a single controller time interval, individual bodies can progress according to their own sets of time steps, as illustrated in Figure \ref{fig:domains}. Additionally, each body in contact has unique geometric characteristics and can be represented using varying mesh topologies and sizes, as exemplified by contact boundaries $\Gamma^1$ and $\Gamma^2$ in Figure~\ref{fig:projections_domains}.

Given these considerations, information must be exchanged between spatially distinct interfaces and incompatible mesh grids. To implement the Schwarz contact boundary conditions, data transfer is necessary both temporally (due to differing time steps) and spatially (owing to non-matching grids). Typical fields that need to be transferred include positions, velocity, and acceleration for Dirichlet boundary conditions, as well as traction for Neumann boundary conditions. The details of this information exchange are provided in Algorithm~\ref{alg:schwarz}.

As emphasized in previous work \cite{Mota:2017, Mota:2022}, the selection of appropriate transfer operators for various fields is a crucial aspect of the Schwarz alternating method and can pose challenges. These operators must be defined with great care, as improper choices may result in artificial numerical diffusion, imprecision, and increased computational expense, see also \textcite{Kumar:2015, Bucher:2007, Koliesnikova:2022, Hennig:2018}.

The design of effective transfer operators is of paramount importance in a variety of scientific and engineering contexts, including but not limited to adaptive mesh refinement techniques with evolving or incompatible meshes, multiscale and multiphysics coupling methods, as well as domain decomposition approaches involving both overlapping and non-overlapping domains. Further insights into this subject can be found in \textcite{Bochev.Kuberry:2015, Bai.Brandt:1987, Peric:1996, Dureisseix.Bavestrello:2006, Gosselet:2018, Koliesnikova:2022}.

In the specific scenario where two bodies are in contact, let us define the contact boundaries as $\Gamma^{\mathrm{src}} \in \Omega^{\mathrm{src}}$ and $\Gamma^{\mathrm{dst}} \in \Omega^{\mathrm{dst}}$. Here, $\Omega^{\mathrm{src}}$ is the domain from which quantities are transferred, while $\Omega^{\mathrm{dst}}$ is the domain receiving these quantities. To distinguish them, we use the notation $[\cdot]^{\mathrm{src}}$ for quantities originating from $\Omega^{\mathrm{src}}$ and $[\cdot]^{\mathrm{dst}}$ for those defined in $\Omega^{\mathrm{dst}}$.

Utilizing the above notation, the projections needed for transferring fields and thereby defining the contact boundary conditions are as follows:
\begin{itemize}
    \item For Dirichlet BCs: $\mathcal{P}^{~\varphi}_{(\Gamma^{\mathrm{src}} \to \Gamma^{\mathrm{dst}}) \times  t^{\mathrm{dst}}}[{~\varphi}(\Omega^{\mathrm{src}}, t^{\mathrm{dst}})]$, applicable also to velocity and acceleration.
    \item For Neumann BCs: $\mathcal{P}^{~T}_{(\Gamma^{\mathrm{src}} \to \Gamma^{\mathrm{dst}}) \times  t^{\mathrm{dst}}}[~T(\Omega^{\mathrm{src}}, t^{\mathrm{dst}})]$.
\end{itemize}
Firstly, the requisite variables should be interpolated at the specific time stops $t^{\mathrm{dst}}$ for which the Schwarz contact boundary conditions are intended. For example, the position field ${~\varphi}(\Omega^{\mathrm{src}}, t^{\mathrm{dst}})$ corresponding to domain $\Omega^{\mathrm{src}}$ should be defined at the specific time step $t^{\mathrm{dst}}$, corresponding to the time discretization of $\Omega^{\mathrm{dst}}$. If there are differing time steps between $\Omega^{\mathrm{src}}$ and $\Omega^{\mathrm{dst}}$, temporal information transfer is needed, as outlined in Section~\ref{sec:time_transfer}.
Subsequently, the spatial transfer operators $\mathcal{P}^{~\varphi}_{\Gamma^{\mathrm{src}} \to \Gamma^{\mathrm{dst}}}$ and $\mathcal{P}^{~T}_{\Gamma^{\mathrm{src}} \to \Gamma^{\mathrm{dst}}}$ are employed to convey these fields between the domains, as detailed in Section~\ref{sec:space_transfer}.

For the sake of clarity, Figure~\ref{fig:projections_domains} zooms in on a specific global controller time interval, denoted as $I_{k} := \{t \in [t_{k}, t_{k+1}]\}$. In Figure~\ref{fig:projections_domains_a}, the red domain serves as the source domain $\Omega^{\mathrm{src}}$, while the green domain functions as the destination domain $\Omega^{\mathrm{dst}}$. These domains have respective time stops $\{t^{\mathrm{src}}_l\}_{l=0}^{2}$ and $\{t^{\mathrm{dst}}_l\}_{l=0}^{3}$, as well as corresponding contact boundaries $\Gamma^{\mathrm{src}}$ and $\Gamma^{\mathrm{dst}}$.
In Figure~\ref{fig:projections_domains_b}, the roles of the domains are reversed.

\subsubsection{Temporal information exchange}  \label{sec:time_transfer}

At the left endpoint $t_k$ of the interval $I_k$, information swapping between the local time instances $t^{\mathrm{src}}_0$ and $t^{\mathrm{dst}}_0$ is relatively straightforward, as these time instances are synchronized; refer to Figure~\ref{fig:projections_domains_a}. The same holds true for the right endpoint $t_{k+1},$ with local time stops $t^{\mathrm{src}}_2$ and $t^{\mathrm{dst}}_3$ in Figure~\ref{fig:projections_domains_a}, and stops $t^{\mathrm{src}}_3$ and $t^{\mathrm{dst}}_2$ in Figure~\ref{fig:projections_domains_b}. For these synchronized time stops, the exchange of information in the spatial domain is all that is needed.

When dealing with intermediate local time instances, however, time-based interpolation becomes necessary. For instance, as illustrated in Figure~\ref{fig:projections_domains_a}, information at time stop $t^{\mathrm{dst}}_1$ is derived through interpolation between the time stops $t^{\mathrm{src}}_0$ and $t^{\mathrm{src}}_1$. Similarly, for the time stop $t^{\mathrm{dst}}_2$, an interpolation is conducted between the time stops $t^{\mathrm{src}}_1$ and $t^{\mathrm{src}}_2$.  For the scenario presented in Figure~\ref{fig:projections_domains_b}, the state of a field at the time stop $t^{\mathrm{dst}}_1$ is obtained by interpolating between $t^{\mathrm{src}}_1$ and $t^{\mathrm{src}}_2$.

This temporal information exchange is effectively a straightforward interpolation between the nearest time stops, utilizing the most up-to-date solutions from the Schwarz iterations. The goal of this temporal transfer process is to obtain the fields ${~\varphi}(\Omega^{\mathrm{src}}, t^{\mathrm{dst}})$, $\dot{~\varphi}(\Omega^{\mathrm{src}}, t^{\mathrm{dst}})$, $\ddot{~\varphi}(\Omega^{\mathrm{src}}, t^{\mathrm{dst}})$, and $~T(\Omega^{\mathrm{src}}, t^{\mathrm{dst}})$.

\subsubsection{Spatial information exchange} \label{sec:space_transfer}

After interpolating the fields with respect to time, we proceed to transfer these fields across the contact boundaries. We achieve this using specialized transfer operators and source fields. The computed destination fields are denoted as $~\varphi(\Omega^{\mathrm{dst}}, t^{\mathrm{dst}})$, $\dot{~\varphi}(\Omega^{\mathrm{dst}}, t^{\mathrm{dst}})$, $\ddot{~\varphi}(\Omega^{\mathrm{dst}}, t^{\mathrm{dst}})$, and $~T(\Omega^{\mathrm{dst}}, t^{\mathrm{dst}})$.

To create these transfer operators, we use nodal vectors of finite element interpolation functions, $~N^{\mathrm{src}}$ and $~N^{\mathrm{dst}}$, on the source and destination contact boundaries $\Gamma^{\mathrm{src}}$ and $\Gamma^{\mathrm{dst}}$.

The updated position field $~\varphi^{\mathrm{dst}}(\Omega^{\mathrm{dst}}, t^{\mathrm{dst}})$ is given by
\begin{equation}
    ~\varphi^{\mathrm{dst}}(\Omega^{\mathrm{dst}}, t^{\mathrm{dst}}) = \mathcal{P}^{~\varphi}_{\Gamma^{\mathrm{src}} \to \Gamma^{\mathrm{dst}}} ~\varphi^{\mathrm{src}}(\Omega^{\mathrm{src}}, t^{\mathrm{dst}}),
\end{equation}
where the transfer operator $\mathcal{P}^{~\varphi}_{\Gamma^{\mathrm{src}} \to \Gamma^{\mathrm{dst}}}$ is
\begin{equation}
    \mathcal{P}^{~\varphi}_{\Gamma^{\mathrm{src}} \to \Gamma^{\mathrm{dst}}} =  ~W^{-1} ~L.
\end{equation}
The matrices $~W$ and $~L$ are obtained by integrating the finite element basis functions over the contact boundaries as follows
\begin{equation}
    ~W = \int_{\Gamma^{\mathrm{dst}}} ~N^{\mathrm{dst}} (~N^{\mathrm{dst}})^{\mathrm{T}} \, \diff S,
    \label{eq:Mmatrix}
\end{equation}
and
\begin{equation}
    ~L = \int_{\Gamma^{\mathrm{dst}}} ~N^{\mathrm{dst}} (~N^{\mathrm{src}})^{\mathrm{T}} \, \diff S.
    \label{eq:Lmatrix}
\end{equation}
The velocity and acceleration fields are updated using the same transfer operator. The updated traction field $~T^{\mathrm{dst}}(\Omega^{\mathrm{dst}}, t^{\mathrm{dst}})$ is
\begin{equation}
    ~T^{\mathrm{dst}}(\Omega^{\mathrm{dst}}, t^{\mathrm{dst}}) = \mathcal{P}^{~T}_{\Gamma^{\mathrm{src}} \to \Gamma^{\mathrm{dst}}} ~T^{\mathrm{src}}(\Omega^{\mathrm{src}}, t^{\mathrm{dst}}),
\end{equation}
with the corresponding transfer operator
\begin{equation}
    \mathcal{P}^{~T}_{\Gamma^{\mathrm{src}} \to \Gamma^{\mathrm{dst}}} = ~L ~H^{-1}.
\end{equation}
Matrix $~H$ is defined as
\begin{equation}
    ~H = \int_{\Gamma^{\mathrm{src}}} ~N^{\mathrm{src}} (~N^{\mathrm{src}})^{\mathrm{T}} \, \diff S.
    \label{eq:Hmatrix}
\end{equation}
The benefits of these transfer operators are manifold: they are agnostic to the underlying geometry, easy to implement thanks to the reliance on finite element interpolation functions, and yield highly accurate results. Interestingly, while we developed these transfer operators independently, we discovered that similar approaches have been proposed in previous work. For a comprehensive discussion, refer to \textcite{Bochev.Kuberry:2015}.

\section{Comparative analysis of Schwarz and traditional contact methods on a 1D impact problem} \label{sec:results}

This section focuses on a numerical evaluation of the Schwarz alternating method, as detailed in Section~\ref{sec:schwarz}. We demonstrate the significant advantages of this method by comparing it with several traditional contact approaches. For a brief overview of the evaluated methods, refer to Section~\ref{sec:methods}. Sections~\ref{sec:carpenter} and~\ref{sec:comparison} delve into the formulation of the problem and the analysis of numerical results, respectively. All simulations for this section were carried out using a specialized MATLAB-based software developed for this study \cite{Mota.etal:2023}. 

\subsection{Overview of evaluated contact methods} \label{sec:methods}

In this section, we specifically compare three categories of contact methods: (1) the penalty method \cite{Hunek:1993, Faltus:2020}, (2) the Lagrange multiplier method \cite{Carpenter:1991, Belytschko.Xiao:2003}, and (3) the Schwarz alternating method. Brief descriptions of the conventional methods (penalty and Lagrange multiplier) can be found in Section~\ref{sec:conventional_contact}, while the Schwarz alternating method is outlined in Section~\ref{sec:schwarz}. The variants for each method are summarized in Table~\ref{tab:methods}.

\begin{table}
    \footnotesize
    \centering
    \begin{tabular}{l l}
        \toprule
        Method & Time Integration Scheme
        \\
        \midrule
        \addlinespace[0.5em]  
        \multirow{2}{*}{Penalty} & Implicit Newmark-$\beta$
        \\
        & Explicit Newmark-$\beta$
        \\
        \addlinespace[0.5em]
        \multirow{2}{*}{Lagrange Multiplier} & Implicit Backward Euler
        \\
        & Explicit Newmark-$\beta$
        \\
        \addlinespace[0.5em]
        \multirow{3}{*}{Schwarz} & All-implicit Newmark-$\beta$
        \\
        & Hybrid Implicit-Explicit Newmark-$\beta$
        \\
        & All-explicit Newmark-$\beta$
        \\
        \bottomrule
    \end{tabular}
    \caption{Comparison of contact methods and their time integration schemes. The table outlines both implicit and explicit forms of the penalty and Lagrange multiplier methods. For the Schwarz alternating method, three variations are considered: (1) all subdomains use implicit time integrators, (2) all subdomains use explicit time integrators, and (3) a mixed approach using both explicit and implicit time integrators in different subdomains.}
    \label{tab:methods}
\end{table}

We examine both implicit and explicit versions of the penalty and Lagrange multiplier methods. In \textcite{Carpenter:1991}, the explicit variant of the Lagrange multiplier method (termed forward increment) was noted for its superior characteristics; nonetheless, we present results for both versions. In existing literature, specifically \textcite{Doyen:2011, DiStasio:2019}, the implicit Lagrange multiplier method was reported to exhibit instability, energy dissipation errors, and spurious oscillations in impact problems. These observations align with our findings. For our numerical example employing the implicit Lagrange multiplier method, only the backward Euler time integrator was consistently reliable. Further implementation details for the conventional methods can be found in Section~\ref{sec:conventional_contact}.

Regarding the Schwarz alternating method, we consider three configurations, summarized in Table \ref{tab:methods}: (1) using implicit time integrators across all subdomains, (2) using explicit time integrators across all subdomains, and (3) employing a hybrid approach with both explicit and implicit time integrators in different subdomains.

To ensure an unbiased comparison, all contact methods utilize the Newmark-$\beta$ time integration scheme, except for the implicit Lagrange multiplier method due to the previously stated limitations. For more information on the explicit and implicit Newmark-$\beta$ schemes, see Section~\ref{sec:chatter}.

\subsection{One-dimensional impact benchmark} \label{sec:carpenter}

Here, we focus on a widely recognized 1D impact benchmark problem, which involves two prismatic rods moving toward each other. This test case serves as a standard example for validating numerical methods dealing with contact mechanics and is frequently cited in the literature \cite{Carpenter:1991, Doyen:2011, Wriggers:2006, DiStasio:2019}. From a numerical perspective, this seemingly simple problem presents several challenges. The discretization of the governing equations is known to introduce spurious high-frequency oscillations at the contact surface during the impact \cite{DiStasio:2019}. Furthermore, it has been observed that this specific test case can reveal issues related to poor energy conservation \cite{Doyen:2011}.

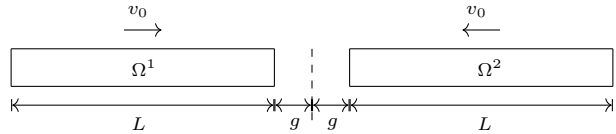
\begin{figure}
    \centering
    \begin{tikzpicture}[scale=1.0]
        \scriptsize
        \draw[-] (-4,-0.25) -- (-0.5,-0.25);
        \draw[-] (0.5,-0.25) -- (4,-0.25);
        \draw[-] (-4,-0.75) -- (-0.5,-0.75);
        \draw[-] (0.5,-0.75) -- (4,-0.75);
        \draw[-] (-4, -0.75) -- (-4, -0.25);
        \draw[-] (4, -0.75) -- (4, -0.25);
        \draw[-] (-0.5, -0.75) -- (-0.5, -0.25);
        \draw[-] (0.5, -0.75) -- (0.5, -0.25);
        \draw [|<->|] (-4,-1) -- (-0.5,-1);
        \draw[|<->|] (-0.5,-1) -- (0,-1);
        \draw[|<->|](0, -1) -- (0.5, -1);
        \draw[|<->|](0.5,-1) -- (4, -1);
        \node[text width=0.5cm] at (-2.15,-1.25) {$L$};
        \node[text width=0.5cm] at (2.45,-1.25) {$L$};
        \node[text width=0.5cm] at (-2.15,-0.5) {$\Omega^1$};
        \node[text width=0.5cm] at (2.45,-0.5) {$\Omega^2$};
        \node[text width=0.5cm] at (-0.05,-1.25) {$g$};
        \node[text width=0.5cm] at (0.45,-1.25) {$g$};
        \draw[->] (-2.5, 0) -- (-2.0,0);
        \node[text width=0.5cm] at (-2.20,0.25) {$v_0$};
        \draw[<-] (2.0, 0) -- (2.5,0);
        \node[text width=0.5cm] at (2.30,0.25) {$v_0$};
        \draw [dashed] (0.0, -1.2) -- (0.0, -0.2);
    \end{tikzpicture}
    \caption{One-dimensional impact benchmark problem. Two prismatic rods ($ \Omega^1 $ and $ \Omega^2 $) moving towards each other. Each rod is characterized by a linear elastic material model. Both rods are symmetrical about the plane of impact and are initially separated by a distance of $ 2g $. Furthermore, they both have an initial velocity of $ v_0 $.}
    \label{fig:1D_impact_illustration}
\end{figure}

The configuration, as illustrated in Figure \ref{fig:1D_impact_illustration}, involves two prismatic rods ($ \Omega^1 $ and $ \Omega^2 $) moving towards each other. Each rod is characterized by a linear elastic material model, with properties such as density $ \rho $, elastic modulus $ E $, and cross-sectional area $ A $. Both rods are symmetrical about the plane of impact and are initially separated by a distance of $ 2g $. Furthermore, they both have an initial velocity of $ v_0 $. The parameters for this benchmark problem are detailed in Table \ref{tab:carpenterParamsHighSpeed}, which includes units and values for variables such as $ \rho $, $ E $, $ A $, $ L $, $ g $, $ v_0 $, $ t_0 $, and $ t_N $.

\begin{table}
    \footnotesize
    \centering
    \begin{tabular}{l r l}
        \toprule
        Parameter & {Value} & {Unit}
        \\
        \midrule
        $\rho$ & 1000 & \unit{\kilo\gram\per\cubic\meter}
        \\
        $E$ &  1 & \unit{\giga\pascal}
        \\
        $A$ & 1 & \unit{\square\milli\meter}
        \\
        $L$ &  250 & \unit{\milli\meter}
        \\
        $g$ &  20 & \unit{\milli\meter}
        \\
        $v_0$ & 100 & \unit{\meter\per\second}
        \\
        $t_{0}$ &  -200 & \unit{\micro\second}
        \\
        $t_{N}$ & 800 & \unit{\micro\second}
        \\
        \bottomrule
    \end{tabular}
    \caption{Parameters for the 1D impact benchmark. Density $\rho$, elastic modulus $E$, cross-sectional area $A$, length $L$, initial semi-distance $g$, initial velocity $v_0$, and initial and final simulation times $t_{0}$ and $t_{N}$, respectively.}
    \label{tab:carpenterParamsHighSpeed}
\end{table}

The analytical solution for this problem is well-defined and provides insights into the behavior of the rods at the time of impact $ t_{\mathrm{imp}} $ and release $ t_{\mathrm{rel}} $, as presented in \eqref{eq:rod_disp} and \eqref{eq:impact_time}. Furthermore, our analysis examines the contact force $ f_{\mathrm{contact}} $, as well as the kinetic $ T $ and potential $ V $ energies for each rod, as defined by \eqref{eq:rod_kinetic_energy} and \eqref{eq:rod_potential_energy}.

Following the derivations presented in \textcite{Carpenter:1991}, we can readily deduce the expressions for the position $ x(t) $ and velocity $ v(t) $ of the right end (contact point) of the left rod in domain $ \Omega^1 $. These are given by
\begin{equation}
    \label{eq:rod_disp}
    x(t) = \begin{cases}
        -g + v_0 (t - t_0), & t < t_{\mathrm{imp}}, \\
        0, &   t_{\mathrm{imp}} \leq t \leq t_{\mathrm{rel}}, \\
        - v_0 (t - t_{\mathrm{rel}}), & t > t_{\mathrm{rel}},
    \end{cases}
    \qquad
    v(t) = \begin{cases}
        v_0, & t < t_{\mathrm{imp}}, \\
        0, &   t_{\mathrm{imp}} \leq t \leq t_{\mathrm{rel}}, \\
        - v_0, & t > t_{\mathrm{rel}},
    \end{cases}
\end{equation}
where $ t_{\mathrm{imp}} $ and $ t_{\mathrm{rel}} $ denote the impact and release times, respectively. The analytical expressions for these times are
\begin{equation}
    \label{eq:impact_time}
    t_{\mathrm{imp}} = t_0 + \frac{g}{v_0}, \quad
    t_{\mathrm{rel}} = t_{\mathrm{imp}} + 2 L \sqrt{\frac{\rho}{E}}.
\end{equation}
The contact force during the impact is given by $ f_{\mathrm{contact}} = v_0 \sqrt{E \rho} A $. Furthermore, the kinetic $ T $ and potential energies $ V $ for each rod are defined as
\begin{equation}
    \label{eq:rod_kinetic_energy}
    T = \begin{cases}
        \frac{1}{2} \rho A L v_0^2 &  t < t_{\mathrm{imp}}, \\
        \frac{1}{2} \rho A L v_0^2 - \frac{1}{2} \sqrt{\rho E} A v_0^2
        (t - t_{\mathrm{imp}}) &  t_{\mathrm{imp}} \leq t\leq t_{\mathrm{m}}, \\
        \frac{1}{2} \sqrt{\rho E} A v_0^2 (t - t_{\mathrm{m}}) &
        t_{\mathrm{m}} \leq t \leq t_{\mathrm{rel}} ,\\
        \frac{1}{2} \rho A L v_0^2 &  t > t_{\mathrm{rel}},
    \end{cases}
\end{equation}
and
\begin{equation}
    \label{eq:rod_potential_energy}
    V  = \begin{cases}
        0 & t < t_{\mathrm{imp}} \\
        \frac{1}{2} \sqrt{\rho E} A v_0^2 (t - t_{\mathrm{imp}}) &
        t_{\mathrm{imp}} \leq t \leq t_{\mathrm{m}}, \\
        \frac{1}{2} \rho A L v_0^2 - \frac{1}{2}
        \sqrt{\rho E} A v_0^2 (t - t_{\mathrm{m}}) &  t_{\mathrm{m}} \leq t\leq t_{\mathrm{rel}}, \\
        0 &  t > t_{\mathrm{rel}}.
    \end{cases}
\end{equation}
Here, $ t_{\mathrm{m}} $ represents the time at which the maximum potential energy and minimum kinetic energy are achieved. It is given by $ t_{\mathrm{m}} = t_{\mathrm{imp}} + L \sqrt{\frac{\rho}{E}} $.

\subsection{Comparative analysis of the Schwarz alternating method and conventional contact approaches} \label{sec:comparison}

In this section, we employ the finite element method for spatial discretization. Here, $ N_x $ denotes the number of elements in each rod, while $ \triangle t^i $ represents the time step used in domain $ \Omega^i $ for a given time integration scheme. Unless specified otherwise, each rod is discretized using $ N_x = 200 $ linear elements. The controller time interval $ I_k $ and local time steps for the two rods are set to $ I_k = \triangle t^1 = \triangle t^2 = \qty{1e-7}{\second}$. It is important to note that, for explicit schemes, the selected time steps are sufficiently small to satisfy the Courant-Friedrichs-Levy (CFL) condition.

An exception is made for the implicit-explicit Schwarz variant, where we use different time steps in each domain: $ \triangle t^1 = \qty{1e-7}{\second} $ in $ \Omega^1 $ and $ \triangle t^2 = \qty{1e-8}{\second} $ in $ \Omega^2 $. The aim is to demonstrate the Schwarz alternating method's capability to not only couple different time-integrators but also different time steps across various subdomains. For the Schwarz iterative procedure, the convergence tolerances for relative and absolute errors defined by \eqref{eq:convergence-errors} are set at $ \num{1e-12} $ and $ \qty{1e-15}{\meter}$, respectively.

In the evaluation of the two penalty methods, a penalty parameter $ \tau = \qty{7.5e4}{\newton\per\meter} $ is chosen, as this value produced the most accurate outcomes.

\begin{figure}
    \centering
    \includegraphics[page=1,trim={0mm 35mm 0mm 25mm},clip,width=1\textwidth]{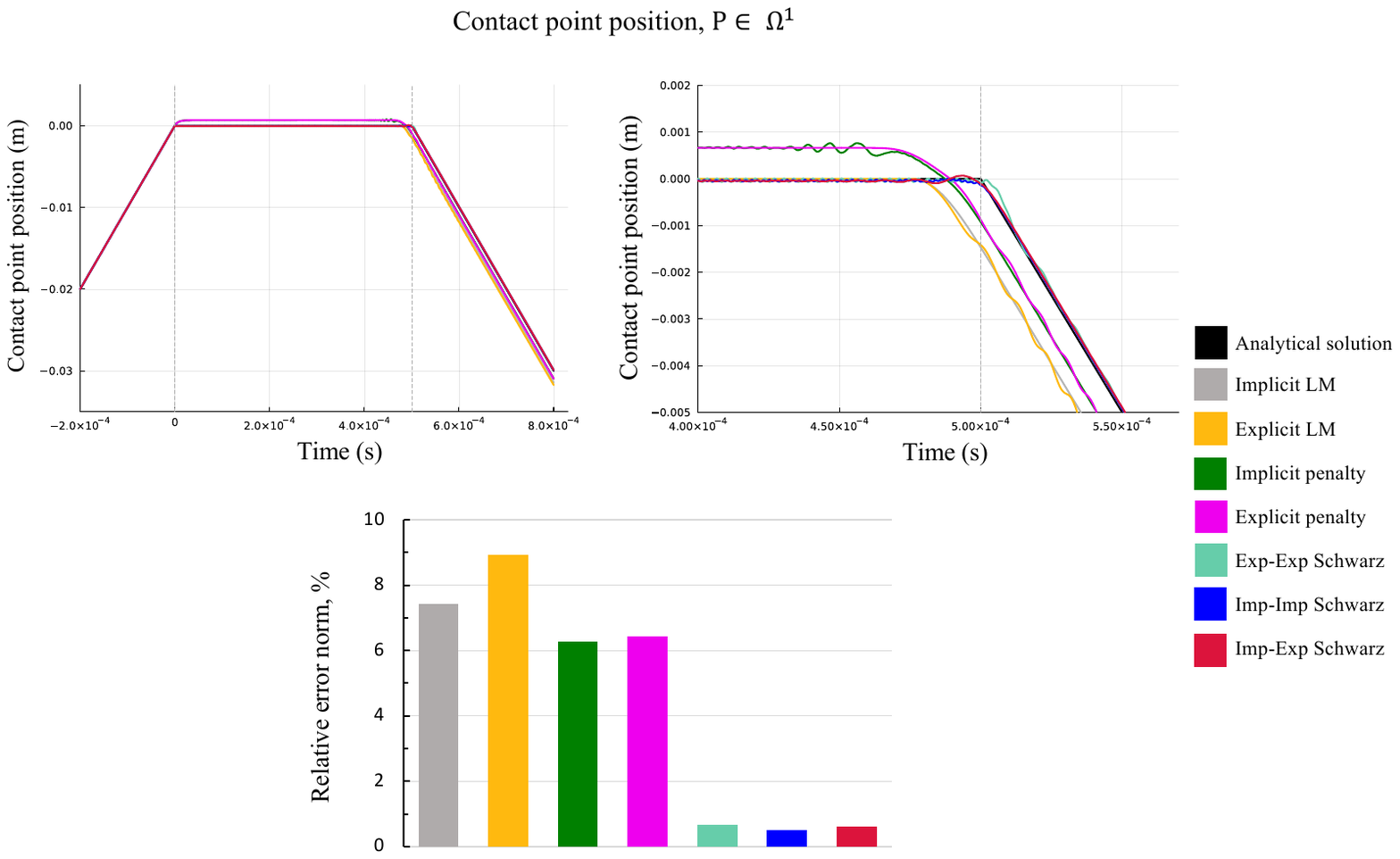}
    \caption{Time evolution of the position of the contact point on the left bar (denoted as domain $\Omega^1$). The lower part of the figure presents total relative errors, comparing traditional contact methods with the Schwarz alternating method. Notably, the penalty and Lagrange multiplier methods tend to underestimate the time of release. Additionally, penalty methods consistently overestimate the contact point's position during the contact phase. Unlike these methods, the Schwarz variants do not exhibit this behavior.}
    \label{fig:hsp_contact_pt_position}
\end{figure}

\begin{figure}
    \centering
    \includegraphics[page=3,trim={0mm 25mm 0mm 25mm},clip,width=1\textwidth]{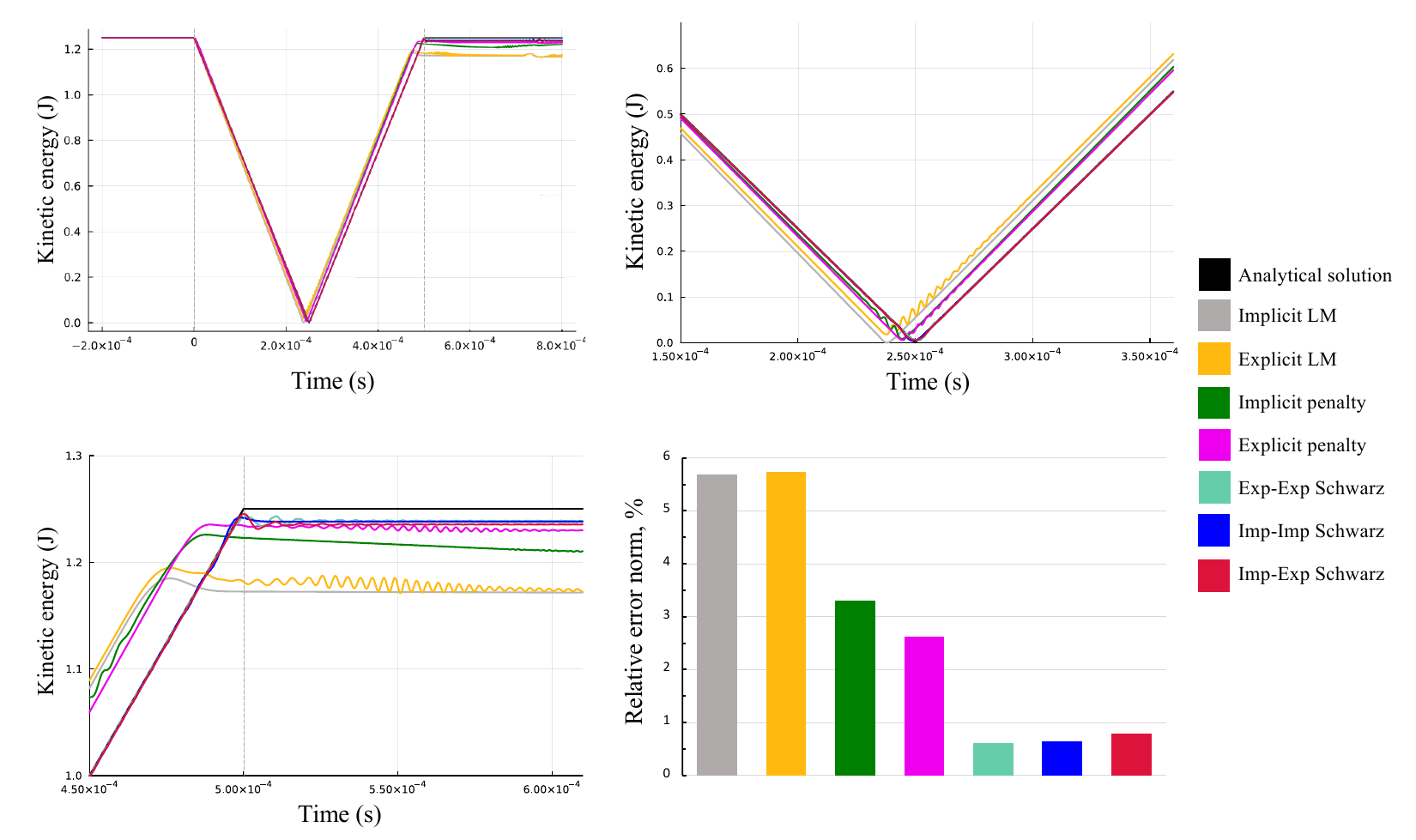}
    \caption{Time evolution of kinetic energy. The bottom-right plot presents total relative errors in kinetic energy for both traditional contact methods and the Schwarz alternating method. Conventional methods suffer from noticeable inaccuracies and oscillations in kinetic energy after contact occurs. Specifically, these methods fail to accurately capture the lowest point of the kinetic energy curve. In contrast, all three Schwarz variants demonstrate minimal errors and provide accurate estimates for kinetic energy. These methods are capable of capturing energy peaks with a relative error of less than $0.1\%$ when compared to the analytical solution.
    }
    \label{fig:hsp_ke}
\end{figure}

\begin{figure}
    \centering
    \includegraphics[page=2,trim={0mm 25mm 0mm 25mm},clip,width=1\textwidth]{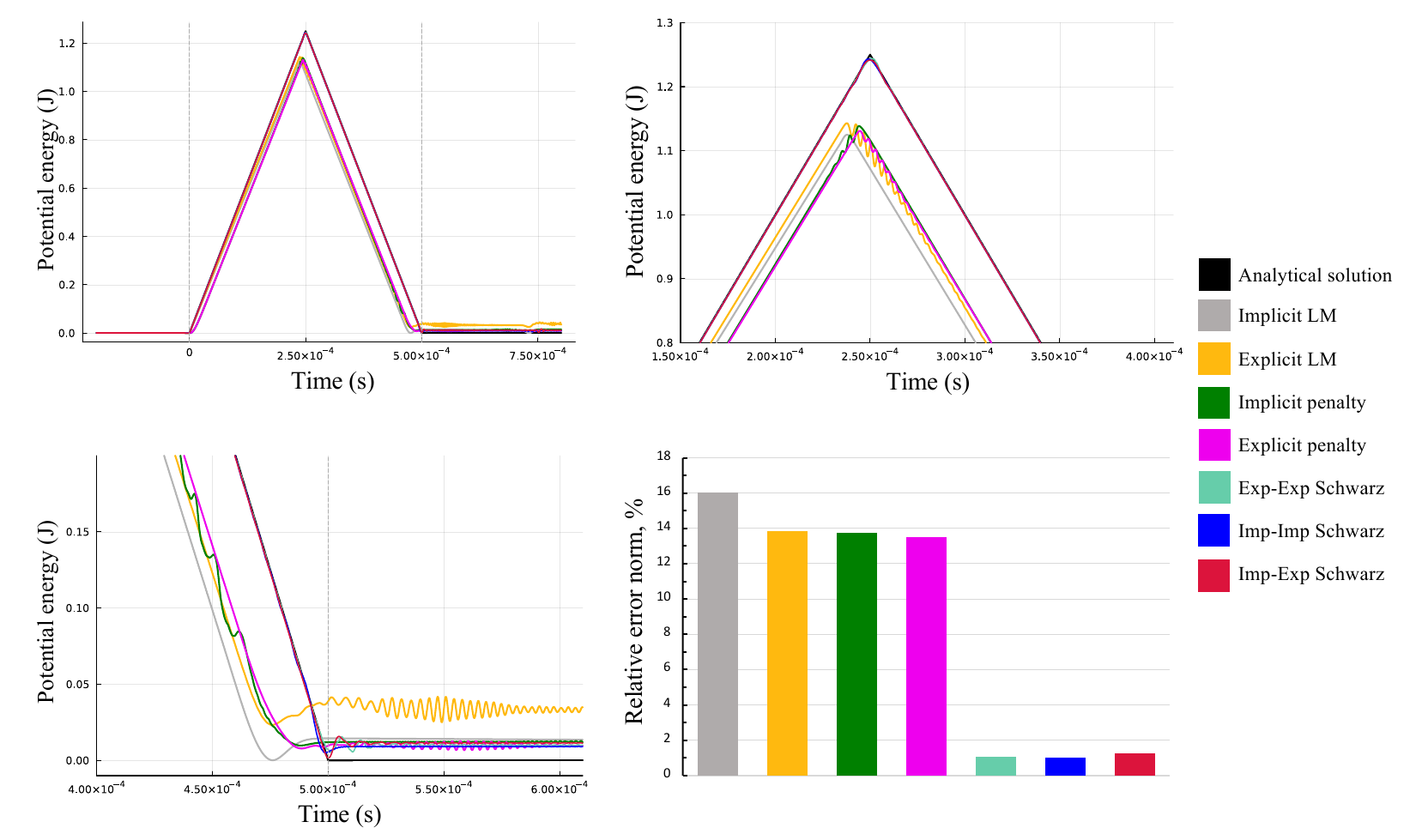}
    \caption{Time evolution of potential energy. The bottom-right plot presents total relative errors in potential energy for both traditional contact methods and the Schwarz alternating method. Conventional methods suffer from noticeable inaccuracies and oscillations in potential energy after contact occurs. Specifically, these methods underestimate the peak value by roughly 10\% and introduce artifacts around the time of release. In contrast, all three Schwarz variants demonstrate minimal errors and provide accurate estimates for potential energy. These methods are capable of capturing energy peaks with a relative error of less than $0.1\%$ when compared to the analytical solution.}
    \label{fig:hsp_pe}
\end{figure}

\begin{figure}
    \centering
    \includegraphics[page=4,trim={0mm 110mm 0mm 30mm},clip,width=1\textwidth]{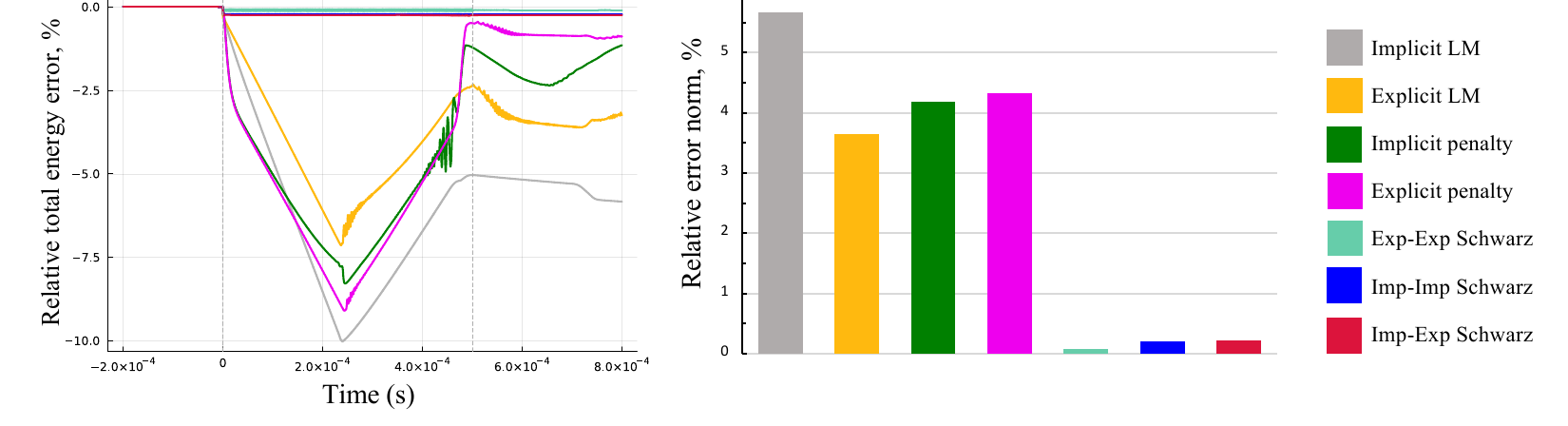}
    \caption{Time evolution of total energy error. The right plot presents total relative errors in the total energy for both traditional contact methods and the Schwarz alternating method. Conventional methods display up to a $10\%$ loss in total energy upon contact initiation. In contrast, the Schwarz method attains a maximum error of merely $0.25\%$ in total energy. Notably, this excellent energy performance is an inherent feature of the Schwarz method in its \emph{classical} form, a trait seldom seen in conventional methods. Further, the total relative errors for the Schwarz methods peak at $0.2\%$, significantly lower than the $3.5-5.5\%$ errors seen with conventional methods.}
    \label{fig:hsp_te_error}
\end{figure}

\begin{figure}
    \centering
    \includegraphics[page=6,trim={0mm 110mm 0mm 25mm},clip,clip,width=1\textwidth]{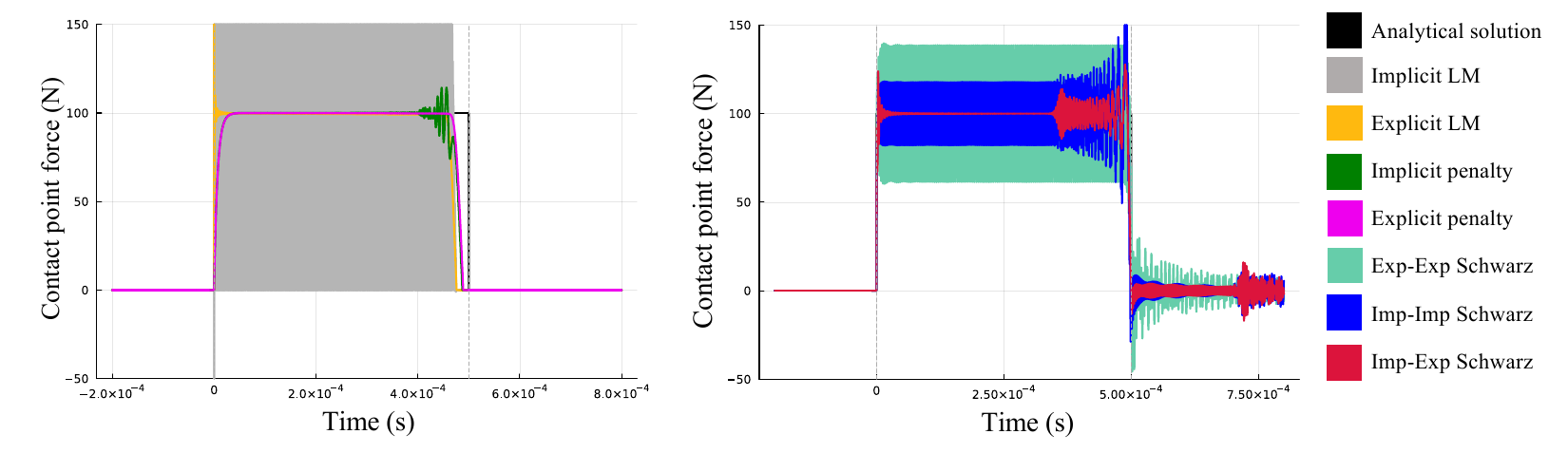}
    \caption{Time evolution of contact point force for the left bar (denoted as domain $\Omega^1$). Overall, conventional methods generally produce a relatively smooth contact force solution, albeit with some noticeable issues. For instance, the explicit Lagrange multiplier method shows a spike at the impact and release times, while the penalty method manifests minor oscillations. Notably, the implicit Lagrange multiplier method exhibits significant artificial oscillations.}
    \label{fig:hsp_contact_force}
\end{figure}

\begin{figure}
    \centering
    \includegraphics[page=5,trim={0mm 110mm 0mm 25mm},clip,width=1\textwidth]{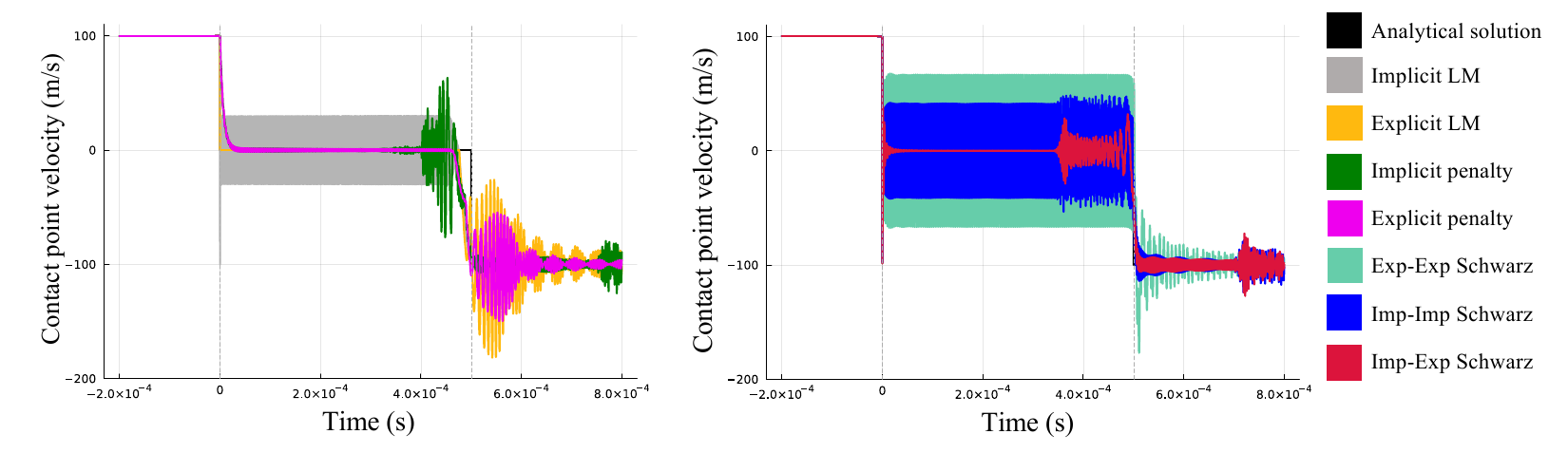}
    \caption{Time evolution of contact point velocity for the left bar (denoted as domain $\Omega^1$). Similar trends to the contact force from Figure \ref{fig:hsp_contact_force} are evident in the behavior of the contact point velocity as well. Especially noteworthy are the artificial oscillations in both contact point force and velocity that emerge after the contact begins. These spurious fluctuations are a recognized challenge in mechanical contact simulations and remain an active area of research within the contact dynamics community.}
    \label{fig:hsp_contact_pt_velocity}
\end{figure}

Our primary findings are presented in Figures \ref{fig:hsp_contact_pt_position}--\ref{fig:hsp_contact_pt_velocity}. We offer both qualitative and quantitative evaluations of our numerical simulations.

To assess the accuracy of the numerical solutions, we calculate relative errors compared to the analytical solution. We denote numerical and analytical values for key variables as $ [\cdot]^{\mathrm{num}} $ and $ [\cdot]^{\mathrm{ana}} $, respectively. Using the Euclidean norm, the formula for the relative error is
\begin{equation}
  \epsilon^{\mathrm{rel}} := \frac{ \|\, [\cdot]^{\mathrm{num}} - [\cdot]^{\mathrm{ana}} \|\, }{ \| [\cdot]^{\mathrm{ana}} \| } \times 100\%.
  \label{eq:total_error}
\end{equation}
We employ this metric to evaluate various key variables, including the position and force at the contact point, as well as the kinetic, potential, and total energies.

Figure \ref{fig:hsp_contact_pt_position} depicts the time-dependent position of the right-most node in the left bar ($\Omega^1$).
Notably, the penalty and Lagrange multiplier methods tend to underestimate the time of release. Additionally, penalty methods consistently overestimate the contact point's position during the contact phase, as highlighted by \textcite{Carpenter:1991}.
Unlike these methods, the Schwarz variants do not exhibit this behavior.
Though minor oscillations are visible in the contact point positions determined by the Schwarz methods, they represent an insignificant deviation from the exact location. Remarkably, the Schwarz methods approximate the release time with an impressive accuracy of approximately $0.01\%$ relative to the analytical solution.

The bottom plot in Figure \ref{fig:hsp_contact_pt_position} provides the relative errors. It is evident that all Schwarz contact methods achieve remarkably low relative errors, less than $1\%$, when compared to the analytical solution. In contrast, traditional contact methods result in errors ranging between $6\%$ and $9\%$.

Figures \ref{fig:hsp_ke} and \ref{fig:hsp_pe} display the time-dependent behavior of kinetic and potential energies for the left bar. These figures show that conventional methods suffer from noticeable inaccuracies and oscillations in both kinetic and potential energies after contact occurs. Specifically, these methods fail to accurately capture the lowest point of the kinetic energy curve, as indicated in Figure \ref{fig:hsp_ke}. For potential energy, conventional approaches underestimate the peak value by roughly 10\% and introduce artifacts around the time of release, as seen in Figure \ref{fig:hsp_pe}.

In stark contrast, all three Schwarz variants demonstrate minimal errors and provide accurate estimates for both kinetic and potential energies. These methods are capable of capturing energy peaks with a relative error of less than $0.1\%$ when compared to the analytical solution.

Regarding total relative errors, conventional methods result in errors ranging from $2.5\%$ to $5.5\%$ for kinetic energy and from $13\%$ to $16\%$ for potential energy. Schwarz contact methods, however, achieve much higher precision, with errors less than $1\%$, as substantiated by the bottom-right plots in Figures \ref{fig:hsp_ke} and \ref{fig:hsp_pe}.

Next, we turn our attention to the total energy behavior as demonstrated by the various contact methods under investigation. Figure~\ref{fig:hsp_te_error} shows the time-dependent relative errors in total energy in comparison to the analytical solution. The total energy is the sum of kinetic and potential energies.

The importance of energy conservation in contact problems cannot be overstated, particularly for ensuring accurate long-term simulations, as emphasized by \textcite{Doyen:2011}. Nevertheless, many conventional computational methods are suboptimal in conserving energy.

Our results substantiate this claim. Conventional methods display up to a $10\%$ loss in total energy upon contact initiation. In contrast, the Schwarz method attains a maximum error of merely $0.25\%$ in total energy. Notably, this excellent energy performance is an inherent feature of the Schwarz method in its \emph{classical} form, a trait seldom seen in conventional methods.

Further, the total relative errors for the Schwarz methods peak at $0.2\%$, significantly lower than the $3.5-5.5\%$ errors seen with conventional methods, as illustrated in Figure~\ref{fig:hsp_te_error} on the right.

It is worth pointing out that achieving energy conservation with conventional methods often requires fine spatial and temporal discretizations, adding to the computational burden, as stated by \textcite{Doyen:2011, DiStasio:2019}. In contrast, our Schwarz approach maintains high performance even with relatively coarse discretizations.

In summary, the Schwarz method excels in various metrics, offering precise estimations for key physical parameters like impact/release times, contact positions, and energies, while also maintaining remarkable energy conservation. Nonetheless, its performance slightly deviates when examining the contact point force and velocity, as detailed in Figures \ref{fig:hsp_contact_force} and \ref{fig:hsp_contact_pt_velocity}.

Overall, conventional methods generally produce a relatively smooth contact force solution, albeit with some noticeable issues. For instance, the explicit Lagrange multiplier method shows a spike at the impact and release times, while the penalty method manifests minor oscillations. These observations are evident from Figure \ref{fig:hsp_contact_force}. Notably, the implicit Lagrange multiplier method exhibits significant artificial oscillations, corroborating the findings by \textcite{Carpenter:1991}.

Similar trends are evident in the behavior of the contact point velocity, as illustrated in Figure \ref{fig:hsp_contact_pt_velocity}. Especially noteworthy are the artificial oscillations in both contact point force and velocity that emerge after the contact begins. These spurious fluctuations are a recognized challenge in mechanical contact simulations and remain an active area of research within the contact dynamics community \cite{Doyen:2011, Wriggers:2006, DiStasio:2019}. For further discussions on this subject, refer to Section \ref{sec:chatter_intro}.

Our numerical tests indicate that the oscillations are largely insensitive to the convergence tolerances set for the Schwarz method. Notably, employing an implicit-explicit coupling in the Schwarz framework significantly attenuates these artificial oscillations.

\begin{figure}
    \centering
    \includegraphics[page=7,trim={0mm 120mm 0mm 25mm},clip,width=1\textwidth]{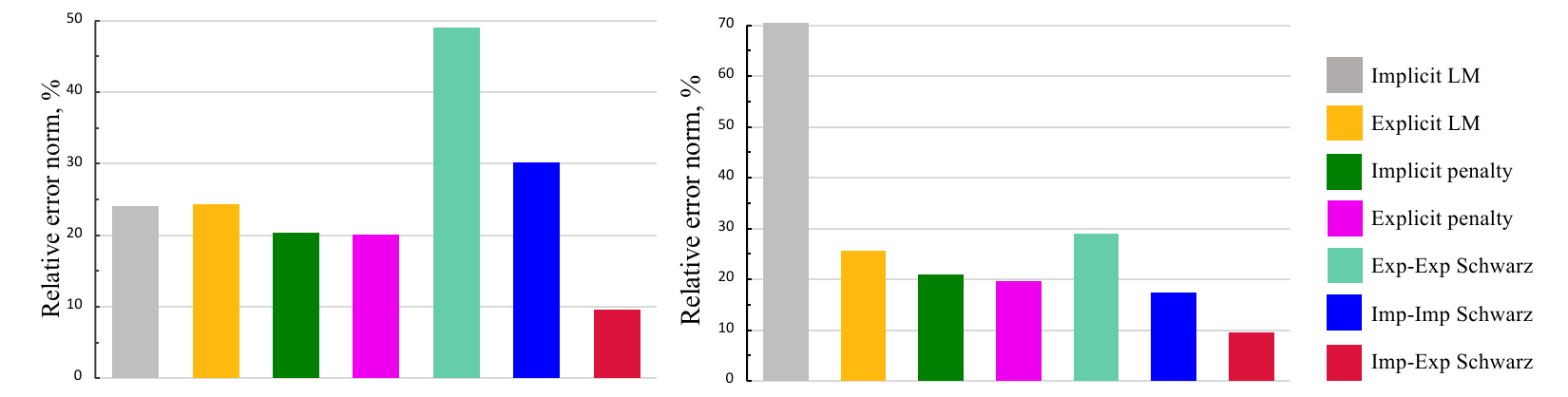}
    \caption{Total relative errors for the contact point velocity (left) and contact point force (right) using conventional contact methods and the Schwarz alternating method. There is a noteworthy relationship between energy loss and the degree of oscillations. By comparing Figure \ref{fig:hsp_te_error} with Figures \ref{fig:hsp_contact_force}, \ref{fig:hsp_contact_pt_velocity}, and \ref{fig:velocity_force_errors}, it becomes apparent that the method experiencing the greatest loss of total energy shows the fewest oscillations. This observation aligns well with existing literature, which argues that some level of energy dissipation is essential for maintaining persistent contact \cite{Solberg:1998}.}
    \label{fig:velocity_force_errors}
\end{figure}

A significant correlation between energy dissipation and oscillatory behavior is evident in Figure \ref{fig:hsp_te_error}. When contrasted with Figures \ref{fig:hsp_contact_force}, \ref{fig:hsp_contact_pt_velocity}, and \ref{fig:velocity_force_errors}, a general trend surfaces: methods with greater energy loss typically display fewer oscillations. It is important to note, however, that the Implicit Lagrange Multiplier method is an exception, exhibiting both the highest energy loss and the most oscillations, particularly in the contact force. This observation is consistent with existing literature, which emphasizes the necessity of some level of energy dissipation for maintaining persistent contact~\cite{Solberg:1998}.

In Section \ref{sec:chatter}, we outline strategies to minimize or completely remove these oscillations. Importantly, these proposed solutions are minimally intrusive and do not compromise the accuracy and energy-conserving characteristics of the Schwarz algorithm.

\subsubsection{Evaluating the convergence of the Schwarz alternating method} \label{sec:convergence}

We next assess the convergence performance of the Schwarz alternating method for contact problems, focusing on a single quantity of interest (QOI): the kinetic energy of the left bar.

\begin{figure}
    \centering
    \subfigure[Mesh convergence for $\triangle t = \qty{1e-8}{\second}$]
        {\includegraphics[trim={5mm 48mm 10mm 61mm},clip,width=0.42\textwidth]{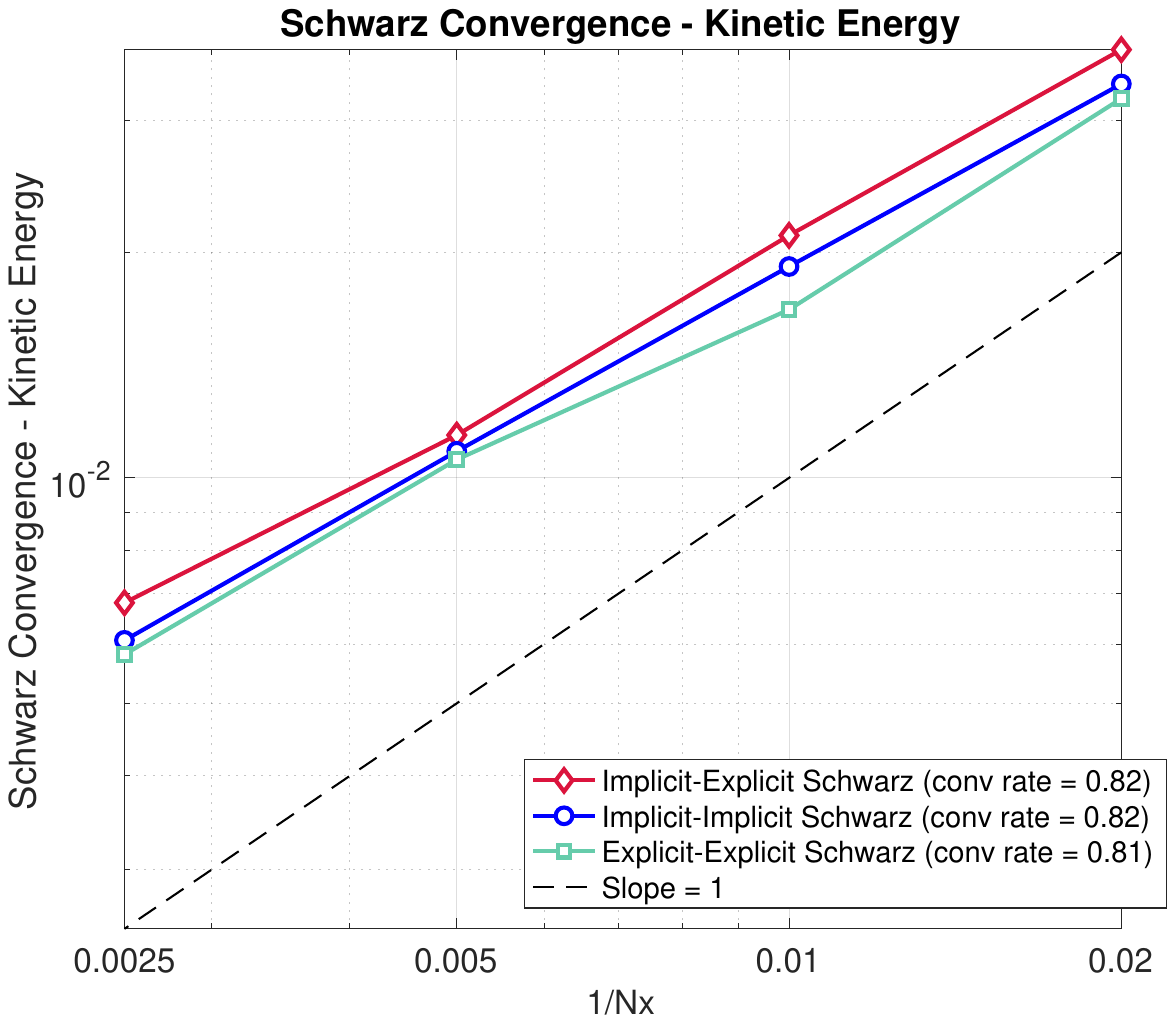}}
    \subfigure[Number of Schwarz iterations for $N_x = 200$, $\triangle t =\qty{1e-7}{\second}$]
        {\includegraphics[trim={0mm 0mm 0mm 10mm},clip,width=0.57\textwidth]{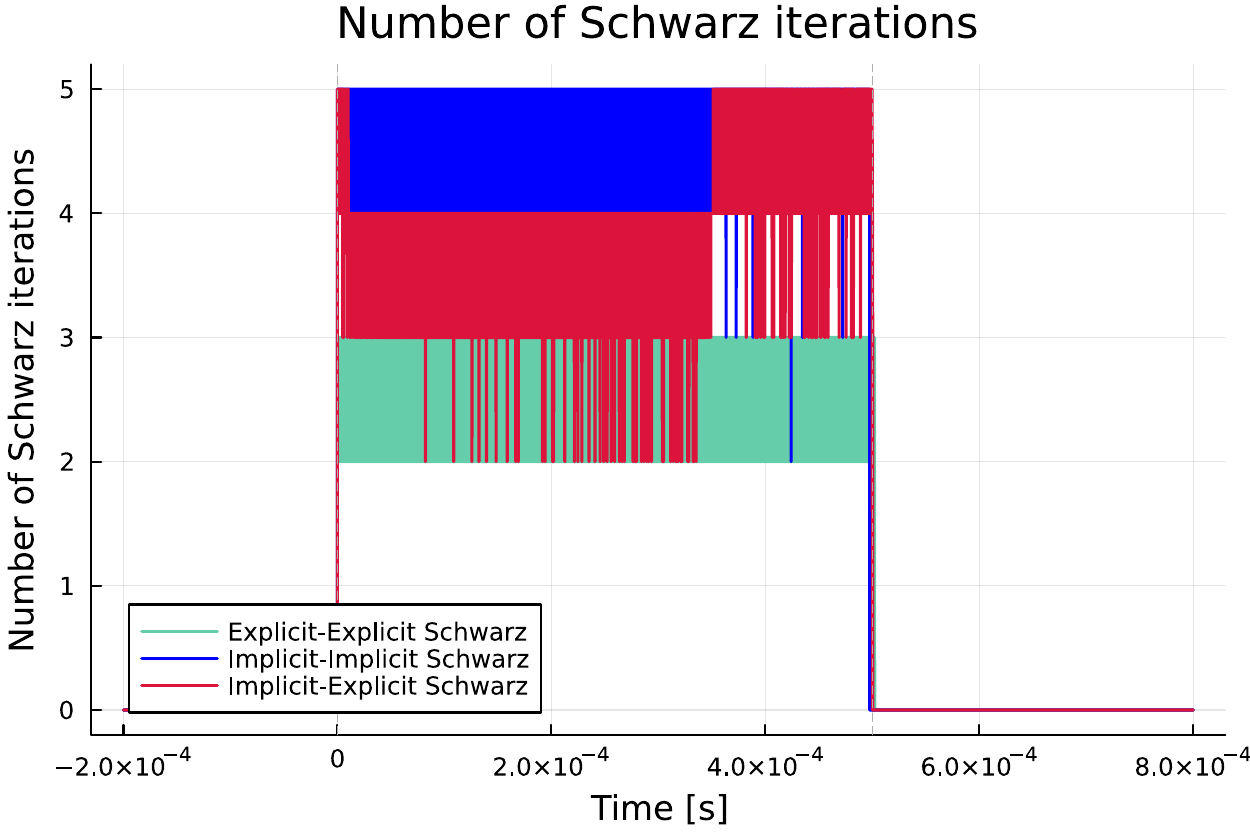}}
    \caption{Convergence metrics for various Schwarz couplings. (a) The mesh is incrementally refined, starting from $N_x = 50$ elements and going up to $N_x = 400$, with a constant time step of $\triangle t = \qty{1e-8}{\second}$ employed in both subdomains. Across all variants, we observe a convergence rate of approximately $0.82$. (b) Between two and five Schwarz iterations are required for convergence, depending on the coupling type. This is observed when a mesh resolution of $N_x = 200$ and a time step of $\triangle t = \qty{1e-7}{\second}$ are used. Interestingly, explicit-explicit Schwarz requires the least number of iterations for convergence at this resolution, ranging between two and three.}
    \label{fig:hsp_mesh_conv}
\end{figure}

In Figure \ref{fig:hsp_mesh_conv}(a), we present the mesh convergence rates for the three Schwarz method variants. The mesh is incrementally refined, starting from $N_x = 50$ elements and going up to $N_x = 400$, with a constant time step of $\triangle t = \qty{1e-8}{\second}$ employed in both subdomains. Across all variants, we observe a convergence rate of approximately $0.82$. This rate is similar to what has been observed for a low-speed version of this problem using Sandia's ALEGRA code, which employs the forward increment explicit Lagrange multiplier method~\cite{Tezaur:2021}.

Finally, we offer insights into the computational efficiency by discussing the number of Schwarz iterations needed for convergence. Figure \ref{fig:hsp_mesh_conv}(b) illustrates that between two and five Schwarz iterations are required for convergence, depending on the coupling type. This is observed when a mesh resolution of $N_x = 200$ and a time step of $\triangle t = \qty{1e-7}{\second}$ are used. Interestingly, explicit-explicit Schwarz requires the least number of iterations for convergence at this resolution, ranging between two and three. It should be noted that Schwarz iterations are unnecessary before the bars make contact and after they separate.

Tables \ref{tab:schwarz_iters_Nx_conv} and \ref{tab:schwarz_iters_dt_conv} present the maximum and average number of Schwarz iterations. These are reported both as functions of the spatial mesh resolution $N_x$ (keeping the time step fixed at $\triangle t = \qty{1e-8}{\second}$) and as functions of the time step $\triangle t$ (with a constant mesh resolution of $N_x = 200$).

As the reader will note, the number of Schwarz iterations generally increases slightly with finer spatial mesh resolutions. Among the different coupling strategies, implicit-implicit coupling typically demands the most Schwarz iterations. Interestingly, the number of iterations remains largely stable when varying the time step, except in the case of explicit-explicit Schwarz coupling. A noticeable change in the required number of iterations occurs between time steps of $\qty{1e-7}{\second}$ and $\qty{1e-8}{\second}$, as indicated in Table \ref{tab:schwarz_iters_dt_conv}.

\begin{table}
    \footnotesize
    \centering
    \begin{tabular}{r c c c}
        \toprule
        $N_x$ & Implicit-Implicit & Implicit-Explicit & Explicit-Explicit
        \\
        \midrule
        50 & 4/1.7 & 4/1.7 & 4/1.8
        \\
        100 & 5/1.9 & 4/1.8 & 4/1.8
        \\
        200 & 5/2.2 & 5/2.0 & 5/2.2
        \\
        400 & 6/2.6 & 5/2.3 & 5/2.3
        \\
        \bottomrule
    \end{tabular}
    \caption{Maximum/average number of Schwarz iterations as a function of the spatial mesh resolution $N_x$ keeping the time step fixed at $\triangle t = \qty{1e-8}{\second}$ for various Schwarz couplings. The number of Schwarz iterations generally increases slightly with finer spatial mesh resolutions. Among the different coupling strategies, implicit-implicit coupling typically demands the most Schwarz iterations.}
    \label{tab:schwarz_iters_Nx_conv}
\end{table}

\begin{table}
    \footnotesize
    \centering
    \begin{tabular}{c c c c}
        \toprule
        $\triangle t$ & Implicit-Implicit & Implicit-Explicit & Explicit-Explicit
        \\
        \midrule
        $\qty{1e-7}{\second}$ & 5/2.4 & 5/1.9 & 3/1.2
        \\
        $\qty{1e-8}{\second}$ & 5/2.2 & 5/2.0 & 5/2.2
        \\
        $\qty{1e-9}{\second}$ & 5/2.2 & 5/2.1 & 5/2.2
        \\
        \bottomrule
    \end{tabular}
    \caption{Maximum/average number of Schwarz iterations as a function of the time step $\triangle t$ with a constant mesh resolution of $N_x = 200$ for various Schwarz couplings. The number of iterations remains largely stable when varying the time step, except in the case of explicit-explicit Schwarz coupling. A noticeable change in the required number of iterations occurs between time steps of $\qty{1e-7}{\second}$ and $\qty{1e-8}{\second}$.}
    \label{tab:schwarz_iters_dt_conv}
\end{table}
\section{Mitigation of spurious oscillations in the Schwarz contact method} \label{sec:chatter}

Section~\ref{sec:results} highlighted the Schwarz alternating method as a compelling alternative to traditional techniques in simulating mechanical contact. Despite its superior performance in terms of accuracy and energy conservation, the method also manifests artificial oscillations in both contact forces and velocities. The focus of this section is to introduce a minimally invasive approach that effectively suppresses these oscillations, without sacrificing the advantages of the Schwarz method.

Specifically, we suggest modifying the time-integration scheme that governs the system of equations as a plausible avenue for alleviating chatter. To accomplish this, we evaluate several time-integration schemes and stabilization methods that have been shown to reduce spurious oscillations in traditional contact algorithms. While we adapt these approaches to the Schwarz method, we refrain from exploring their impacts on conventional methods, as such investigations have already been conducted in existing literature \cite{Chaudhary.Bathe:1986, Tchamwa:1999, Chung.Lee:1994, Doyen:2011, Kane.etal:1999, Deuflhard:2008}.

\subsection{Time-integration schemes and stabilization techniques} \label{sec:time_integr}

Our review of the literature, as outlined in Section \ref{sec:chatter_intro}, reveals that modified versions of the Newmark-$\beta$ time-integrator and stabilization techniques may be particularly fitting for our Schwarz method. Specifically, we evaluate the following schemes, summarized later in Table \ref{tab:compar}:
\begin{enumerate}
    \item Classic Newmark-$\beta$ algorithm with dissipative $\beta$ and $\gamma$ values (implicit/explicit).
    \item Chaudhary-Bathe scheme \cite{Chaudhary.Bathe:1986} (implicit).
    \item Tchamwa-Wielgosz scheme \cite{Tchamwa:1999} (explicit).
    \item Chung-Lee scheme \cite{Chung.Lee:1994} (explicit).
    \item Naïve-stabilized Newmark-$\beta$ scheme \cite{Doyen:2011} (implicit/explicit).
    \item Contact-implicit Newmark-$\beta$ scheme \cite{Kane.etal:1999} (implicit).
    \item Contact-stabilized Newmark-$\beta$ scheme \cite{Deuflhard:2008} (implicit).
\end{enumerate}
Schemes 1-4 represent modified Newmark-$\beta$ approaches, while schemes 5-7 are categorized as stabilization methods.

To ensure the self-contained nature of this paper, we first introduce the classic explicit and implicit Newmark-$\beta$ time integration schemes. We then provide a brief overview of the alternative time-integration schemes and stabilization techniques mentioned above. Readers interested in more in-depth explanations are encouraged to consult the cited references.

\paragraph{Classic Newmark-$\beta$ scheme.} This scheme relies on Taylor expansions of the position $~x$ and velocity $~v$ fields \cite{Wriggers:2006, Rao:2017}. Using the notations from previous sections, let $t_{k}$ and $t_{k+1}$ be two consecutive time steps, where $\triangle t = t_{k+1} - t_{k}$. Ignoring higher-order terms in the Taylor series and employing the traditional parameters $\gamma, 2\beta \in [0,1]$, the following equations emerge
\begin{equation}
    \begin{split}
        ~x_{t_{k+1}} & = ~x_{t_{k}} + \triangle t ~v_{t_{k}} + \frac{(\triangle t)^2}{2} \left[ (1 - 2\beta) ~a_{t_{k}} + 2\beta ~a_{t_{k+1}} \right]
        \\
        ~v_{t_{k+1}} & = ~v_{t_{k}} +  \triangle t  \left[ (1 - \gamma) ~a_{t_{k}} + \gamma ~a_{t_{k+1}} \right]
    \end{split}
    \label{eq_stdNewmark}
\end{equation}
The classic Newmark-$\beta$ scheme can be either first-order or second-order accurate, based on the chosen values of $\beta$ and $\gamma$. Furthermore, it can be either implicit, ensuring unconditional stability, or explicit, its stability contingent upon the parameter values.

The explicit Newmark-$\beta$ scheme, also known as the central difference scheme, can be derived by setting $\gamma = \frac{1}{2}$ and $\beta = 0$. This approach is straightforward to implement since it only uses known quantities to estimate the solution at the subsequent time step $t_{k+1}$. It is also computationally efficient, especially when a diagonalized (lumped) mass matrix $~M$ is utilized. This explicit approach, however, is conditionally stable and governed by the Courant–Friedrichs–Lewy (CFL) criterion. The explicit Newmark-$\beta$ algorithm is detailed in Algorithm~\ref{alg:Class_Exp_Newmark}.

\begin{algorithm}
    \footnotesize
	\begin{algorithmic}[1]
		\State
		$~v^{\mathrm{pred}}_{t_{k+1}} \gets ~v_{t_{k}} + \triangle t  (1 - \gamma) ~a_{t_{k}}  $
		\Comment{Prediction step}

		\State  Solve
		$ ~M \triangle ~a_{t_{k+1}} = ~f^{\mathrm{ext}}_{t_{k+1}}  - ~f^{\mathrm{int}}(~x_{t_{k+1}}) - ~M ~a_{t_{k}} $
        \Comment{Solve linear system}

		\State $~a_{t_{k+1}} \gets ~a_{t_{k}} + \triangle ~a_{t_{k+1}} $
        \Comment{Correction step}

		\State  $~v_{t_{k+1}} \gets  ~v^{\mathrm{pred}}_{t_{k+1}}  + \triangle t \gamma ~a_{t_{k+1}} $

		\State$~x_{t_{k+1}} \gets ~x_{t_{k}} + \triangle t ~v_{t_{k}} + \frac{(\triangle t)^2}{2}  ~a_{t_{k}}$
	\end{algorithmic}
	\caption{Classic explicit Newmark-$\beta$ scheme.} \label{alg:Class_Exp_Newmark}
\end{algorithm}

The implicit Newmark-$\beta$ scheme, often referred to as the trapezoidal rule, utilizes the standard parameters $\gamma = \frac{1}{2}$ and $\beta = \frac{1}{4}$. This scheme is unconditionally stable, regardless of the time step chosen. The implicit approach, however, is computationally more demanding, as it requires solving a nonlinear equation at each time step. This solution is often obtained through a Newton-type iterative method. The corresponding algorithmic details are outlined in Algorithm~\ref{alg:Class_Imp_Newmark}.

\begin{algorithm}
    \footnotesize
	\begin{algorithmic}[1]
		\State  $~x^{\mathrm{pred}}_{t_{k+1}} \gets ~x_{t_{k}} + \triangle t ~v_{t_{k}} + (\triangle t)^2 (\frac{1}{2} - \beta)  ~a_{t_{k}}$
        \Comment{Prediction step}

		\State  $~v^{\mathrm{pred}}_{t_{k+1}} \gets ~v_{t_{k}} + \triangle t  (1 - \gamma) ~a_{t_{k}}  $

		\State  $~a^{\mathrm{pred}}_{t_{k+1}} \gets \frac{1}{(\triangle t)^2 \beta} (~x - ~x^{\mathrm{pred}}_{t_{k+1}}) $

		\State $\ell \gets 0$

		\State $~x^{\ell}_{t_{k+1}} \gets ~x^{\mathrm{pred}}_{t_{k+1}} $, \quad  $~v^{\ell}_{t_{k+1}} \gets ~v^{\mathrm{pred}}_{t_{k+1}} $
		\quad $~a^{\ell}_{t_{k+1}} \gets ~a^{\mathrm{pred}}_{t_{k+1}} $
        \Comment{Initialization of Newton-type iterations}

		\Repeat
        \Comment{Newton-type iterations}
		      \State  Solve
		      $[ \frac{1}{ (\triangle t)^2 \beta} ~M + ~K(~x^{\ell}_{t_{k+1}}) ] \triangle ~x^{\ell+1}_{t_{k+1}}  =   ~f^{\mathrm{ext}}_{t_{k+1}} - ~f^{\mathrm{int}}(~x^{\ell}_{t_{k+1}}) - ~M  ~a^{\ell}_{t_{k+1}} $
		      \Comment{Solve linear system}

		      \State     $~x^{\ell+1}_{t_{k+1}} \gets ~x^{\ell}_{t_{k+1}}  + \triangle~x^{\ell+1}_{t_{k+1}}  $
            \Comment{Correction step}

		      \State  $~a^{\ell+1}_{t_{k+1}} \gets  \frac{1}{(\triangle t)^2 \beta} (~x^{\ell+1}_{t_{k+1}}- ~x^{\mathrm{pred}}_{t_{k+1}})$

		      \State  $~v^{\ell+1}_{t_{k+1}} \gets  ~v^{\mathrm{pred}}_{t_{k+1}} + \triangle t \gamma ~a^{\ell+1}_{t_{k+1}}$
    		\State $\ell \gets \ell + 1$
		\Until{converged}
	\end{algorithmic}
    \caption{Classic implicit Newmark-$\beta$ scheme} \label{alg:Class_Imp_Newmark}
\end{algorithm}

In this context, the mass and stiffness matrices resulting from the finite element discretization of \eqref{eq:semidiscrete} are denoted as $~M$ and $~K$, respectively. The Newton-type iterative process continues until the prescribed error tolerance between two consecutive iterations (measured in a selected norm) is achieved.

\paragraph{Dissipative Newmark-$\beta$ scheme.} \textcite{Fung:2003, Wriggers:2006} introduce a controlled amount of numerical dissipation to filter out high-frequency responses, thus reducing artificial numerical oscillations. This is achieved by varying the parameters $\gamma$ and $\beta$ in Algorithms \ref{alg:Class_Exp_Newmark} and \ref{alg:Class_Imp_Newmark}, respectively, for explicit and implicit integration. The commonly recommended choices for these parameters are $\gamma > \frac{1}{2}$ and $\beta = \frac{1}{4} \left( \gamma + \frac{1}{2} \right)^2$~\cite{Fung:2003}.

\paragraph{Chaudhary-Bathe scheme.} As proposed by \textcite{Chaudhary.Bathe:1986}, this scheme improves upon the implicit Newmark-$\beta$ scheme for impact problems by using $\gamma = \beta = \frac{1}{2}$. The authors demonstrate that this choice of parameters results in an energy and momentum balance for the contacting bodies, particularly when a reasonably small time step is employed.

\paragraph{Tchamwa-Wielgosz scheme.} Introduced by \textcite{Tchamwa:1999, Rio:2005}, this explicit scheme offers controllable high-frequency dissipation. While it is only first-order accurate, the scheme results in smaller numerical errors in the low-frequency range, leading to a less perturbed response in structural dynamic analysis. Using a parameter $\Phi \geq 1$, the authors propose fully explicit expressions for positions and velocities, detailed in Algorithm~\ref{alg:Tchamwa_Wielgosz}.

\begin{algorithm}
    \footnotesize
	\begin{algorithmic}[1]
		\State Solve
		$ ~M \triangle ~a_{t_{k+1}} = ~f^{\mathrm{ext}}_{t_{k+1}}  - ~f^{\mathrm{int}}(~x_{t_{k+1}}) - ~M ~a_{t_{k}} $
        \Comment{Solve linear system}

		\State $~a_{t_{k+1}} \gets ~a_{t_{k}} + \triangle ~a_{t_{k+1}} $
        \Comment{Correction step}

		\State$~x_{t_{k+1}} \gets ~x_{t_{k}} + \triangle t ~v_{t_{k}} +    (\triangle t)^2 \Phi ~a_{t_{k}}$

		\State  $~v_{t_{k+1}} \gets ~v_{t_{k}} + \triangle t  ~a_{t_{k}}  $
	\end{algorithmic}
	\caption{Tchamwa-Wielgosz scheme}
	\label{alg:Tchamwa_Wielgosz}
\end{algorithm}

\paragraph{Chung-Lee scheme.} \textcite{Chung.Lee:1994} introduce a family of explicit, single-step time-integration schemes with adjustable high-frequency dissipation. The authors demonstrate the second-order accuracy and stability of the method across linear examples, both damped and undamped. This scheme is encapsulated in Algorithm~\ref{alg:Chung_Lee}, with recommended parameter choices as follows: $\gamma = \frac{3}{2}$, $1 \leq \beta \leq \frac{28}{27}$, $\hat{\beta} = \frac{1}{2} - \beta$, and $\hat{\gamma} = 1 - \gamma$.

\begin{algorithm}
    \footnotesize
	\begin{algorithmic}[1]
		\State $~x^{\mathrm{pred}}_{t_{k+1}} \gets ~x_{t_{k}} + \triangle t ~v_{t_{k}} +  (\triangle t)^2 \hat{\beta} ~a_{t_{k}} $
        \Comment{Prediction step}

		\State $~v^{\mathrm{pred}}_{t_{k+1}} \gets ~v_{t_{k}} +  \triangle t  \hat{\gamma} ~a_{t_{k}} $

		\State Solve
		$ ~M \triangle ~a_{t_{k+1}} =  ~f^{\mathrm{ext}}_{t_{k+1}}  - ~f^{\mathrm{int}}(~x_{t_{k+1}}) - ~M ~a_{t_{k}} $
        \Comment{Solve linear system}

		\State $~a_{t_{k+1}} = ~a_{t_{k}} + \triangle ~a_{t_{k+1}} $
        \Comment{Correction step}

		\State$~x_{t_{k+1}} = ~x^{\mathrm{pred}}_{t_{k+1}} +  (\triangle t)^2 \beta ~a_{t_{k+1}}$

		\State  $~v_{t_{k+1}} = ~v^{\mathrm{pred}}_{t_{k+1}} + \triangle t  \gamma ~a_{t_{k+1}} $
	\end{algorithmic}
	\caption{Chung-Lee scheme}
	\label{alg:Chung_Lee}
\end{algorithm}

Another class of methods relies on stabilization techniques \cite{Kane.etal:1999, Deuflhard:2008, Suwannachit:2012}. As discussed in Section~\ref{sec:chatter_intro}, these authors suggest that the primary cause of spurious oscillations stems from the discrete boundary mass being translated into forces at the contact boundary. Such oscillations are purely numerical artifacts, as the boundary is assigned mass solely due to spatial discretization. In the continuous case, the boundary would have a measure of zero. Therefore, the rationale behind stabilization techniques involves eliminating the non-physical components of the boundary forces by nullifying the acceleration on the contact boundary.

\paragraph{Contact-implicit method.} As described by \textcite{Kane.etal:1999}, the contact-implicit method focuses on a fully implicit treatment of the contact forces. In the original formulation, the authors suggest partitioning the accelerations into two separate terms, $~a^{\mathrm{int}}$ and $~a^{\mathrm{con}}$, associated with the internal forces and contact forces, respectively. This division allows for the individual treatment of each component as
\begin{equation}
   ~a = ~a^{\mathrm{int}} + ~a^{\mathrm{con}}
   \label{eq:accel_split}
\end{equation}
where the internal acceleration is expressed as $ ~a^{\mathrm{int}} = ~M^{-1} ( ~f^{\mathrm{ext}} - ~f^{\mathrm{int}}(~x)) $ and the contact acceleration is written as $ ~a^{\mathrm{con}} = - ~M^{-1}  ~f^{\mathrm{con}}(~x) $.
Subsequently, the equations of motion are adapted to include the partitioning of the inertia term \eqref{eq:accel_split}, as detailed in \textcite{Kane.etal:1999, Suwannachit:2012}.

When implementing this algorithm within our Schwarz contact method, we observed that only modifications to the correction step are pertinent. This is due to the alternation between position and traction boundary conditions used for managing contact. Therefore, our version introduces the acceleration decomposition, as detailed in \eqref{eq:accel_split}, and incorporates it into the computation of the velocity. This augmented correction step is outlined in Algorithm~\ref{alg:Contact_imp_Newmark} and is executed during the contact phase as part of the standard Newmark implicit Algorithm~\ref{alg:Class_Imp_Newmark}. It is worth noting that during time steps when contact is not detected, the standard implicit integration scheme remains applicable. This adaptation maintains the prediction and solution steps of the classical Newton Algorithm~\ref{alg:Class_Imp_Newmark}, modifying only the correction step.

\begin{algorithm}
    \footnotesize
	\begin{algorithmic}[1]
		\State $~a^{\mathrm{int}}_{t_{k+1}} \gets ~M^{-1}\left[~f^{\mathrm{ext}}_{t_{k+1}} - ~f^{\mathrm{int}}(~x_{t_{k+1}})\right] $
        \Comment{Internal acceleration}

		\State $~a^{\mathrm{con}}_{t_{k+1}} \gets \frac{1}{(\triangle t)^2}  (~x_{t_{k+1}}   - {~x}^{\mathrm{pred}}_{t_{k+1}} ) -  2 \beta ~a^{\mathrm{int}}_{t_{k+1}}$
        \Comment{Contact acceleration}

		\State $~a_{t_{k+1}} \gets ~a^{\mathrm{int}}_{t_{k+1}} + ~a^{\mathrm{con}}_{t_{k+1}}$
        \Comment{Total acceleration}

		\State $~v_{t_{k+1}} \gets ~v_{t_{k}} +  \triangle t  (1 - \gamma) ~a^{\mathrm{int}}_{t_{k+1}} +  \triangle t \gamma ~a^{\mathrm{int}}_{t_{k+1}} +   \triangle t ~a^{\mathrm{con}}_{t_{k+1}} $
        \Comment{Velocity update}
	\end{algorithmic}
    \caption{Contact-implicit scheme: correction step with acceleration decomposition \eqref{eq:accel_split}.}
    \label{alg:Contact_imp_Newmark}
\end{algorithm}

\paragraph{Contact-stabilized method.} This method is introduced in \textcite{Deuflhard:2008} as an enhancement over the contact-implicit approach of \textcite{Kane.etal:1999}. The contact-stabilized algorithm employs a stabilized predictor-corrector scheme, incorporating an extra nonlinear prediction step based on an $ L_2 $ projection during the prediction phase. Therefore, the method comprises the additional nonlinear prediction step, as described in Algorithm~\ref{alg:Contact_stab_Newmark}, along with the correction step from the contact-implicit scheme (found in Algorithm~\ref{alg:Contact_imp_Newmark}) integrated into the conventional implicit Newmark framework.
In Algorithm~\ref{alg:Contact_stab_Newmark}, matrices $\tilde{~M}$ and $~K^{\mathrm{con}} $ represent the continuum mass matrix and the contact stiffness matrix, respectively, while $ ~r(~x) $ denotes the equivalent contact forces. Further details about these quantities are available in \textcite{Deuflhard:2008, Suwannachit:2012}.
Importantly, it should be noted that the additional nonlinear prediction is effectively equivalent to solving an extra nonlinear contact problem per time step, which may influence the total computational time~\cite{Hauret:2010}.

\begin{algorithm}
    \footnotesize
	\begin{algorithmic}[1]
		\State $m \gets 0$, $~x^{\mathrm{pred}, m}_{t_{k+1}} = ~x_{t_{k}} $
        \Comment{Initialization of Newton iterations}

		\State $m \gets 1$

		\Repeat
        \Comment{Newton iterations}

		      \State Solve $[ \tilde{~M} + ~K^{\mathrm{con}}(~x^{\mathrm{pred},m}_{t_{k+1}}) ] \triangle ~x^{\mathrm{pred}}_{t_{k+1}}  =  -\tilde{~M} ~x^{\mathrm{pred},m}_{t_{k+1}} + \tilde{~M} ~x_{t_{k}} + \triangle t  \tilde{~M} ~v_{t_{k}} - ~r(~x^{\mathrm{pred},m}_{t_{k+1}})  $
            \Comment{Solve linear system}

		      \State
		      $~x^{\mathrm{pred},m+1}_{t_{k+1}} \gets ~x^{\mathrm{pred},m}_{t_{k+1}} + \triangle ~x^{\mathrm{pred}}_{t_{k+1}}$
            \Comment{Update position}
		\Until{converged}
	\end{algorithmic}
	\caption{Contact-stabilized scheme: nonlinear prediction step}
	\label{alg:Contact_stab_Newmark}
\end{algorithm}

The contact-implicit and contact-stabilized methods are both implicit, as they entail a fully implicit treatment of the contact forces.

\paragraph{Naïve-stabilized scheme.} Drawing from the stabilization techniques discussed earlier and the premise that acceleration on the contact boundary should be nullified, the authors in \textcite{Doyen:2011} introduce the naïve-stabilized scheme. This method strives to enforce zero acceleration on the contact boundary during contact events. In the context of our Schwarz method, this leads to setting the accelerations on the contact boundary to zero as part of the Schwarz boundary condition definition step while in the contact phase. Owing to its versatile design, this method is compatible with both implicit and explicit integrators.

\subsection{Numerical evaluation of chatter-reduction techniques in the Schwarz alternating method} \label{sec:chatter_num}

This section presents the main findings from applying the various schemes outlined in Section \ref{sec:time_integr}. Table \ref{tab:compar} summarizes the schemes examined in this analysis, along with the selection of associated parameters. Our primary aim is to demonstrate the efficacy of our chatter-reduction techniques in cases that are most susceptible to oscillations. Given that implicit-explicit coupling inherently shows minimal chatter, it would not be the most illuminating test case for the methods we propose. Hence, we concentrate on explicit-explicit and implicit-implicit schemes for the purpose of this evaluation.

\begin{table}
	\footnotesize
	\centering
	\begin{tabular}{l l l}
		\toprule
		Scheme & Explicit & Implicit                                                                                                              \\
		\midrule
		Classic Newmark &
        Alg. \ref{alg:Class_Exp_Newmark}, $\gamma = 0.5, \beta =0$ &
        Alg. \ref{alg:Class_Imp_Newmark}, $\gamma = 0.5, \beta =0.25$
        \smallskip
        \\
		Dissipative Newmark &
        Alg. \ref{alg:Class_Exp_Newmark}, $\gamma = 0.9, \beta = 0$ &
        Alg. \ref{alg:Class_Imp_Newmark}, $\gamma = 0.9, \beta =0.49$
        \smallskip
        \\
		Chaudhary-Bathe & $-$ & Alg. \ref{alg:Class_Imp_Newmark}, $\gamma = 0.5, \beta =0.5$
        \smallskip
        \\
		Tchamwa-Wielgosz & Alg. \ref{alg:Tchamwa_Wielgosz}, $\Phi = 1.05$ & $-$
        \smallskip
        \\
		Chung-Lee &
        Alg. \ref{alg:Chung_Lee}, $\gamma = \frac{3}{2}, \beta=1, \hat{\gamma} = \hat{\beta} = -\frac{1}{2}$ &
        $-$
        \smallskip
        \\
		\multirow{2}{*}{Naïve-stabilized} &
        Alg. \ref{alg:Class_Exp_Newmark}, zero contact acceleration &
        Alg. \ref{alg:Class_Imp_Newmark}, zero contact accelerations
        \\
        & and $\gamma = 0.5, \beta =0$ &
        and $\gamma = 0.5, \beta =0.25$
        \smallskip
        \\
		\multirow{2}{*}{Contact-implicit} &
        \multirow{2}{*}{$-$} &
        Alg. \ref{alg:Class_Imp_Newmark}, additional correction \ref{alg:Contact_imp_Newmark}
        \\
        & & and $\gamma = 0.5, \beta =0.25$
        \smallskip
        \\
		\multirow{2}{*}{Contact-stabilized} &
        \multirow{2}{*}{$-$} &
        Alg. \ref{alg:Class_Imp_Newmark}, additional prediction \ref{alg:Contact_stab_Newmark},
        \\
        & & correction \ref{alg:Contact_imp_Newmark} and $\gamma = 0.5, \beta =0.25$
        \\
		\bottomrule
	\end{tabular}
    \caption{Summary of methods to mitigate artificial oscillations: variants of the Newmark-$\beta$ time integrator and stabilization schemes.}
	\label{tab:compar}
\end{table}

Our analysis focuses on contact point position, velocity, force, and overall energy error. We investigate the temporal evolution of these key metrics for the Schwarz contact method, when paired with different techniques aimed at reducing artificial oscillations. Consistent spatial and temporal discretizations are employed for both interacting bodies, using $N_x = 200$ elements and time steps of $\triangle t = \qty{1e-7}{\second}$.

\begin{figure}
    \centering
    \subfigure[Contact point velocity]
        {\includegraphics[trim={0mm 0mm 0mm 10mm},clip,width=0.48\textwidth]{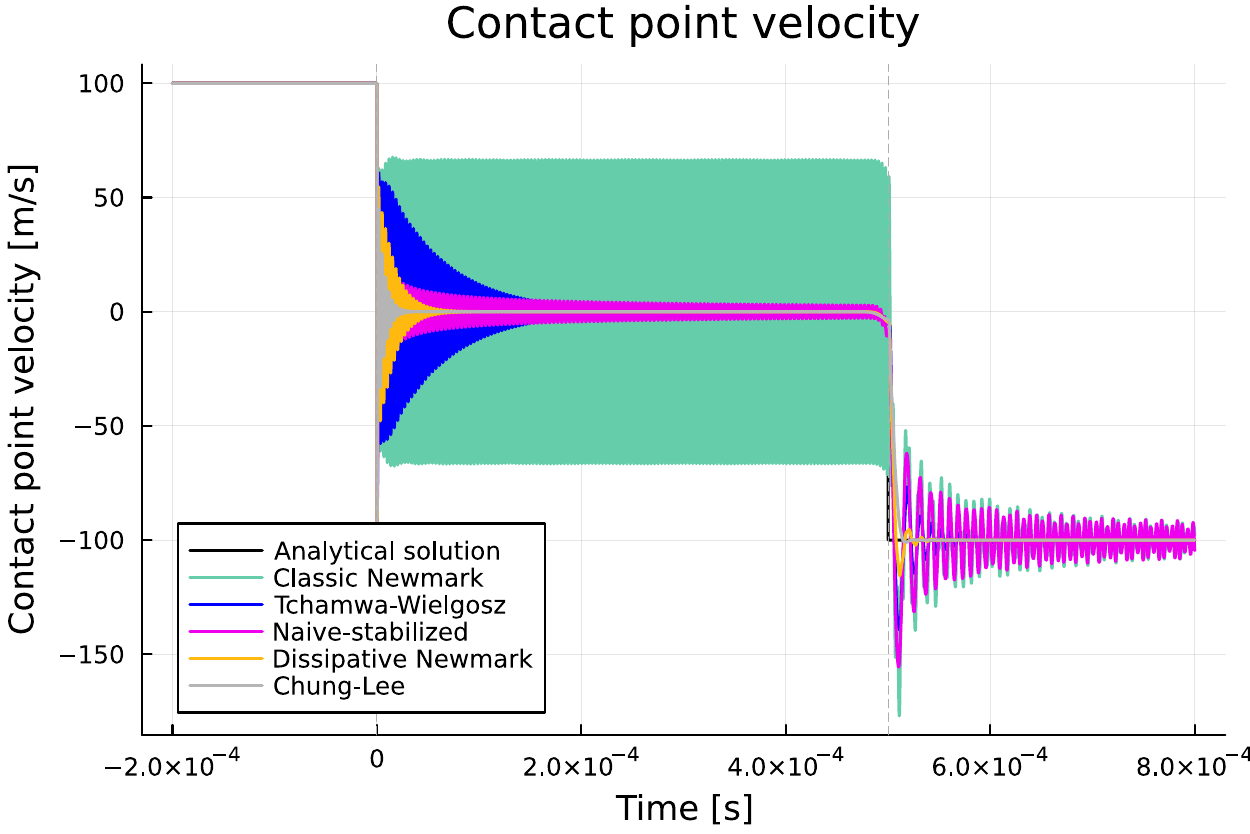} \label{fig:chatter_velo_expl}}
    \subfigure[Contact point force]
        {\includegraphics[trim={0mm 0mm 0mm 9mm},clip,width=0.48\textwidth]{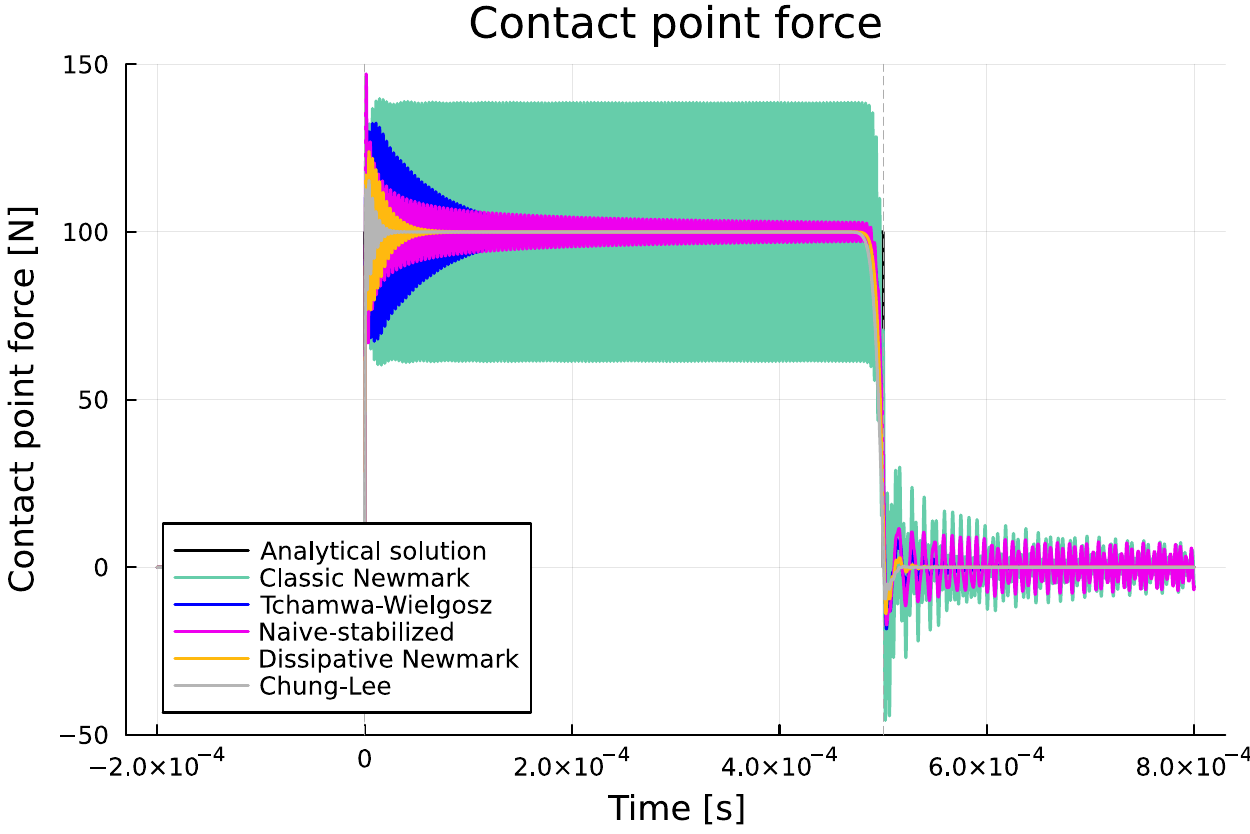} \label{fig:chatter_force_expl}}
    \subfigure[Contact point position]
        {\includegraphics[trim={13mm 13mm 12mm 16mm},clip,width=0.48\textwidth]{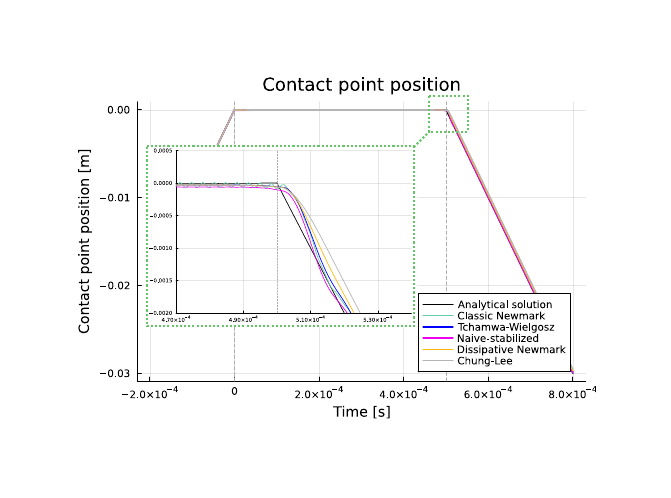} \label{fig:displacement_expl}}
    \subfigure[Total energy error]
        {\includegraphics[trim={0mm 0mm 0mm 10mm},clip,width=0.48\textwidth]{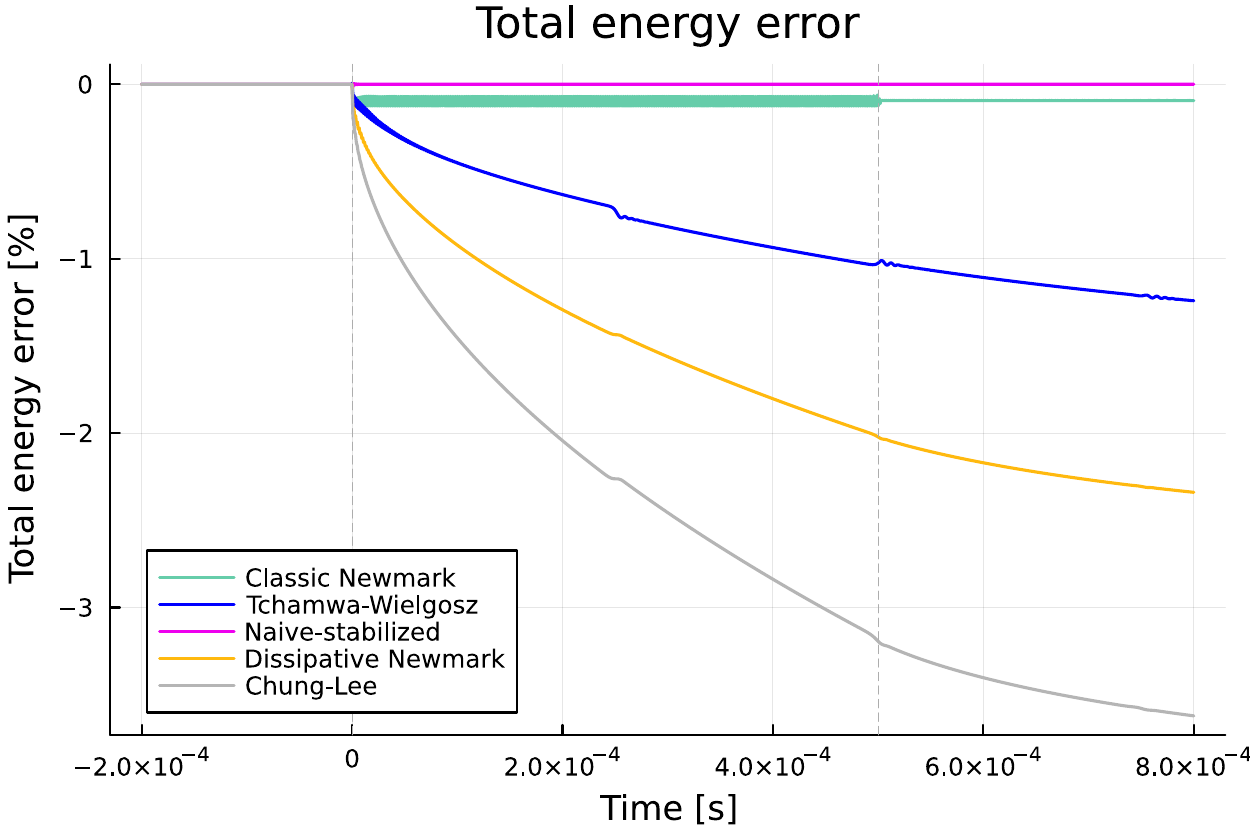} \label{fig:chatter_energy_expl}}
    \caption{Techniques to reduce artificial oscillations for the explicit-explicit Schwarz contact method. Modified versions of time integrators significantly reduce artificial oscillations in both contact velocity and forces. As anticipated, this reduction comes at the cost of energy dissipation, ranging from $1\%$ to $4\%$ for modified Newmark schemes, in contrast to $0.25\%$ for the standard Newmark method. A notable exception is the naïve-stabilized scheme, which effectively conserves energy throughout the simulation while efficiently dampening spurious oscillations. An additional observation is that the naïve-stabilized approach also minimizes energy oscillations compared to the standard Newmark scheme.}
    \label{fig:chatter_expl}
\end{figure}

\begin{figure}
    \centering
    \subfigure[Contact point velocity]
        {\includegraphics[trim={0mm 0mm 0mm 10mm},clip,width=0.48\textwidth]{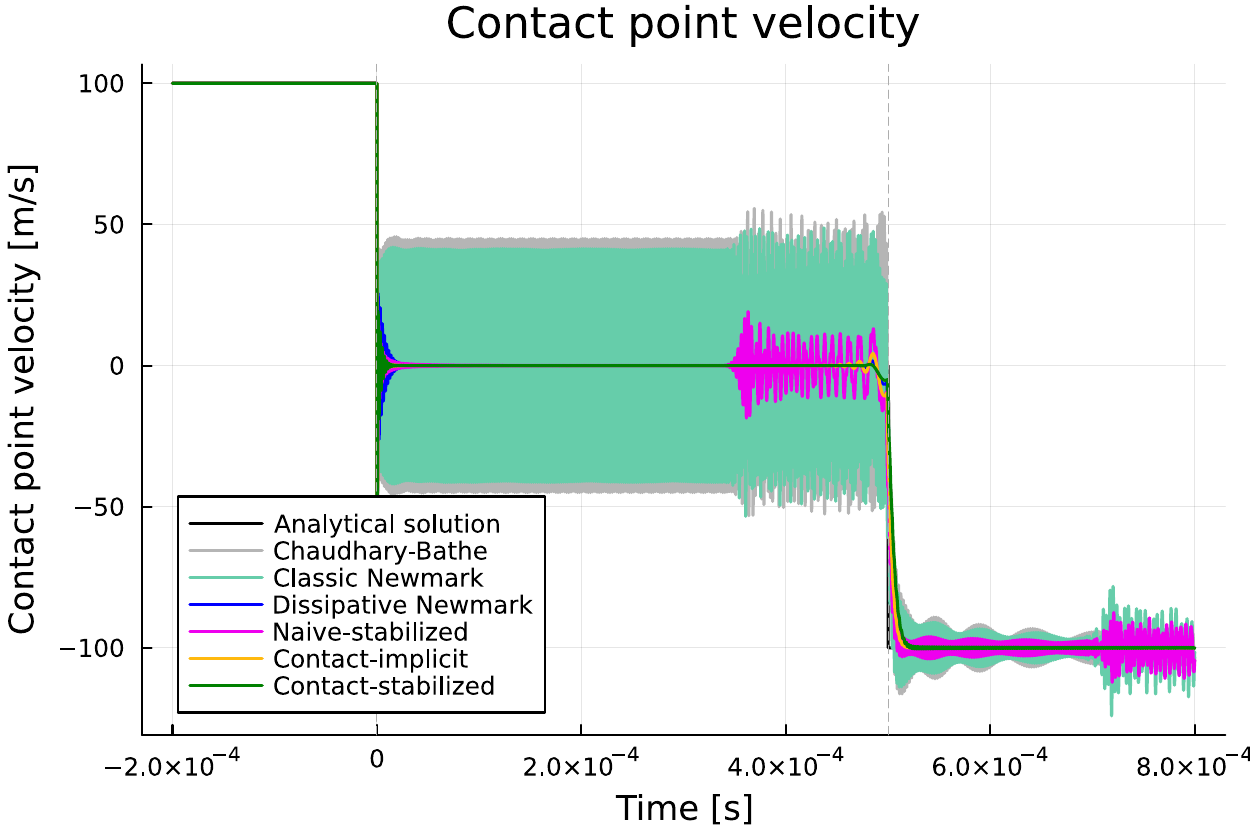} \label{fig:chatter_velo_impl}}
    \subfigure[Contact point force]
        {\includegraphics[trim={0mm 0mm 0mm 9mm},clip,width=0.48\textwidth]{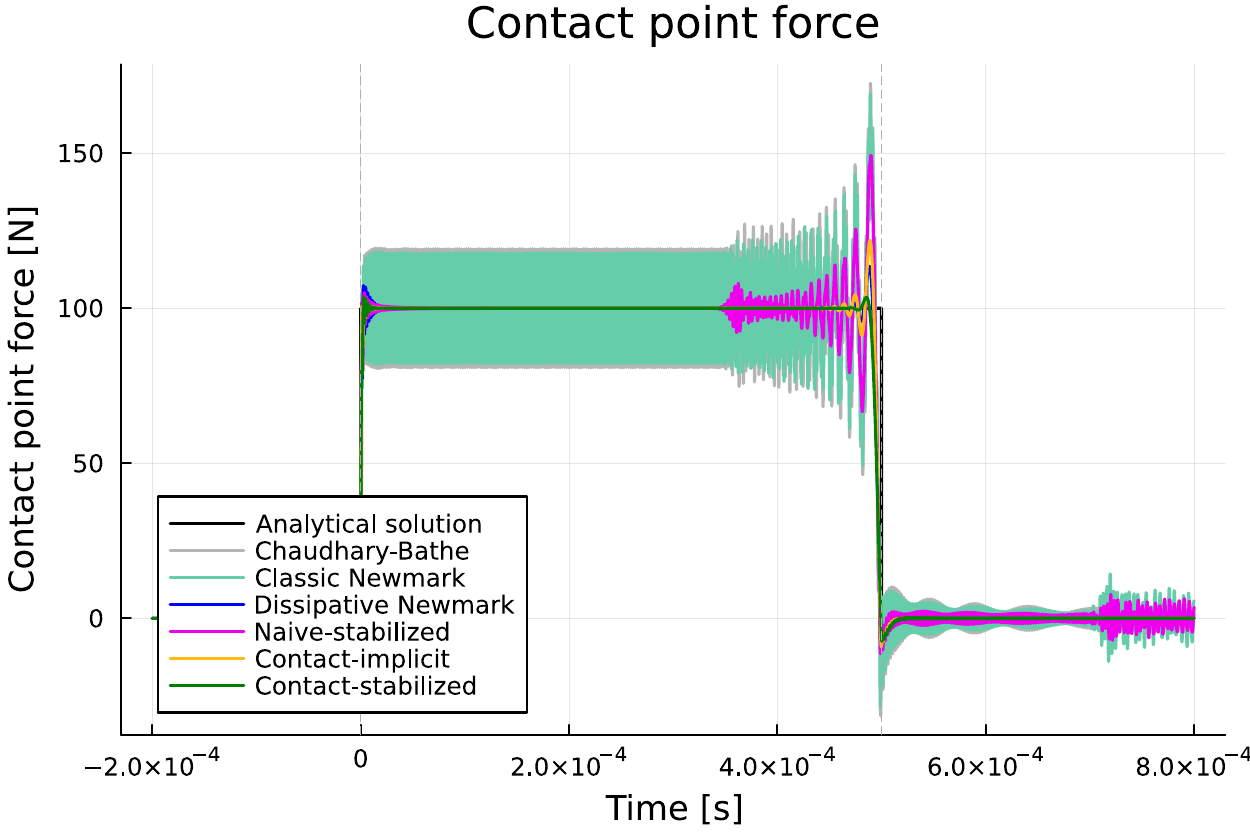} \label{fig:chatter_force_impl}}
    \subfigure[Contact point position]
        {\includegraphics[trim={13mm 13mm 12mm 16mm},clip,width=0.48\textwidth]{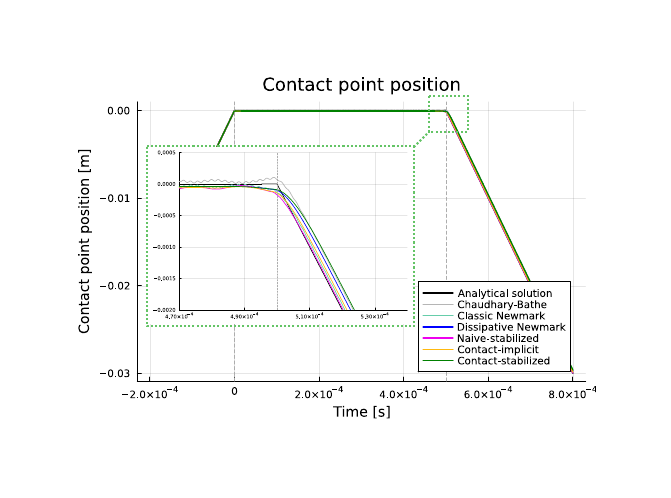} \label{fig:displacement_impl}}
    \subfigure[Total energy error]
        {\includegraphics[trim={0mm 0mm 0mm 10mm},clip,width=0.48\textwidth]{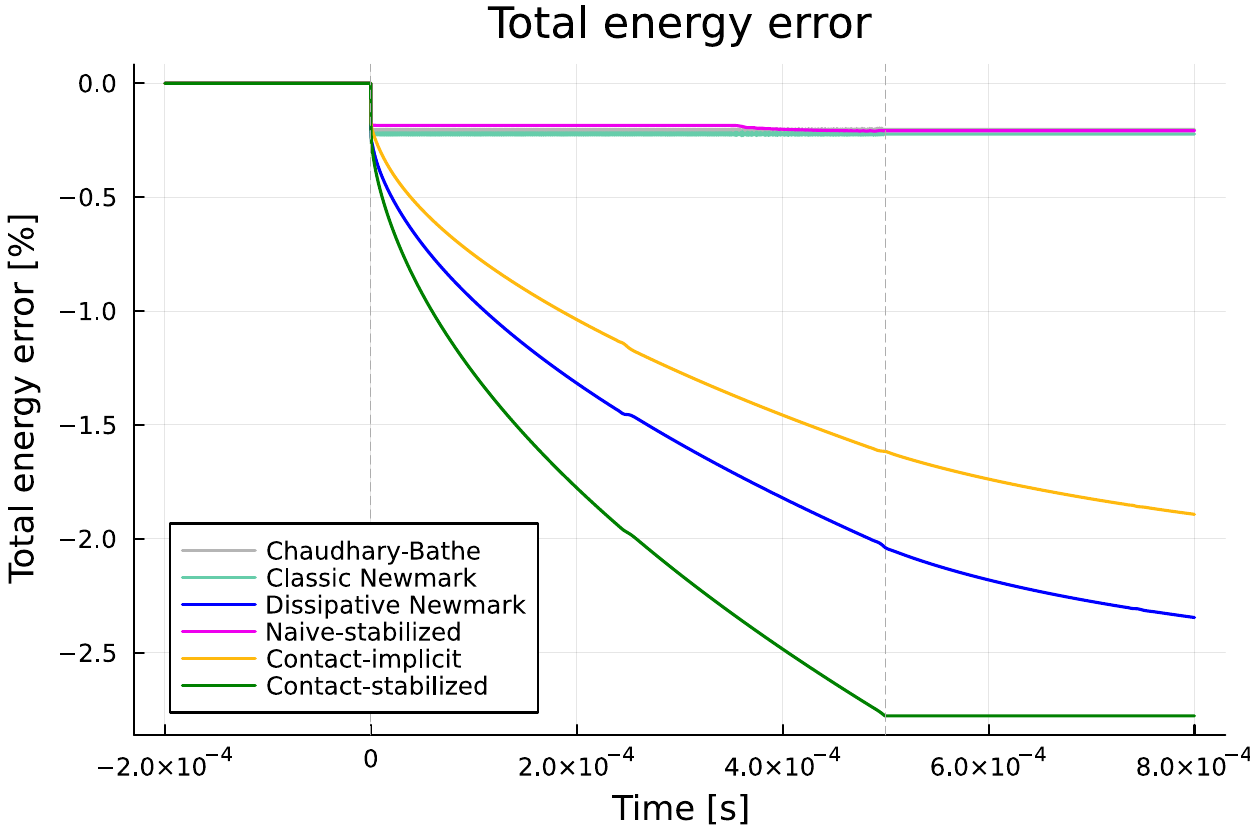}\label{fig:chatter_energy_impl}}
        \caption{Techniques to reduce artificial oscillations for the implicit-implicit Schwarz contact method. Most strategies—excluding the Chaudhary-Bathe scheme—effectively mitigate oscillations in both velocity and force. In the context of energy behavior, the schemes that minimize oscillations most effectively, such as the dissipative Newmark, contact-implicit, and contact-stabilized methods, also result in a more substantial energy loss of up to $2.5\%$. Once again, the naïve-stabilized approach yields the most satisfactory performance, both in terms of reducing oscillations and conserving energy, with a maximum energy loss of approximately $0.2\%$.}
        \label{fig:chatter_impl}
\end{figure}

Figures \ref{fig:chatter_expl} and \ref{fig:chatter_impl} show the results for the explicit-explicit and implicit-implicit coupling schemes, respectively. Initially, it is evident that all tested techniques produce outcomes closely aligned with analytical solutions.

For the explicit-explicit Schwarz method, modified versions of time integrators significantly reduce artificial oscillations in both contact velocity and forces, as seen in Figures \ref{fig:chatter_velo_expl} and \ref{fig:chatter_force_expl}. As anticipated, this reduction comes at the cost of energy dissipation, ranging from $1\%$ to $4\%$ for modified Newmark schemes, in contrast to $0.25\%$ for the standard Newmark method; see Figure \ref{fig:chatter_energy_expl}. A notable exception is the naïve-stabilized scheme, which effectively conserves energy throughout the simulation while efficiently dampening spurious oscillations. An additional observation is that the naïve-stabilized approach also minimizes energy oscillations compared to the standard Newmark scheme, as shown in Figure \ref{fig:chatter_energy_expl}. Lastly, it should be noted that regardless of the technique employed, the contact point position is captured accurately, as illustrated in Figure \ref{fig:displacement_expl}.

Regarding the implicit-implicit Schwarz approach, Figures \ref{fig:chatter_velo_impl} and \ref{fig:chatter_force_impl} show that most strategies—excluding the Chaudhary-Bathe scheme—effectively mitigate oscillations in both velocity and force. In the context of energy behavior, the schemes that minimize oscillations most effectively, such as the dissipative Newmark, contact-implicit, and contact-stabilized methods, also result in a more substantial energy loss of up to $2.5\%$. It is worth noting that this aspect of energy dissipation in stabilization techniques has been previously acknowledged by \textcite{Deuflhard:2008, Kane.etal:1999} and is corroborated by our findings. Once again, the naïve-stabilized approach yields the most satisfactory performance, both in terms of reducing oscillations and conserving energy, with a maximum energy loss of approximately $0.2\%$.

In order to rigorously evaluate the accuracy and precision of our results, we present detailed metrics in Tables~\ref{tab:schwarz_mean_std_chatter_explicit}, \ref{tab:schwarz_mean_std_chatter_implicit}, \ref{tab:schwarz_total_chatter_explicit}, and \ref{tab:schwarz_total_chatter_implicit}. The initial two tables detail the mean $\mu$ and standard deviation $\sigma$ of the absolute errors for various quantities, namely contact point position, velocity, force, and total energy. The absolute error is defined as the discrepancy between our computational solution and the analytical benchmark over the entire time interval from the initial to the final time. While the mean gauges accuracy, the standard deviation serves as an indicator of precision. Notably, the naïve-Newmark method exhibits a minimal energy gain.

The succeeding tables delve into the total relative errors—expressed as percentages—for the aforementioned quantities and further encompass potential, kinetic, and total energies. These data points accentuate the naïve-stabilized scheme's superior performance, especially when integrated with the Schwarz contact algorithm we advocate. In almost all the evaluated metrics, this method shows superior performance, thereby confirming its effectiveness.

\begin{table}
    \footnotesize
    \centering
    \begin{tabular}{l r r r r r r r r}
        \toprule
        \multirow{2}{*}{Time integrator} &  \multicolumn{2}{c}{Position error} &  \multicolumn{2}{c}{Velocity error} &
        \multicolumn{2}{c}{Force error} & \multicolumn{2}{c}{Total energy error}
        \\
        & \multicolumn{1}{c}{$\mu (\unit{\meter})$} & \multicolumn{1}{c}{$\sigma (\unit{\meter})$}
        & \multicolumn{1}{c}{$\mu (\unit{\meter\per\second})$} & \multicolumn{1}{c}{$\sigma (\unit{\meter\per\second})$}
        & \multicolumn{1}{c}{$\mu (\unit{\newton})$} & \multicolumn{1}{c}{$\sigma (\unit{\newton})$}
        & \multicolumn{1}{c}{$\mu (\unit{\joule})$} & \multicolumn{1}{c}{$\sigma (\unit{\joule})$}
        \\
        \midrule
        Classic Newmark & $2.0 \times 10^{-5}$ & $7.0 \times 10^{-5}$ & 0.13 & 34.7 & -0.36 & 20.4 & $-2.0 \times 10^{-3}$ & $1.0 \times 10^{-3}$ \\
        Tchamwa-Wielgosz & $3.4 \times 10^{-5}$ &$8.8 \times 10^{-5}$ & 0.17 & 10.4 & -0.41 & 6.6  & $-1.7 \times 10^{-2}$ & $1.1 \times 10^{-2}$ \\
        Naïve-stabilized & $-2.4 \times 10^{-5}$ & $4.4 \times 10^{-5}$ & 0.04 & 9.3 & -0.38 & 6.5 & $1.7 \times 10^{-5}$ & $1.3 \times 10^{-5}$ \\
        Dissipative Newmark & $7.0 \times 10^{-5}$ & $1.4 \times 10^{-4}$& 0.30 & 6.9 & -0.53 & 5.7 & $-3.4 \times 10^{-2}$ & $2.1 \times 10^{-2}$ \\
        Chung-Lee & $1.2 \times 10^{-4}$ & $2.3 \times 10^{-4}$ & 0.49 &  6.5 & -0.72 & 6.3  & $-5.3 \times 10^{-2}$ & $3.3 \times 10^{-2}$ \\
        \bottomrule
    \end{tabular}
    \caption{Explicit time integrators. Mean values $\mu$ and standard deviations $\sigma$ of the absolute error functions, comparing numerical solutions to analytical ones for contact point position, velocity, force, and total energy. These metrics further validate our observations regarding the performance of the naïve-stabilized scheme when combined with the proposed Schwarz contact algorithm. For almost all the quantities of interest examined, this technique results in better performance metrics.}
    \label{tab:schwarz_mean_std_chatter_explicit}
\end{table}

\begin{table}
    \footnotesize
    \centering
    \begin{tabular}{l r r r r r r r r}
        \toprule
        \multirow{2}{*}{Time integrator} &  \multicolumn{2}{c}{Position error} &  \multicolumn{2}{c}{Velocity error} &
        \multicolumn{2}{c}{Force error} & \multicolumn{2}{c}{Total energy error}
        \\
        & \multicolumn{1}{c}{$\mu (\unit{\meter})$} & \multicolumn{1}{c}{$\sigma (\unit{\meter})$}
        & \multicolumn{1}{c}{$\mu (\unit{\meter\per\second})$} & \multicolumn{1}{c}{$\sigma (\unit{\meter\per\second})$}
        & \multicolumn{1}{c}{$\mu (\unit{\newton})$} & \multicolumn{1}{c}{$\sigma (\unit{\newton})$}
        & \multicolumn{1}{c}{$\mu (\unit{\joule})$} & \multicolumn{1}{c}{$\sigma (\unit{\joule})$}
        \\
        \midrule
        Classic Newmark & $4.3 \times 10^{-6}$ & $5.6 \times 10^{-5}$ & 0.09 & 21.4 & -0.36 & 12.3 & $-4.4 \times 10^{-3}$ & $2.2 \times 10^{-3}$ \\
        Chaudhary-Bathe  & $1.2 \times 10^{-4}$ & $1.4 \times 10^{-4}$ & 0.34 & 23.1 & -0.32 & 12.5 & $-4.1 \times 10^{-3}$ & $2.1 \times 10^{-3}$ \\
        Naïve-stabilized & $-2.1 \times 10^{-5}$ & $4.1 \times 10^{-5}$ & 0.04 & 5.1 & -0.39 & 8.9 & $-3.9 \times 10^{-3}$ & $2.0 \times 10^{-3}$ \\
        Dissipative Newmark & $5.4 \times 10^{-5}$ & $1.3 \times 10^{-4}$ & 0.26 & 4.5 & -0.54 & 6.5 & $-3.4 \times 10^{-2}$ & $2.1 \times 10^{-2}$ \\
        Contact-implicit  & $3.7 \times 10^{-5}$ & $1.0 \times 10^{-4}$ & 0.20 & 4.4 & -0.51 & 6.8 & $-2.7 \times 10^{-2}$ & $1.7 \times 10^{-2}$ \\
        Contact-stabilized  & $1.1 \times 10^{-4}$ & $2.0 \times 10^{-4}$ & 0.42 & 5.3 & -0.48 & 6.6 & $-4.4 \times 10^{-2}$ & $2.7 \times 10^{-2}$ \\
        \bottomrule
    \end{tabular}
    \caption{Implicit time integrators. Mean values $\mu$ and standard deviations $\sigma$ of the absolute error functions, comparing numerical solutions to analytical ones for contact point position, velocity, force, and total energy. These metrics further validate our observations regarding the performance of the naïve-stabilized scheme when combined with the proposed Schwarz contact algorithm. For almost all the quantities of interest examined, this technique results in better performance metrics.}
    \label{tab:schwarz_mean_std_chatter_implicit}
\end{table}

\begin{table}
    \footnotesize
    \centering
    \begin{tabular}{l r r r r r r}
        \toprule
        Time integrator & Position & Velocity & Force & Potential energy & Kinetic energy & Total energy \\
        \midrule
        Classic Newmark  & 0.67 & 49.06& 28.48 & 1.08 &  0.60 & 0.08 \\
        Tchamwa-Wielgosz & 0.86 & 14.68 & 8.30 & 0.88 & 1.12 & 0.81 \\
        Naïve-stabilized & 0.46 & 13.18 & 8.00 & 1.13 & 0.56 & 0.00 \\ 
        Dissipative Newmark & 1.46 & 9.78 & 6.92 & 1.36 & 1.86 & 1.57 \\
        Chung-Lee & 2.39 & 9.24 & 8.10 & 2.09 & 2.84 & 2.47 \\
        \bottomrule
    \end{tabular}
    \caption{Explicit time integrators. Percentage of total relative error, comparing numerical solutions to analytical ones for contact point position, velocity, force, potential, kinetic and total energies.}
    \label{tab:schwarz_total_chatter_explicit}
\end{table}

\begin{table}
    \footnotesize
    \centering
    \begin{tabular}{l r r r r r r}
        \toprule
        Time integrator & Position & Velocity & Force & Potential energy & Kinetic energy & Total energy \\
        \midrule
        Classic Newmark & 0.51 & 30.25 & 17.13 & 0.99 & 0.64 & 0.19 \\
        Chaudhary-Bathe  & 1.72 & 32.60 &  18.31 & 0.99 & 0.62 & 0.18 \\
        Naïve-stabilized & 0.42 & 7.20 & 10.89 & 1.01 & 0.64 & 0.17 \\
        Dissipative Newmark & 1.30 & 6.43 & 8.59 & 1.37 & 1.86 & 1.59 \\
        Contact-implicit  & 0.81 & 5.97 & 8.71 & 1.14 & 1.54 & 1.27 \\
        Contact-stabilized & 1.88 & 7.29 & 8.72 & 1.89 & 2.41& 2.06 \\
        \bottomrule
    \end{tabular}
    \caption{Implicit time integrators. Percentage of total relative error, comparing numerical solutions to analytical ones for contact point position, velocity, force, potential, kinetic and total energies.}
    \label{tab:schwarz_total_chatter_implicit}
\end{table}

\begin{table}
    \footnotesize
    \centering
        \begin{tabular}{l c c}
            \toprule
            Time Integrator & Max Iterations & Avg Iterations \\
            \midrule
            Classic Newmark & 3 & 2.50 \\
            Tchamwa-Wielgosz & 3 & 2.50 \\
            Naïve-stabilized & 3 & 2.50 \\
            Dissipative Newmark & 3 & 2.24 \\
            Chung-Lee & 6 & 2.44 \\
            \bottomrule
        \end{tabular}
    \caption{Maximum/average number of Schwarz iterations for Schwarz contact method with explicit time integrators.}
    \label{tab:schwarz_iters_chatter_explicit}
\end{table}

\begin{table}
    \footnotesize
    \centering
        \begin{tabular}{l c c}
            \toprule
            Time Integrator & Max Iterations & Avg Iterations \\
            \midrule
            Classic Newmark & 5 & 4.50 \\
            Chaudhary-Bathe & 6 & 5.35 \\
            Naïve-stabilized & 5 & 3.82 \\
            Dissipative Newmark & 6 & 2.55 \\
            Contact-implicit & 5 & 3.82 \\
            Contact-stabilized & 8 & 7.48 \\
            \bottomrule
        \end{tabular}
    \caption{Maximum/average number of Schwarz iterations for Schwarz contact method with implicit time integrators.}
    \label{tab:schwarz_iters_chatter_implicit}
\end{table}

Tables~\ref{tab:schwarz_iters_chatter_explicit} and \ref{tab:schwarz_iters_chatter_implicit} display both the maximum and average counts of Schwarz iterations required by each method. In this context, the number of Schwarz iterations serves as an indicator of numerical efficiency and is closely tied to computational time. It is evident that all tested methods require a comparable number of Schwarz iterations for convergence. As anticipated, the implicit schemes consume more computational resources. Among all the approaches, the contact-stabilized method proves to be the most resource-intensive, given its requirement to solve an additional nonlinear problem within the prediction phase. The naïve-stabilized technique is found to be comparably efficient to the classic Newmark scheme, requiring an equal or fewer number of Schwarz iterations.

In summary, our Schwarz contact algorithm proves to be highly versatile and adaptable, seamlessly integrating with a variety of time integrators and stabilization methods. Particularly, the naïve-stabilized approach effectively curtails artificial oscillations while maintaining the energy-conserving properties of the Schwarz alternating method. Furthermore, the naïve-stabilized method outperforms other techniques examined in this study in several respects: it is compatible with both explicit and implicit time integrators, it is straightforward to implement, and it is efficient in terms of the number of Schwarz iterations required for convergence.

\section{Three-dimensional impact benchmark} \label{sec:results_3D}

The aim of this section is to evaluate the effectiveness of our Schwarz contact algorithm within a three-dimensional setting. Crucial to this assessment is the transfer of both spatial and temporal information, as detailed in Sections \ref{sec:space_transfer} and \ref{sec:time_transfer}. To illustrate the algorithm's performance, we generalize the one-dimensional impact problem originally presented in Section~\ref{sec:results} to a 3D environment. All simulations for this section were carried out using Norma, a specialized Julia-based software developed for this study \cite{Mota.Koliesnikova:2023}.

As before, this involves two prismatic rods, denoted by $\Omega^1$ and $\Omega^2$, colliding with each other. Each rod is characterized by a linear elastic material model with properties like density $\rho$, elastic modulus $E$, and cross-sectional area $A$. They are symmetrical about the impact plane, initially separated by $2g$, and have an initial velocity of $v_0$. Relevant parameters are listed in Table \ref{tab:3Dproblem}, and the setup is illustrated in Figure~\ref{fig:1D_impact_illustration}.

The analytically determined impact and release times are $t_{\mathrm{imp}} = \qty{0}{\second}$ and $t_{\mathrm{rel}} = \qty{2e-6}{\second}$, as defined by \eqref{eq:impact_time}.

\begin{table}
    \footnotesize
    \centering
    \begin{tabular}{l r l}
        \toprule
        Parameter & {Value} & {Unit}
        \\
        \midrule
        $\rho$ & 1000 & \unit{\kilo\gram\per\cubic\meter}
        \\
        $E$ &  1 & \unit{\giga\pascal}
        \\
        $A$ & 10000 & \unit{\square\micro\meter}
        \\
        $L$ &  1 & \unit{\milli\meter}
        \\
        $g$ &  100 & \unit{\micro\meter}
        \\
        $v_0$ & 100 & \unit{\meter\per\second}
        \\
        $t_{0}$ &  -1 & \unit{\micro\second}
        \\
        $t_{N}$ & 3 & \unit{\micro\second}
        \\
        \bottomrule
    \end{tabular}
    \caption{Parameters for the 3D impact benchmark. Density $\rho$, elastic modulus $E$, cross-sectional area $A$, length $L$, initial semi-distance $g$, initial velocity $v_0$, and initial and final simulation times $t_{0}$ and $t_{N}$, respectively.}
    \label{tab:3Dproblem}
\end{table}

\subsection{Numerical results}

\begin{figure}
    \centering
    \subfigure[Implicit-Implicit Schwarz contact method]
        {\includegraphics[page=1, trim={17mm 30mm 17mm 30mm}, width=0.48\textwidth]{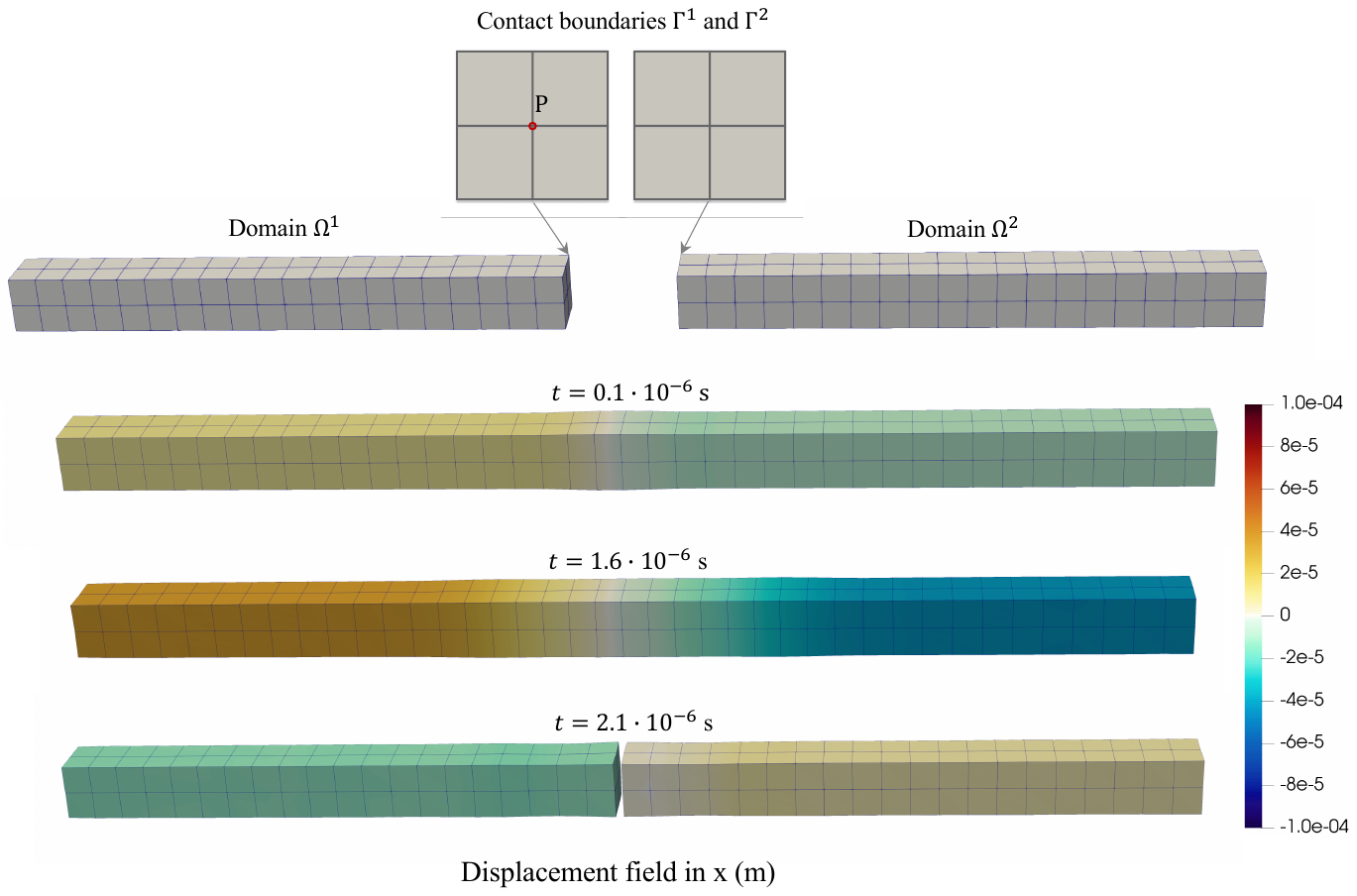} \label{fig:examples_meshes_3Da}}
    \subfigure[Explicit-Explicit Schwarz  contact method]
        {\includegraphics[page=2, trim={17mm 30mm 17mm 30mm}, width=0.48\textwidth]{plots/examples_meshes_3D.pdf} \label{fig:examples_meshes_3Db}}
    \subfigure[Implicit-Explicit Schwarz  contact method]
        {\includegraphics[page=3, trim={17mm 30mm 17mm 30mm}, width=0.48\textwidth]{plots/examples_meshes_3D.pdf} \label{fig:examples_meshes_3Dc}}
    \subfigure[Explicit-Implicit Schwarz  contact method]
        {\includegraphics[page=4, trim={17mm 30mm 17mm 30mm}, width=0.48\textwidth]{plots/examples_meshes_3D.pdf} \label{fig:examples_meshes_3Dd}}
    \caption{Meshes, contact boundaries and displacement fields for specified times for the different versions of the Schwarz contact method. At impact, one can observe contraction in the bars along the $x$-axis and expansion in the directions perpendicular to it. Wave propagation along the $x$-axis is evident in both bars. Additionally, deformations in the $y$ and $z$ axes occur as waves travel along the length of each bar, creating vibrations. These waves continue to travel until they reach the free ends of the bars and then return to the point of impact. The bars remain in contact until this wave makes its return, marking the release time. Throughout the contact phase, the boundaries $\Gamma^1$ and $\Gamma^2$ maintain contact without either separating or penetrating each other. Their displacement along the $x$-axis remains nearly zero. Point $ P $ is the centroid of the contact boundary $ \Gamma^1 $ in the left domain.}
    \label{fig:examples_meshes_3D}
\end{figure}

Table~\ref{tab:3D_std_Schwarz} lists the various configurations of the Schwarz method evaluated in this section.
\begin{itemize}
    \item \textbf{Implicit-Implicit Version:} Both domains $\Omega^1$ and $\Omega^2$ use identical settings, including implicit time integrators, time steps, and mesh types (See Figure~\ref{fig:examples_meshes_3Da}).
    \item \textbf{Explicit-Explicit Version:} Here too, both domains $\Omega^1$ and $\Omega^2$ share the same settings, but with explicit time integrators (See Figure~\ref{fig:examples_meshes_3Db}).
    \item \textbf{Implicit-Explicit Version:} This configuration employs different time integrators, time steps, and mesh types for $\Omega^1$ and $\Omega^2$ (See Figure~\ref{fig:examples_meshes_3Dc}).
    \item \textbf{Explicit-Implicit Version:} In this setup, all settings, including mesh sizes, differ between $\Omega^1$ and $\Omega^2$ (See Figure~\ref{fig:examples_meshes_3Dd}).
\end{itemize}

\begin{table}
    \footnotesize
    \centering
    \begin{tabular}{ c c c c c c c c c }
        \toprule
        \multirow{2}{*}{Schwarz Versions} & \multicolumn{2}{c}{Mesh Type} & \multicolumn{2}{c}{Mesh Size} & \multicolumn{2}{c}{Number of Nodes} & \multicolumn{2}{c}{Time Step} \\
        & $\Omega^1$ & $\Omega^2$ &
        $\Omega^1$ (\unit{\micro\meter}) & $\Omega^2$ (\unit{\micro\meter}) &
        $\Omega^1$ & $\Omega^2$ &
        $\Omega^1$ (\unit{\nano\second}) & $\Omega^2$ (\unit{\nano\second}) \\
        \midrule
        Implicit-Implicit &  HEX8 & HEX8 &  50 & 50 & 189 & 189 & 10 & 10 \\
        Explicit-Explicit &  TET4& TET4 &  50 & 50 & 199 & 199 & 1& 1 \\
        Implicit-Explicit & HEX8 & TET4 &  50 & 50 & 189 & 199 & 5 &  1 \\
        Explicit-Implicit & TET4 & HEX8 & 25 & 33 & 1025 & 745 & 1 &  5 \\
        \bottomrule
    \end{tabular}
    \caption{Comparison of different Schwarz contact method versions in the three-dimensional benchmark: variations in time integrator, element type, mesh size, and time step.}
    \label{tab:3D_std_Schwarz}
\end{table}

The mesh configurations for each Schwarz method version are depicted in Figure~\ref{fig:examples_meshes_3D}. In both the implicit-implicit and explicit-explicit versions, each domain utilizes identical meshes, resulting in matching discretizations at the contact boundaries $\Gamma^1$ and $\Gamma^2$. This makes the transfer of boundary condition data trivial, requiring neither spatial nor temporal interpolation.

Conversely, the implicit-explicit and explicit-implicit configurations use different types of meshes and time steps for $\Omega^1$ and $\Omega^2$ (as shown in Figures~\ref{fig:examples_meshes_3Dc} and~\ref{fig:examples_meshes_3Dd}). Thus, data transfer involves both spatial and temporal interpolations. The spatial transfer is detailed in Section~\ref{sec:space_transfer}, while the temporal interpolation is covered in Section~\ref{sec:time_transfer}.

Note that explicit integrators demand smaller time steps compared to their implicit counterparts. In those cases, time steps meeting the CFL condition are selected for explicit integration, as outlined in Table~\ref{tab:3D_std_Schwarz}.

\subsubsection{Standard Schwarz contact method}

This subsection presents key findings derived from applying the Schwarz contact method to the three-dimensional impact problem. Figure~\ref{fig:examples_meshes_3D} depicts the displacement fields in the $x$-direction at specific times for all four Schwarz method variants.

At the moment of impact, one can observe contraction in the bars along the $x$-axis and expansion in the directions perpendicular to it. Wave propagation along the $x$-axis is evident in both bars. Additionally, deformations in the $y$ and $z$ axes occur as waves travel along the length of each bar, creating vibrations. These waves continue to travel until they reach the free ends of the bars and then return to the point of impact. The bars remain in contact until this wave makes its return, marking the release time. Throughout the contact phase, the boundaries $\Gamma^1$ and $\Gamma^2$ maintain contact without either separating or penetrating each other. Their displacement along the $x$-axis remains nearly zero, consistent with expectations.

All of these intricate three-dimensional behaviors are effectively captured by each variant of the Schwarz contact method examined here.

\begin{figure}
    \centering
    \includegraphics[trim={5mm 20mm 5mm 19mm}, width=1\textwidth]{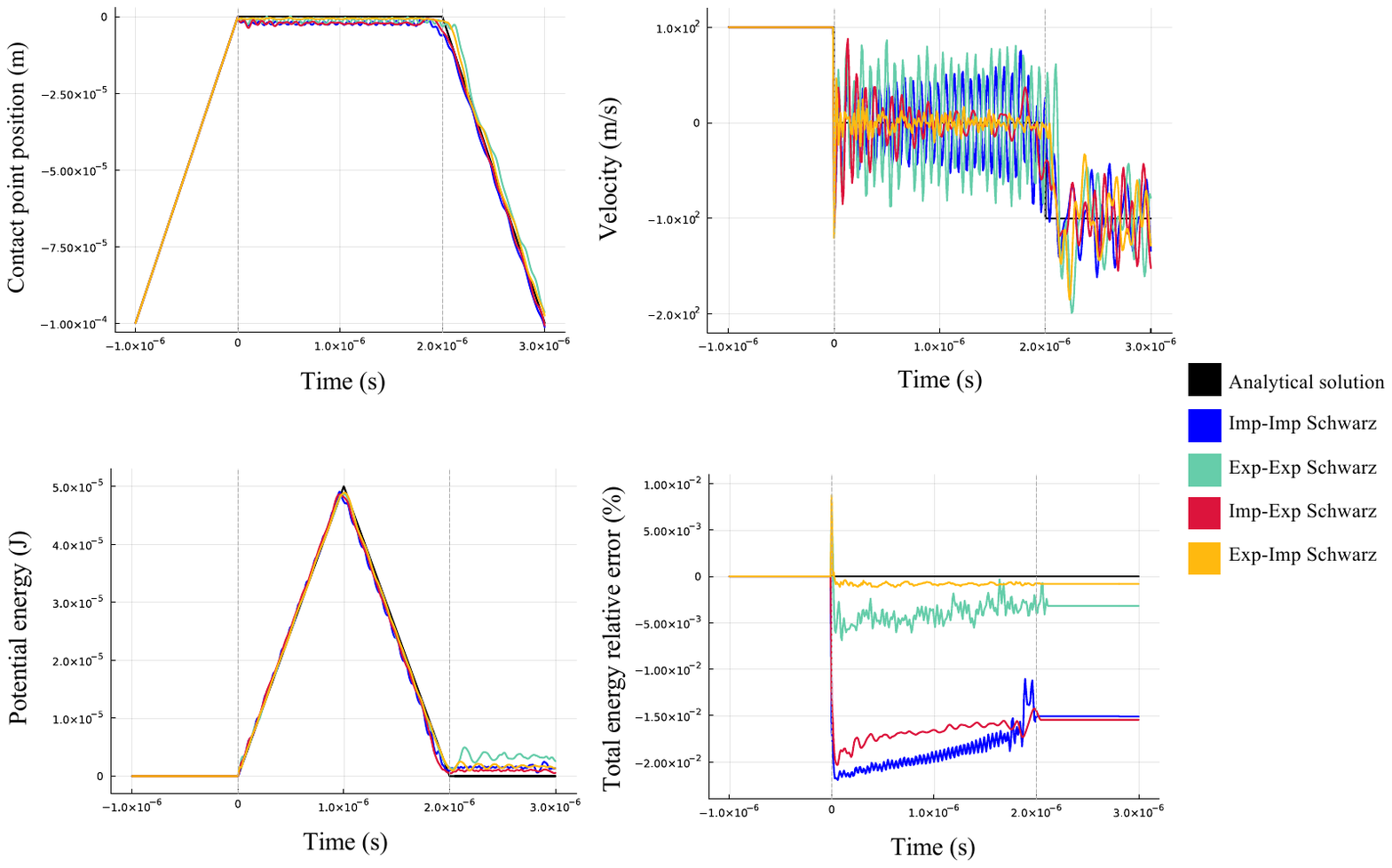}
    \caption{Three-dimensional benchmark with standard Schwarz alternating contact method. Evolution of contact point position, velocity, potential energy, and total energy relative error. Point $ P $ is the centroid of the contact boundary $ \Gamma^1 $ in the left domain. Across all Schwarz method variants, our findings indicate accurate predictions for these key metrics. The algorithm also effectively identifies the specific times of impact and release. In line with observations from the 1D benchmark of Section~\ref{sec:results}, all Schwarz method variants exhibit excellent energy conservation; the maximum energy loss recorded is $ 0.02\% $ in the implicit-implicit case. The method does introduce artificial oscillations during the contact and post-release phases, but these oscillations seem to be alleviated when different time integrators are used in distinct domains.}
    \label{fig:3D_2bars_std}
\end{figure}

Figure~\ref{fig:3D_2bars_std} shows the temporal evolution of various key quantities, including the position and velocity at a point on the contact interface, denoted by $ P $, the total energy relative error, and the potential energy. Point $ P $ is chosen to be the centroid of the contact boundary $ \Gamma^1 $ in the left domain, as illustrated in Figures \ref{fig:examples_meshes_3Da}-\ref{fig:examples_meshes_3Dd}. The potential energy is specific to the left domain $ \Omega^1 $, whereas the total energy error represents the deviation between the computed total energy for both bars and the corresponding analytical value, detailed in Section~\ref{sec:comparison}.

Across all Schwarz method variants, our findings indicate accurate predictions for these key metrics. The algorithm also effectively identifies the specific times of impact and release. In line with observations from the 1D benchmark discussed in Section~\ref{sec:results}, all Schwarz method variants exhibit excellent energy conservation; the maximum energy loss recorded is $ 0.02\% $ in the implicit-implicit case. The method does introduce artificial oscillations during the contact and post-release phases, but these oscillations seem to be alleviated when different time integrators are used in distinct domains.

\subsubsection{Schwarz contact method with stabilization}

This section discusses the implementation of the naïve-stabilized approach in the three-dimensional Schwarz contact method. As previously demonstrated in our one-dimensional benchmark (see Section~\ref{sec:chatter}), the naïve-stabilized technique effectively minimizes artificial oscillations.

\begin{figure}
    \centering
    \includegraphics[trim={5mm 20mm 5mm 20mm}, width=1\textwidth]{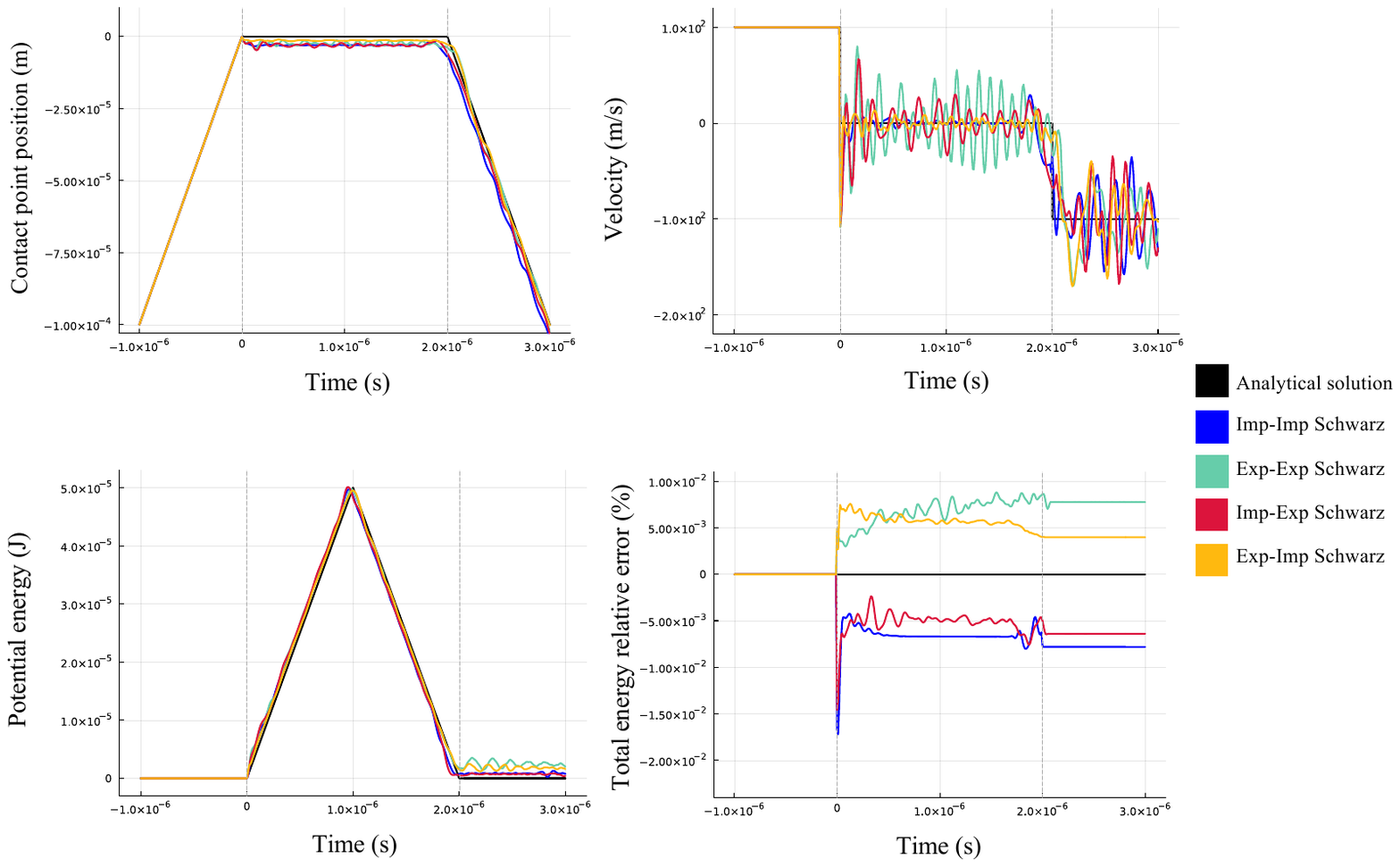}
    \caption{Three-dimensional benchmark with naïve-stabilized Schwarz alternating contact method. Evolution of contact point position, velocity, potential energy, and total energy relative error. Point $ P $ is the centroid of the contact boundary $ \Gamma^1 $ in the left domain. The naïve-stabilized technique yields similar benefits in both 1D and 3D scenarios, notably in reducing the oscillatory behavior of the velocity. Specifically, oscillations during the contact phase are substantially mitigated, particularly for the implicit-implicit and explicit-explicit cases. Additionally, the method maintains high accuracy in capturing other key metrics. An interesting point to note is that the explicit-explicit and explicit-implicit methods showed a very small energy gain, less than $0.01\%$ when compared to the analytical solution. Such minor deviations in energy are negligible and can be further reduced through finer spatial or temporal discretization.}
    \label{fig:3D_2bars_naive}
\end{figure}

Figures~\ref{fig:3D_2bars_naive} display the results of applying this stabilization to all four versions of the Schwarz method. The naïve-stabilized technique yields similar benefits in both 1D and 3D scenarios, notably in reducing the oscillatory behavior of the velocity. Specifically, oscillations during the contact phase are substantially mitigated, particularly for the implicit-implicit and explicit-explicit cases.

Additionally, the method demonstrates high accuracy across various key metrics. Notably, both the explicit-explicit and explicit-implicit schemes exhibit an energy gain of less than $0.01\%$ compared to the analytical solution. Such minor deviations in energy are negligible and can be further mitigated through more refined spatial or temporal discretization. 

The average number of iterations required for convergence with both the standard and naïve-stabilized Schwarz methods are summarized in Table~\ref{tab:iter_number}. Interestingly, the naïve-stabilized version typically requires fewer iterations than the standard method. This observation is consistent with the one-dimensional case discussed in Section~\ref{sec:chatter_num}. Our theory for this reduced iteration count is that Schwarz methods have been found to require more iterations when coupling less accurate models. This is because the method spends additional iterations attempting to correct and align the solutions. We hypothesize that the same principle is applicable to the contact problem; the presence of chatter likely poses convergence challenges, leading to more iterations for the standard Schwarz method compared to its naïve-stabilized counterpart.

\begin{table}
    \footnotesize
    \centering
    \begin{tabular}{c r r}
        \toprule
        \multirow{2}{*}{Schwarz versions} & \multicolumn{2}{c}{Average number of Schwarz iterations} \\
        & Standard & Naïve-stabilized \\
        \midrule
        Implicit-Implicit & 7.2 & 6.4\\
        Explicit-Explicit & 5.9 & 5.5\\
        Implicit-Explicit & 6.1 & 5.2 \\
        Explicit-Implicit & 8.7 & 7.8 \\
        \bottomrule
    \end{tabular}
    \caption{Three-dimensional benchmark. Average number of Schwarz iterations required for both the standard and naïve-stabilized Schwarz methods. Note that the naïve-stabilized version typically requires one fewer iteration than its standard counterpart. This finding aligns with the results from the one-dimensional case discussed in Section~\ref{sec:chatter_num}.}
    \label{tab:iter_number}
\end{table}


\section{Summary} \label{sec:conc}

This paper presents a new computational framework for simulating mechanical contact based on the Schwarz alternating method. In a multi-body contact context, the Schwarz alternating method aims to treat each body separately and to handle contact with an iterative process transferring the information through alternating Dirichlet-Neumann boundary conditions. Introduced initially in a domain decomposition framework, this strategy is equipped with a strong theoretical ground and provides numerous advantages over existing contact approaches, i.e. the use of different meshes, material models, solvers, time integration schemes, etc. for each body, which can be efficiently exploited in a multiscale and/or multiphysics contact context. Moreover, an important advantage of this Dirichlet-Neumann iterative procedure is that it is in general straightforward to implement into existing software infrastructures.

An in-depth numerical comparative study allows us to highlight the potentialities of the
Schwarz methodology compared to conventional contact algorithms, namely the penalty method and the Lagrange multiplier method. Our results demonstrate that the Schwarz alternating method delivers a solution with substantially better accuracy than the conventional approaches for quantities of interest (i.e., impact/release times, contact positions, velocities, forces, kinetic and potential energies), and, moreover, offers a remarkable total energy conserving behavior.

In its original version, the Schwarz method was shown to suffer from artificial oscillations in contact velocities and forces as a side effect of the total energy conservation. We have proposed an efficient way to mitigate those spurious oscillations. The propose remedy is based on the naïve-stabilized approach which aims to suppress the inertia of the contact boundary by making the acceleration on the contact boundary vanish. This technique is suitable for explicit and implicit integrators, and results in significant chatter reduction while preserving the Schwarz algorithm’s accuracy, energy conservation property and efficiency.

We have also extended our approach to multiple spatial dimensions. The numerical results obtained for a 3D impact benchmark confirm the efficiency and accuracy of the Schwarz contact approach in handling contact. This 3D test case underscores the Schwarz method's inherent ability to employ different time integrators and time steps, as well as different mesh topologies and mesh sizes, in the various domains involved in contact. This is a significant advantage rarely afforded by conventional contact methods.

Generic questions related to the construction of transfer operators for the information transfer between different time steps, as well as non-matching meshes, have been addressed. We have proposed a reliable strategy for the information transfer between incompatible interfaces and have confirmed its efficiency on a real example.

Future research will aim to broaden the scope of the proposed contact algorithm to encompass more realistic and intricate geometries. In addition, we intend to investigate the applicability of the Schwarz contact method to challenging scenarios, such as persistent and non-smooth contact. A further key area of focus will be the incorporation of diverse contact conditions—specifically, friction, rolling, and sliding—into the existing Schwarz framework.

\section{Acknowledgments} \label{sec:ack}

This article has been authored by an employee of National Technology \& Engineering Solutions of Sandia, LLC under Contract No. DE-NA0003525 with the U.S. Department of Energy (DOE). The employee owns all right, title and interest in and to the article and is solely responsible for its contents. The United States Government retains and the publisher, by accepting the article for publication, acknowledges that the United States Government retains a non-exclusive, paid-up, irrevocable, world-wide license to publish or reproduce the published form of this article or allow others to do so, for United States Government purposes. The DOE will provide public access to these results of federally sponsored research in accordance with the DOE Public Access Plan \url{https://www.energy.gov/downloads/doe-public-access-plan}.

This paper describes objective technical results and analysis. Any subjective views or opinions that might be expressed in the paper do not necessarily represent the views of the U.S. Department of Energy or the United States Government.

This work was supported by the Advanced Simulation and Computing (ASC) and Laboratory Directed Research and Development (LDRD) programs at Sandia National Laboratories. Additional funding was graciously provided by the Presidential Early Career Award for Scientists and Engineers (PECASE), awarded to the third author, Dr. Irina Tezaur.

We extend our heartfelt gratitude to Dr. Reese Jones and Dr. Thomas Voth for their invaluable contributions to this work through stimulating and insightful discussions.

\printbibliography

@article{Acary:2016,
  author =       {Acary, Vincent},
  doi =          {https://doi.org/10.1002/zamm.201400231},
  journal =      {ZAMM - Journal of Applied Mathematics and Mechanics
                  / Zeitschrift f{\"u}r Angewandte Mathematik und
                  Mechanik},
  number =       5,
  pages =        {585-603},
  title =        {Energy conservation and dissipation properties of
                  time-integration methods for nonsmooth
                  elastodynamics with contact},
  url =
                  {https://onlinelibrary.wiley.com/doi/abs/10.1002/zamm.201400231},
  volume =       96,
  year =         2016
}

@article{Alart:1991,
  author =       {P. Alart and A. Curnier},
  doi =          {https://doi.org/10.1016/0045-7825(91)90022-X},
  issn =         {0045-7825},
  journal =      {Computer Methods in Applied Mechanics and
                  Engineering},
  number =       3,
  pages =        {353-375},
  title =        {A mixed formulation for frictional contact problems
                  prone to Newton like solution methods},
  url =
                  {https://www.sciencedirect.com/science/article/pii/004578259190022X},
  volume =       92,
  year =         1991
}

@Article{Bai.Brandt:1987,
  author =       {Bai, D. and Brandt, A.},
  journal =      {SIAM Journal on Scientific and Statistical
                  Computing},
  title =        {Local Mesh Refinement Multilevel Techniques},
  year =         1987,
  number =       2,
  pages =        {109-134},
  volume =       8,
  doi =          {10.1137/0908025},
  url =          {https://doi.org/10.1137/0908025},
}

@misc{Barnett:2022,
  title =        {The Schwarz alternating method for the seamless
                  coupling of nonlinear reduced order models and full
                  order models},
  author =       {Joshua Barnett and Irina Tezaur and Alejandro Mota},
  year =         2022,
  eprint =       {2210.12551},
  archivePrefix ={arXiv},
  primaryClass = {math.NA}
}

@article{Bathe:1985,
  author =       {K.-J. Bathe and A. Chaudhary},
  journal =      {International Journal for Numerical Methods in
                  Engineering},
  pages =        {65--88},
  title =        {A solution method for planar and axisymmetric
                  contact problems},
  volume =       21,
  year =         1985
}

@article{Belytschko.Xiao:2003,
  address =      {{50 CROSS HIGHWAY, REDDING, CT 06896 USA}},
  affiliation =  {{Belytschko, T (Reprint Author), Northwestern Univ,
                  Dept Mech Engn, 2145 N Sheridan Rd, Evanston, IL
                  60208 USA. Belytschko, T.; Xiao, S. P., Northwestern
                  Univ, Dept Mech Engn, Evanston, IL 60208 USA.}},
  author =       {Belytschko, T. and Xiao, S. P.},
  da =           {{2019-10-31}},
  doc-delivery-number ={{V26NQ}},
  funding-acknowledgement ={{NASA University Research, Engineering and
                  Technology Institute on Bio Inspired Materials
                  (BIMat)National Aeronautics \& Space Administration
                  (NASA) {[}NCC-1-02037]}},
  funding-text = {{We gratefully acknowledge the grant support of NASA
                  University Research, Engineering, and Technology
                  Institute on Bio Inspired Materials (BIMat), award
                  No. NCC-1-02037.}},
  issn =         {{1543-1649}},
  journal =      {{International Journal for Multiscale Computational
                  Engineering}},
  journal-iso =  {{Int. J. Multiscale Comput. Eng.}},
  number =       {{1}},
  number-of-cited-references ={{12}},
  pages =        {{115-126}},
  publisher =    {{Begell House Inc.}},
  research-areas ={{Engineering; Mathematics}},
  times-cited =  {{177}},
  title =        {{Coupling Methods for Continuum Model with Molecular
                  Model}},
  type =         {{Article}},
  unique-id =    {{ISI:000208552700010}},
  usage-count-last-180-days ={{0}},
  usage-count-since-2013 ={{11}},
  volume =       {{1}},
  year =         2003
}

@article{Blanze:1996,
  author =       {C. Blanze and L. Champaney and J.Y. Cognard and
                  P. Ladeveze},
  journal =      {Engineering Computations},
  title =        {{A modular approach to structure assembly
                  computations: application to contact problems}},
  year =         1996
}

@inproceedings{Bochev.Kuberry:2015,
  author =       {Kuberry, Paul and Bochev, Pavel},
  booktitle =    {1st Pan-American Congress on Computational
                  Mechanics},
  title =        {A variational flux recovery approach for
                  elastodynamics problems with interfaces},
  year =         2015
}

@article{Bochev:2017,
  author =       {James Cheung and M. Perego and P. Bochev and
                  M. Gunzburger},
  doi =          {https://doi.org/10.1016/j.cam.2022.115027},
  issn =         {0377-0427},
  journal =      {Journal of Computational and Applied Mathematics},
  pages =        115027,
  title =        {A coupling approach for linear elasticity problems
                  with spatially non-coincident discretized
                  interfaces},
  url =
                  {https://www.sciencedirect.com/science/article/pii/S0377042722006252},
  volume =       425,
  year =         2023
}

@book{Boffi.Brezzi.Fortin:2013,
  author =       {D. Boffi and F. Brezzi and M. Fortin},
  publisher =    {Springer},
  title =        {Mixed finite element methods and applications},
  year =         2013
}

@article{Bucher:2007,
  author =       {Bucher, A. and Meyer, A. and G{\"o}rke, U. -J. and
                  Krei{\ss}ig, R.},
  doi =          {10.1007/s00466-006-0051-z},
  id =           {Bucher2007},
  journal =      {Computational Mechanics},
  number =       4,
  pages =        {521--536},
  title =        {A Comparison of Mapping Algorithms for Hierarchical
                  Adaptive FEM in Finite Elasto-Plasticity},
  url =          {https://doi.org/10.1007/s00466-006-0051-z},
  volume =       39,
  year =         2007
}

@article{Carpenter:1991,
  author =       {Carpenter, Nicholas J. and Taylor, Robert L. and
                  Katona, Michael G.},
  doi =          {https://doi.org/10.1002/nme.1620320107},
  journal =      {International Journal for Numerical Methods in
                  Engineering},
  number =       1,
  pages =        {103-128},
  title =        {Lagrange constraints for transient finite element
                  surface contact},
  url =
                  {https://onlinelibrary.wiley.com/doi/abs/10.1002/nme.1620320107},
  volume =       32,
  year =         1991
}

@article{Chaudhary.Bathe:1986,
  author =       {Anil B. Chaudhary and Klaus-J{\"u}rgen Bathe},
  doi =          {https://doi.org/10.1016/0045-7949(86)90294-4},
  issn =         {0045-7949},
  journal =      {Computers and Structures},
  number =       6,
  pages =        {855-873},
  title =        {A solution method for static and dynamic analysis of
                  three-dimensional contact problems with friction},
  url =
                  {https://www.sciencedirect.com/science/article/pii/0045794986902944},
  volume =       24,
  year =         1986
}

@article{Chouly:2015,
  author =       {F. Chouly and P. Hild and Y. Renard},
  journal =      {Mathematics of Computation},
  number =       293,
  pages =        {1089--1112},
  publisher =    {American Mathematical Society},
  title =        {{Symmetric and non-symmetric variants of Nitsche's
                  method for contact problems in elasticity: theory
                  and numerical experiments}},
  url =          {http://www.jstor.org/stable/24488882},
  volume =       84,
  year =         2015
}

@article{Chouly:2019,
  author =       {F. Chouly and M. Fabre and P. Hild and R. Mlika and
                  J. Pousin and Y. Renard},
  journal =      {UCL Workshop 2016, UCL (University College London),
                  Jan 2016, London, United Kingdom},
  pages =        {94--141},
  title =        {An overview of recent results on Nitsche's method
                  for contact problems},
  year =         2019
}

@article{Chung.Hulbert:1993,
  author =       {Jintai Chung and Gregory M. Hulbert},
  journal =      {Journal of Applied Mechanics},
  pages =        {371-375},
  title =        {A Time Integration Algorithm for Structural Dynamics
                  With Improved Numerical Dissipation: The
                  Generalized-$\alpha$ Method},
  volume =       60,
  year =         1993
}

@article{Chung.Lee:1994,
  author =       {Chung, Jintai and Lee, Jang Moo},
  doi =          {https://doi.org/10.1002/nme.1620372303},
  journal =      {International Journal for Numerical Methods in
                  Engineering},
  number =       23,
  pages =        {3961-3976},
  title =        {A new family of explicit time integration methods
                  for linear and non-linear structural dynamics},
  url =
                  {https://onlinelibrary.wiley.com/doi/abs/10.1002/nme.1620372303},
  volume =       37,
  year =         1994
}

@inproceedings{Cote:2005,
  address =      {Berlin, Heidelberg},
  author =       {C{\^o}t{\'e}, J. and Gander, M. J. and Laayouni,
                  L. and Loisel, S.},
  booktitle =    {Domain Decomposition Methods in Science and
                  Engineering},
  editor =       {Barth, Timothy J. and Griebel, Michael and Keyes,
                  David E. and Nieminen, Risto M. and Roose, Dirk and
                  Schlick, Tamar and Kornhuber, Ralf and Hoppe, Ronald
                  and P{\'e}riaux, Jacques and Pironneau, Olivier and
                  Widlund, Olof and Xu, Jinchao},
  isbn =         {978-3-540-26825-3},
  pages =        {235--242},
  publisher =    {Springer Berlin Heidelberg},
  title =        {Comparison of the Dirichlet-Neumann and Optimal
                  Schwarz Method on the Sphere},
  year =         2005
}

@article{Deng:2003,
  author =       {Q. Deng},
  journal =      {Communications on Pure and Applied Analysis},
  pages =        {297-310},
  title =        {A nonoverlapping domain decomposition method for
                  nonconforming finite element problems},
  volume =       2,
  year =         2003
}

@article{Deuflhard:2008,
  author =       {Deuflhard, Peter and Krause, Rolf and Ertel,
                  Susanne},
  doi =          {https://doi.org/10.1002/nme.2119},
  journal =      {International Journal for Numerical Methods in
                  Engineering},
  number =       9,
  pages =        {1274-1290},
  title =        {A contact-stabilized Newmark method for dynamical
                  contact problems},
  url =
                  {https://onlinelibrary.wiley.com/doi/abs/10.1002/nme.2119},
  volume =       73,
  year =         2008
}

@article{DiStasio:2019,
  author =       {Di Stasio, Jean and Dureisseix, David and Gravouil,
                  Anthony and Georges, Gabriel and Homolle, Thomas},
  doi =          {10.1186/s40323-019-0126-y},
  id =           {Di Stasio2019},
  journal =      {Advanced Modeling and Simulation in Engineering
                  Sciences},
  number =       1,
  pages =        2,
  title =        {Benchmark cases for robust explicit time integrators
                  in non-smooth transient dynamics},
  url =          {https://doi.org/10.1186/s40323-019-0126-y},
  volume =       6,
  year =         2019
}

@article{Dostal:2009,
  author =       {Zdenek Dost{\'a}l and Tom{\'a}s Kozubek and V{\'\i}t
                  Vondr{\'a}k and Tom{\'a}s Brzobohat{\'y} and
                  Alexandros Markopoulos},
  journal =      {International Journal for Numerical Methods in
                  Engineering},
  pages =        {1384-1405},
  title =        {{Scalable TFETI algorithm for the solution of
                  multibody contact problems of elasticity}},
  volume =       82,
  year =         2009
}

@article{Dostal:2019,
  author =       {Zdenek Dost{\'a}l and Oldrich Vlach and Tomas
                  Brzobohaty},
  doi =          {https://doi.org/10.1016/j.finel.2019.01.002},
  issn =         {0168-874X},
  journal =      {Finite Elements in Analysis and Design},
  pages =        {34-43},
  title =        {{Scalable TFETI based algorithm with adaptive
                  augmentation for contact problems with variationally
                  consistent discretization of contact conditions}},
  url =
                  {https://www.sciencedirect.com/science/article/pii/S0168874X18306978},
  volume =       156,
  year =         2019
}

@article{Doyen:2009,
  author =       {D. Doyen and A. Ern},
  doi =          {cms/1264434145},
  journal =      {Communications in Mathematical Sciences},
  number =       4,
  pages =        {1063 -- 1072},
  publisher =    {International Press of Boston},
  title =        {{Convergence of a space semi-discrete modified mass
                  method for the dynamic Signorini problem}},
  url =          {https://doi.org/},
  volume =       7,
  year =         2009
}

@Article{Doyen:2011,
  author =       {Doyen, David and Ern, Alexandre and Piperno, Serge},
  journal =      {SIAM Journal on Scientific Computing},
  title =        {Time-Integration Schemes for the Finite Element
                  Dynamic Signorini Problem},
  year =         2011,
  number =       1,
  pages =        {223-249},
  volume =       33,
  doi =          {10.1137/100791440},
  url =          {https://doi.org/10.1137/100791440},
}

@article{Dureisseix.Bavestrello:2006,
  author =       {David Dureisseix and Henri Bavestrello},
  doi =          {https://doi.org/10.1016/j.cma.2006.02.003},
  issn =         {0045-7825},
  journal =      {Computer Methods in Applied Mechanics and
                  Engineering},
  number =       44,
  pages =        {6523-6541},
  title =        {Information transfer between incompatible finite
                  element meshes: Application to coupled
                  thermo-viscoelasticity},
  url =
                  {https://www.sciencedirect.com/science/article/pii/S0045782506000673},
  volume =       195,
  year =         2006
}

@article{Eck:2002,
  author =       {R.H. Krause and B.I. Wohlmuth},
  journal =      {Computing and Visualization in Science},
  number =       3,
  pages =        {139--148},
  title =        {{A Dirichlet--Neumann type algorithm for contact
                  problems with friction. Computing and visualization
                  in science}},
  volume =       5,
  year =         2002
}

@phdthesis{Faltus:2020,
  address =      {Prague, Czech Republic},
  author =       {O. Faltus},
  school =       {Czech Technical University},
  title =        {Object-oriented design and implemention of the
                  contact mechanics into finite element code "OOFEM"},
  year =         2020
}

@incollection{Fichera:1972,
  author =       {Gaetano Fichera},
  editor =       {Siegfried Fl{\"u}gge and Clifford A. Truesdell},
  title =        {Boundary value problems of elasticity with
                  unilateral constraints},
  booktitle =    {Festk{\"o}rpermechanik/Mechanics of Solids},
  series =       {Handbuch der Physik (Encyclopedia of Physics)},
  volume =       {VIa/2},
  year =         1972,
  edition =      {paperback 1984},
  pages =        {391--424},
  publisher =    {Springer-Verlag},
  address =      {Berlin--Heidelberg--New York},
  isbn =         {0-387-13161-2},
  zbl =          {0277.73001}
}

@article{Francavilla:1975,
  author =       {Francavilla, A. and Zienkiewicz, O. C.},
  doi =          {https://doi.org/10.1002/nme.1620090410},
  journal =      {International Journal for Numerical Methods in
                  Engineering},
  number =       4,
  pages =        {913-924},
  title =        {A note on numerical computation of elastic contact
                  problems},
  url =
                  {https://onlinelibrary.wiley.com/doi/abs/10.1002/nme.1620090410},
  volume =       9,
  year =         1975
}

@Article{Funaro:1988,
  author =       {D. Funaro and A. Quarteroni and P. Zanolli},
  journal =      {SIAM Journal on Numerical Analysis},
  title =        {An Iterative Procedure with Interface Relaxation for
                  Domain Decomposition Methods},
  year =         1988,
  number =       6,
  pages =        {1213--1236},
  volume =       25,
}

@article{Fung:2003,
  author =       {Fung, T C},
  doi =          {https://doi.org/10.1002/pse.149},
  journal =      {Progress in Structural Engineering and Materials},
  number =       3,
  pages =        {167-180},
  title =        {Numerical dissipation in time-step integration
                  algorithms for structural dynamic analysis},
  url =
                  {https://onlinelibrary.wiley.com/doi/abs/10.1002/pse.149},
  volume =       5,
  year =         2003
}

@article{Gallego:1989,
  author =       {F. Gallego and J. Anza},
  journal =      {International Journal for Numerical Methods in
                  Engineering},
  pages =        {1249--1264},
  title =        {A mixed finite element model for the elastic contact
                  problem},
  volume =       28,
  year =         1989
}

@article{Gander:2008,
  author =       {M. Gander},
  journal =      {Electronic Transactions on Numerical Analysis},
  pages =        {228-255},
  title =        {Schwarz Methods over the Course of Time},
  volume =       31,
  year =         2008
}

@article{Gerardo-Giorda:2013,
  author =       {Gerardo-Giorda, Luca and Perego, Mauro},
  doi =          {10.1051/m2an/2012040},
  journal =      {ESAIM: Mathematical Modelling and Numerical Analysis
                  - Mod\'elisation Math\'ematique et Analyse
                  Num\'erique},
  language =     {en},
  mrnumber =     3021699,
  number =       2,
  pages =        {583-608},
  publisher =    {EDP-Sciences},
  title =        {Optimized Schwarz Methods for the Bidomain system in
                  electrocardiology},
  url =          {http://www.numdam.org/item/M2AN_2013__47_2_583_0},
  volume =       47,
  year =         2013,
  zbl =          {1274.92021}
}

@article{Gilardi:2002,
  author =       {G. Gilardi and I. Sharf},
  doi =          {https://doi.org/10.1016/S0094-114X(02)00045-9},
  issn =         {0094-114X},
  journal =      {Mechanism and Machine Theory},
  number =       10,
  pages =        {1213-1239},
  title =        {Literature survey of contact dynamics modelling},
  url =
                  {https://www.sciencedirect.com/science/article/pii/S0094114X02000459},
  volume =       37,
  year =         2002
}

@book{Glowinski:1989,
  author =       {R. Glowinski and P. Le Tallec},
  publisher =    {Society for Industrial and Applied Mathematics},
  title =        {{Augmented Lagrangian and Operator-Splitting Methods
                  in Nonlinear Mechanics}},
  year =         1989
}

@article{Gosselet:2018,
  author =       {P. Gosselet and M. Blanchard and O. Allix and
                  G. Guguin},
  journal =      {Advanced Modeling and Simulation in Engineering
                  Sciences},
  number =       1,
  pages =        {1-23},
  title =        {{Non-invasive global--local coupling as a Schwarz
                  domain decomposition method: acceleration and
                  generalization}},
  volume =       5,
  year =         2018
}

@ARTICLE{Haddadin.etal:2017,
  author =       {Haddadin, Sami and De Luca, Alessandro and
                  Albu-Schäffer, Alin},
  journal =      {IEEE Transactions on Robotics},
  title =        {Robot Collisions: A Survey on Detection, Isolation,
                  and Identification},
  year =         2017,
  volume =       33,
  number =       6,
  pages =        {1292-1312},
  doi =          {10.1109/TRO.2017.2723903}
}

@article{Hager:2008,
  author =       {Hager, C. and H{\"u}eber, S. and Wohlmuth, B. I.},
  title =        {A stable energy-conserving approach for frictional
                  contact problems based on quadrature formulas},
  journal =      {International Journal for Numerical Methods in
                  Engineering},
  volume =       73,
  number =       2,
  pages =        {205-225},
  doi =          {Hager:2008},
  url =
                  {https://onlinelibrary.wiley.com/doi/abs/10.1002/nme.2069},
  year =         2008,
}

@article{Hauret:2010,
  author =       {Patrice Hauret},
  doi =          {https://doi.org/10.1016/j.cma.2010.06.004},
  issn =         {0045-7825},
  journal =      {Computer Methods in Applied Mechanics and
                  Engineering},
  number =       45,
  pages =        {2941-2957},
  title =        {Mixed interpretation and extensions of the
                  equivalent mass matrix approach for elastodynamics
                  with contact},
  url =
                  {https://www.sciencedirect.com/science/article/pii/S0045782510001714},
  volume =       199,
  year =         2010
}

@article{Hennig:2018,
  author =       {P. Hennig and M. Ambati and L. {De Lorenzis} and
                  M. K{\"a}stner},
  doi =          {https://doi.org/10.1016/j.cma.2018.01.017},
  issn =         {0045-7825},
  journal =      {Computer Methods in Applied Mechanics and
                  Engineering},
  pages =        {313-336},
  title =        {Projection and transfer operators in adaptive
                  isogeometric analysis with hierarchical B-splines},
  url =
                  {https://www.sciencedirect.com/science/article/pii/S0045782518300203},
  volume =       334,
  year =         2018
}

@article{Hughes:1978,
  author =       {Hughes, T. J. R. and Caughey, T. K. and Liu, W. K.},
  doi =          {10.1115/1.3424303},
  issn =         {0021-8936},
  journal =      {Journal of Applied Mechanics},
  month =        06,
  number =       2,
  pages =        {366-370},
  title =        {Finite-Element Methods for Nonlinear Elastodynamics
                  Which Conserve Energy},
  url =          {https://doi.org/10.1115/1.3424303},
  volume =       45,
  year =         1978
}

@book{Hughes:2000,
  author =       {Hughes, T. J. R.},
  publisher =    {Dover Publications, New York},
  title =        {Finite Element Method: Linear Static and Dynamic
                  Finite Element Analysis},
  year =         2000
}

@article{Hunek:1993,
  author =       {I. Hunek},
  doi =          {https://doi.org/10.1016/0045-7949(93)90412-7},
  issn =         {0045-7949},
  journal =      {Computers and Structures},
  number =       2,
  pages =        {193-203},
  title =        {On a penalty formulation for contact-impact
                  problems},
  url =
                  {https://www.sciencedirect.com/science/article/pii/0045794993904127},
  volume =       48,
  year =         1993
}

@article{Jean:1999,
  author =       {M. Jean},
  doi =          {https://doi.org/10.1016/S0045-7825(98)00383-1},
  issn =         {0045-7825},
  journal =      {Computer Methods in Applied Mechanics and
                  Engineering},
  number =       3,
  pages =        {235-257},
  title =        {The non-smooth contact dynamics method},
  url =
                  {https://www.sciencedirect.com/science/article/pii/S0045782598003831},
  volume =       177,
  year =         1999
}

@article{Jimenez.etal:2001,
  title =        {3D collision detection: a survey},
  journal =      {Computers \& Graphics},
  volume =       25,
  number =       2,
  pages =        {269-285},
  year =         2001,
  issn =         {0097-8493},
  doi =          {https://doi.org/10.1016/S0097-8493(00)00130-8},
  url =
                  {https://www.sciencedirect.com/science/article/pii/S0097849300001308},
  author =       {P. Jiménez and F. Thomas and C. Torras}
}

@article{Jones:2001,
  author =       {Jones, Reese E. and Papadopoulos, Panayiotis},
  doi =          {https://doi.org/10.1002/nme.163},
  journal =      {International Journal for Numerical Methods in
                  Engineering},
  number =       7,
  pages =        {791-811},
  title =        {A novel three-dimensional contact finite element
                  based on smooth pressure interpolations},
  url =
                  {https://onlinelibrary.wiley.com/doi/abs/10.1002/nme.163},
  volume =       51,
  year =         2001
}

@article{Kane.etal:1999,
  author =       {C. Kane and E.A. Repetto and M. Ortiz and
                  J.E. Marsden},
  doi =          {https://doi.org/10.1016/S0045-7825(99)00034-1},
  issn =         {0045-7825},
  journal =      {Computer Methods in Applied Mechanics and
                  Engineering},
  number =       1,
  pages =        {1-26},
  title =        {Finite element analysis of nonsmooth contact},
  url =
                  {https://www.sciencedirect.com/science/article/pii/S0045782599000341},
  volume =       180,
  year =         1999
}

@article{Katona:1985,
  author =       {Katona, Michael C and Zienkiewicz, OC},
  journal =      {International Journal for Numerical Methods in
                  Engineering},
  number =       7,
  pages =        {1345--1359},
  publisher =    {Wiley Online Library},
  title =        {A unified set of single step algorithms part 3: the
                  beta-m method, a generalization of the Newmark
                  scheme},
  volume =       21,
  year =         1985
}

@article{Khenous:2008,
  author =       {Houari Boumedi{\`e}ne Khenous and Patrick Laborde
                  and Yves Renard},
  doi =          {https://doi.org/10.1016/j.euromechsol.2008.01.001},
  issn =         {0997-7538},
  journal =      {European Journal of Mechanics - A/Solids},
  number =       5,
  pages =        {918-932},
  title =        {Mass redistribution method for finite element
                  contact problems in elastodynamics},
  url =
                  {https://www.sciencedirect.com/science/article/pii/S0997753808000107},
  volume =       27,
  year =         2008
}

@inproceedings{Kockare.etal:2007,
  author =       {Kockara, S. and Halic, T. and Iqbal, K. and Bayrak,
                  C. and Rowe, Richard},
  booktitle =    {2007 IEEE International Conference on Systems, Man
                  and Cybernetics},
  title =        {Collision detection: A survey},
  year =         2007,
  pages =        {4046-4051},
  doi =          {10.1109/ICSMC.2007.4414258}
}

@article{Koliesnikova:2022,
  author =       {Daria Koliesnikova and Isabelle Rami{\`e}re and
                  Fr{\'e}d{\'e}ric Lebon},
  doi =          {https://doi.org/10.1016/j.cma.2022.115505},
  issn =         {0045-7825},
  journal =      {Computer Methods in Applied Mechanics and
                  Engineering},
  pages =        115505,
  title =        {Fully automatic multigrid adaptive mesh refinement
                  strategy with controlled accuracy for nonlinear
                  quasi-static problems},
  url =
                  {https://www.sciencedirect.com/science/article/pii/S0045782522005151},
  volume =       400,
  year =         2022
}

@article{Krause:2012,
  author =       {Rolf Krause and Mirjam Walloth},
  doi =          {https://doi.org/10.1016/j.apnum.2012.06.014},
  issn =         {0168-9274},
  journal =      {Applied Numerical Mathematics},
  note =         {Selected Papers from NUMDIFF-12},
  number =       10,
  pages =        {1393-1410},
  title =        {Presentation and comparison of selected algorithms
                  for dynamic contact based on the Newmark scheme},
  url =
                  {https://www.sciencedirect.com/science/article/pii/S0168927412001006},
  volume =       62,
  year =         2012
}

@article{Kumar:2015,
  author =       {Sushil Kumar and Lionel Fourment and Simon Guerdoux},
  doi =          {https://doi.org/10.1016/j.finel.2014.09.002},
  issn =         {0168-874X},
  journal =      {Finite Elements in Analysis and Design},
  pages =        {70-84},
  title =        {Parallel, second-order and consistent remeshing
                  transfer operators for evolving meshes with
                  superconvergence property on surface and volume},
  url =
                  {https://www.sciencedirect.com/science/article/pii/S0168874X14001735},
  volume =       93,
  year =         2015
}

@inproceedings{Kwok:2014,
  address =      {Cham},
  author =       {Kwok, Felix},
  booktitle =    {Domain Decomposition Methods in Science and
                  Engineering XXI},
  editor =       {Erhel, Jocelyne and Gander, Martin J. and Halpern,
                  Laurence and Pichot, G{\'e}raldine and Sassi,
                  Taoufik and Widlund, Olof},
  isbn =         {978-3-319-05789-7},
  pages =        {189--198},
  publisher =    {Springer International Publishing},
  title =        {Neumann--Neumann Waveform Relaxation for the
                  Time-Dependent Heat Equation},
  year =         2014
}

@article{Laursen:1997,
  author =       {T. A. Laursen and V. Chawla},
  journal =      {International Journal for Numerical Methods in
                  Engineering},
  number =       5,
  pages =        {863-886},
  title =        {Design of energy conserving algorithms for
                  frictionless dynamic contact problems},
  volume =       40,
  year =         1997
}

@article{Laursen:2002,
  author =       {Laursen, T. A. and Love, G. R.},
  doi =          {https://doi.org/10.1002/nme.264},
  journal =      {International Journal for Numerical Methods in
                  Engineering},
  number =       2,
  pages =        {245-274},
  title =        {Improved implicit integrators for transient impact
                  problems---geometric admissibility within the
                  conserving framework},
  volume =       53,
  year =         2002
}

@book{Laursen:2003,
  author =       {Laursen, T. A.},
  publisher =    {Springer Science \& Business Media},
  title =        {Computational contact and impact mechanics:
                  fundamentals of modeling interfacial phenomena in
                  nonlinear finite element analysis},
  year =         2003
}

@inproceedings{LionsNonOverlap:1988,
  author =       {Lions, P. L.},
  booktitle =    {Third International Symposium on Domain
                  Decomposition Methods for Partial Differential
                  Equations},
  chapter =      11,
  comment =      {(private-note)cited by Mathew (2008, p. 32)},
  editor =       {Chan, Tony F. and Glowinski, R. and P\'{e}riaux,
                  J. and Widlund, O. B.},
  isbn =         9780898712537,
  pages =        {202--223},
  publisher =    {Society for Industrial and Applied Mathematics},
  title =        {On the {Schwarz} Alternating Method {III}: A Variant
                  for Nonoverlapping Subdomains},
  url =
                  {http://www.ddm.org/DD03/On\_the\_Schwarz\_Alternating\_Method\_III\_A\_Variant\_for\_Nonoverlapping\_Subdomains\_(Lions).pdf},
  year =         1990
}

@article{Lui:2001,
  author =       {S.H. Lui},
  doi =          {https://doi.org/10.1016/S0377-0427(99)00374-X},
  issn =         {0377-0427},
  journal =      {Journal of Computational and Applied Mathematics},
  number =       1,
  pages =        {309--321},
  title =        {On accelerated convergence of nonoverlapping Schwarz
                  methods},
  url =
                  {http://www.sciencedirect.com/science/article/pii/S037704279900374X},
  volume =       130,
  year =         2001
}

@phdthesis{Mlika:2018,
  address =      {Lyon, France},
  author =       {R. Mlika},
  school =       {Institut national des sciences appliquees de Lyon},
  title =        {Nitsche method for frictional contact and
                  self-contact: mathematical and numerical study},
  year =         2018
}

@article{Monjaraz:2022,
  author =       {C.D. Monjaraz Tec and J. Gross and M. Krack},
  doi =          {https://doi.org/10.1016/j.compstruc.2021.106698},
  issn =         {0045-7949},
  journal =      {Computers and Structures},
  pages =        106698,
  title =        {A massless boundary component mode synthesis method
                  for elastodynamic contact problems},
  url =
                  {https://www.sciencedirect.com/science/article/pii/S0045794921002200},
  volume =       260,
  year =         2022
}

@article{Moreau:1999,
  author =       {J.J. Moreau},
  doi =          {https://doi.org/10.1016/S0045-7825(98)00387-9},
  issn =         {0045-7825},
  journal =      {Computer Methods in Applied Mechanics and
                  Engineering},
  number =       3,
  pages =        {329-349},
  title =        {Numerical aspects of the sweeping process},
  url =
                  {https://www.sciencedirect.com/science/article/pii/S0045782598003879},
  volume =       177,
  year =         1999
}

@misc{Mota.Koliesnikova:2023,
  author =       {Mota, A. and Koliesnikova, D.},
  title =        {Norma},
  year =         2023,
  publisher =    {GitHub},
  journal =      {GitHub repository},
  howpublished = {\url{https://github.com/lxmota/norma}},
  version =      {0.1}
}

@article{Mota.etal:2003,
  title =        {Finite-element simulation of firearm injury to the
                  human cranium},
  author =       {Mota, A. and Klug, W. and Ortiz, M. and Pandolfi,
                  A.},
  journal =      {Computational Mechanics},
  volume =       31,
  pages =        {115--121},
  year =         2003,
  doi =          {10.1007/s00466-002-0398-8}
}

@misc{Mota.etal:2023,
  author =       {Mota, A. and Hoy, J. and Tezaur, I. and
                  Koliesnikova, D.},
  title =        {1D Matlab dynamic contact code},
  year =         2023,
  publisher =    {GitHub},
  journal =      {GitHub repository},
  howpublished =
                  {\url{https://github.com/ikalash/Schwarz-4-Multiscale}},
  version =      {0.1}
}

@article{Mota:2017,
  author =       {Mota, A. and Tezaur, I. and Alleman, C.},
  journal =      {Computer Methods in Applied Mechanics and
                  Engineering},
  pages =        {19--51},
  title =        {The {S}chwarz alternating method in solid mechanics},
  volume =       319,
  year =         2017
}

@article{Mota:2022,
  author =       {Mota, Alejandro and Tezaur, Irina and Phlipot,
                  Gregory},
  title =        {The Schwarz alternating method for transient solid
                  dynamics},
  journal =      {International Journal for Numerical Methods in
                  Engineering},
  volume =       123,
  number =       21,
  pages =        {5036-5071},
  doi =          {https://doi.org/10.1002/nme.6982},
  url =
                  {https://onlinelibrary.wiley.com/doi/abs/10.1002/nme.6982},
  year =         2022
}

@article{Newmark:1959,
  author =       {Newmark, N.M.},
  journal =      {Journal of Engineering Mechanics, ASCE},
  number =       3,
  pages =        {67--94},
  title =        {A method of computation for structural dynamics},
  volume =       85,
  year =         1959
}

@article{Nitsche:1971,
  author =       {J. Nitsche},
  journal =      {Abhandlungen aus dem Mathematischen Seminar der
                  Universitat Hamburg},
  number =       36,
  pages =        {9--15},
  title =        {{Uber ein Variationsprinzip zur Losung von
                  Dirichlet-Problemen bei Verwendung von Teilraumen,
                  die keinen Randbedingungen unterworfen sind}},
  volume =       84,
  year =         2015
}

@article{Oden:1984,
  author =       {J.T. Oden and E.B. Pires},
  doi =          {https://doi.org/10.1016/0045-7949(84)90212-8},
  issn =         {0045-7949},
  journal =      {Computers and Structures},
  note =         {Special Memorial Issue},
  number =       1,
  pages =        {137-147},
  title =        {Algorithms and numerical results for finite element
                  approximations of contact problems with
                  non-classical friction laws},
  url =
                  {https://www.sciencedirect.com/science/article/pii/0045794984902128},
  volume =       19,
  year =         1984
}

@article{Oumaziz:2018,
  author =       {Oumaziz, Paul and Gosselet, Pierre and Boucard,
                  Pierre-Alain and Abbas, Mickael},
  doi =          {https://doi.org/10.1002/nme.5830},
  journal =      {International Journal for Numerical Methods in
                  Engineering},
  number =       8,
  pages =        {893-912},
  title =        {A parallel noninvasive multiscale strategy for a
                  mixed domain decomposition method with frictional
                  contact},
  url =
                  {https://onlinelibrary.wiley.com/doi/abs/10.1002/nme.5830},
  volume =       115,
  year =         2018
}

@Article{Paoli:2001,
  author =       {Paoli, Laetitia},
  journal =      {Philosophical Transactions of the Royal Society of
                  London. Series A: Mathematical, Physical and
                  Engineering Sciences},
  title =        {Time discretization of vibro-impact},
  year =         2001,
  number =       1789,
  pages =        {2405-2428},
  volume =       359,
  doi =          {10.1098/rsta.2001.0858},
  url =
                  {https://royalsocietypublishing.org/doi/abs/10.1098/rsta.2001.0858},
}

@Article{Paoli:2002,
  author =       {Paoli, Laetitia and Schatzman, Michelle},
  journal =      {SIAM Journal on Numerical Analysis},
  title =        {A Numerical Scheme for Impact Problems I: The
                  One-Dimensional Case},
  year =         2002,
  number =       2,
  pages =        {702-733},
  volume =       40,
  doi =          {10.1137/S0036142900378728},
  url =          {https://doi.org/10.1137/S0036142900378728},
}

@Article{Paoli:2002a,
  author =       {Paoli, Laetitia and Schatzman, Michelle},
  journal =      {SIAM Journal on Numerical Analysis},
  title =        {A Numerical Scheme for Impact Problems II: The
                  Multidimensional Case},
  year =         2002,
  number =       2,
  pages =        {734-768},
  volume =       40,
  doi =          {10.1137/S003614290037873X},
  url =          {https://doi.org/10.1137/S003614290037873X},
}

@article{Peric:1996,
  author =       {D. Peri{\'c} and Ch. Hochard and M. Dutko and
                  D.R.J. Owen},
  doi =          {https://doi.org/10.1016/S0045-7825(96)01070-5},
  issn =         {0045-7825},
  journal =      {Computer Methods in Applied Mechanics and
                  Engineering},
  number =       3,
  pages =        {331-344},
  title =        {Transfer operators for evolving meshes in small
                  strain elasto-placticity},
  url =
                  {https://www.sciencedirect.com/science/article/pii/S0045782596010705},
  volume =       137,
  year =         1996
}

@article{Popp.etal:2012,
  author =       {Popp, A. and Wohlmuth, B. I. and Gee, M. W. and
                  Wall, W. A.},
  journal =      {{SIAM Journal on Scientific Computing}},
  number =       {{4}},
  pages =        {{B421-B446}},
  title =        {{Dual quadratic mortar finite element methods for 3D
                  finite deformation contact}},
  volume =       {{34}},
  year =         2012
}

@article{Puso.Laursen:2004,
  author =       {Michael A. Puso and Tod A. Laursen},
  doi =          {https://doi.org/10.1016/j.cma.2004.06.001},
  issn =         {0045-7825},
  journal =      {Computer Methods in Applied Mechanics and
                  Engineering},
  number =       45,
  pages =        {4891-4913},
  title =        {A mortar segment-to-segment frictional contact
                  method for large deformations},
  url =
                  {https://www.sciencedirect.com/science/article/pii/S0045782504002518},
  volume =       193,
  year =         2004
}

@article{Puso.etal:2008,
  author =       {Puso, Michael A. and Laursen, T. A. and Solberg,
                  Jerome},
  journal =      {{Computer Methods in Applied Mechancis and
                  Engineering}},
  number =       {{6-8}},
  pages =        {{555-566}},
  title =        {{A segment-to-segment mortar contact method for
                  quadratic elements and large deformations}},
  volume =       {{197}},
  year =         2008
}

@article{Puso:2004,
  author =       {Puso, Michael A.},
  doi =          {https://doi.org/10.1002/nme.865},
  journal =      {International Journal for Numerical Methods in
                  Engineering},
  number =       3,
  pages =        {315-336},
  title =        {A 3D mortar method for solid mechanics},
  url =
                  {https://onlinelibrary.wiley.com/doi/abs/10.1002/nme.865},
  volume =       59,
  year =         2004
}

@book{Rao:2017,
  author =       {Rao, Singiresu S},
  publisher =    {Butterworth-heinemann},
  title =        {The finite element method in engineering},
  year =         2017
}

@article{Rio:2005,
  author =       {G{\'e}rard Rio and Anthony Soive and Vincent
                  Grolleau},
  doi =          {https://doi.org/10.1016/j.advengsoft.2004.10.011},
  issn =         {0965-9978},
  journal =      {Advances in Engineering Software},
  number =       4,
  pages =        {252-265},
  title =        {Comparative study of numerical explicit time
                  integration algorithms},
  url =
                  {https://www.sciencedirect.com/science/article/pii/S0965997804002029},
  volume =       36,
  year =         2005
}

@article{Schwarz:1870,
  author =       {H. Schwarz},
  journal =      {Vierteljahrsschriftder Naturforschenden Gesellschaft
                  in Zurich},
  pages =        {272--286},
  title =        {\"{U}ber einen {G}renz\"{u}bergang durch
                  alternierendes Verfahren},
  volume =       15,
  year =         1870
}

@techreport{Sierra:2011,
  author =       {{SIERRA Solid Mechanics Team}},
  institution =  {Sandia National Laboratories Report},
  month =        6,
  number =       {SAND2023-04163},
  title =        {Sierra/SolidMechanics 5.14 User's Guide},
  year =         2023
}

@article{Signorini:1959,
  author =       {Signorini, A.},
  title =        {Questioni di elasticit{\`a} non linearizzata e
                  semilinearizzata},
  journal =      {Rendiconti di Matematica e delle sue Applicazioni,
                  Quinta Serie},
  ISSN =         {0034-4427},
  volume =       18,
  pages =        {95--139},
  year =         1959,
  language =     {Italian},
}

@article{Simo:1985,
  author =       {Juan C. Simo and Peter Wriggers and Robert
                  L. Taylor},
  doi =          {https://doi.org/10.1016/0045-7825(85)90088-X},
  issn =         {0045-7825},
  journal =      {Computer Methods in Applied Mechanics and
                  Engineering},
  number =       2,
  pages =        {163-180},
  title =        {A perturbed Lagrangian formulation for the finite
                  element solution of contact problems},
  url =
                  {https://www.sciencedirect.com/science/article/pii/004578258590088X},
  volume =       50,
  year =         1985
}

@article{Simo:1992,
  author =       {J.C. Simo and T.A. Laursen},
  doi =          {https://doi.org/10.1016/0045-7949(92)90540-G},
  issn =         {0045-7949},
  journal =      {Computers and Structures},
  number =       1,
  pages =        {97-116},
  title =        {An augmented lagrangian treatment of contact
                  problems involving friction},
  url =
                  {https://www.sciencedirect.com/science/article/pii/004579499290540G},
  volume =       42,
  year =         1992
}

@article{Solberg.Jones.Papadopoulos:2007,
  author =       {Solberg, Jerome M. and Jones, Reese E. and
                  Papadopoulos, Panayiotis},
  journal =      {{Computer Methods in Applied Mechanics and
                  Engineering}},
  number =       {{4-6}},
  pages =        {{782-802}},
  title =        {{A family of simple two-pass dual formulations for
                  the finite element solution of contact problems}},
  volume =       {{196}},
  year =         2007
}

@article{Solberg:1998,
  author =       {Jerome M Solberg and Panayiotis Papadopoulos},
  doi =          {https://doi.org/10.1016/S0168-874X(98)00041-9},
  issn =         {0168-874X},
  journal =      {Finite Elements in Analysis and Design},
  number =       4,
  pages =        {297-311},
  title =        {A finite element method for contact/impact},
  url =
                  {https://www.sciencedirect.com/science/article/pii/S0168874X98000419},
  volume =       30,
  year =         1998
}

@article{Suwannachit:2012,
  author =       {Suwannachit, A. and Nackenhorst, U. and Chiarello,
                  R.},
  doi =          {10.1007/s00466-012-0722-x},
  id =           {Suwannachit2012},
  journal =      {Computational Mechanics},
  number =       6,
  pages =        {769--788},
  title =        {Stabilized numerical solution for transient dynamic
                  contact of inelastic solids on rough surfaces},
  url =          {https://doi.org/10.1007/s00466-012-0722-x},
  volume =       49,
  year =         2012
}

@inproceedings{Tchamwa:1999,
  author =       {Tchamwa, Bertrand and Conway, Ted and Wielgosz,
                  Christian},
  title =        "{An Accurate Explicit Direct Time Integration Method
                  for Computational Structural Dynamics}",
  series =       {Recent Advances in Solids and Structures},
  booktitle =    {ASME International Mechanical Engineering Congress
                  and Exposition},
  pages =        {77-84},
  year =         1999,
  month =        11,
  doi =          {10.1115/IMECE1999-0617},
  url =          {https://doi.org/10.1115/IMECE1999-0617},
}

@unpublished{Tezaur:2021,
  author =       {Tezaur, Irina and Voth, Thomas and Niederhaus, John
                  and Robbins, Joshua and Sanchez, Jason},
  title =        {An eXtended Finite Element Method (XFEM) Formulation
                  for Multi-Material Eulerian Solid Mechanics in the
                  ALEGRA Code},
  note =         {Manuscript in preparation},
  year =         2023
}

@article{Wriggers.Rust.Reddy:2016,
  author =       {Wriggers, P. and Rust, W. T. and Reddy, B. D.},
  journal =      {{Computational Mechanics}},
  month =        12,
  number =       {{6}},
  pages =        {{1039-1050}},
  title =        {{A virtual element method for contact}},
  volume =       {{58}},
  year =         2016
}

@book{Wriggers.Zavarise:2004,
  author =       {P. Wriggers and G. Zavarise},
  publisher =    {John Wiley \& Sons, Ltd. Edited by E. Stein, R. de
                  Borst and T.J.R. Hughes},
  title =        {Encyclopedia of Computational Mechanics, Volume 2:
                  Solids and Structures},
  year =         2004
}

@article{Wriggers:1985,
  author =       {P. Wriggers and J. Simo and R. Taylor},
  journal =      {Proceedings of the NUMETA '85 Conference, Elsevier,
                  Amsterdam},
  title =        {Penalty and augmented Lagrangian formulations for
                  contact problems},
  year =         1985
}

@book{Wriggers:2006,
  author =       {Wriggers, P.},
  isbn =         9783540326090,
  lccn =         2006922005,
  publisher =    {Springer Berlin Heidelberg},
  title =        {Computational Contact Mechanics},
  url =          {https://books.google.com/books?id=oBMBy4eMYsQC},
  year =         2006
}

@article{Wriggers:2008,
  author =       {P. Wriggers and G. Zavarise},
  journal =      {Computational Mechanics},
  pages =        {407--420},
  title =        {A formulation for frictionless contact problems
                  using a weak form introduced by Nitsche},
  volume =       41,
  year =         2008
}

@article{Yang:2008,
  author =       {Bin Yang and Tod A. Laursen},
  doi =          {https://doi.org/10.1016/j.cma.2007.09.004},
  issn =         {0045-7825},
  journal =      {Computer Methods in Applied Mechanics and
                  Engineering},
  number =       6,
  pages =        {756-772},
  title =        {A large deformation mortar formulation of self
                  contact with finite sliding},
  url =
                  {https://www.sciencedirect.com/science/article/pii/S0045782507003751},
  volume =       197,
  year =         2008
}

@article{Zanolli:1987,
  author =       {P. Zanolli},
  journal =      {Calcolo},
  pages =        {201--240},
  title =        {Domain decomposition algorithms for spectral
                  methods},
  volume =       24,
  year =         1987
}

@article{Zavarise:2009,
  author =       {Zavarise, Giorgio and De Lorenzis, Laura},
  doi =          {https://doi.org/10.1002/nme.2559},
  journal =      {International Journal for Numerical Methods in
                  Engineering},
  number =       4,
  pages =        {379-416},
  title =        {A modified node-to-segment algorithm passing the
                  contact patch test},
  url =
                  {https://onlinelibrary.wiley.com/doi/abs/10.1002/nme.2559},
  volume =       79,
  year =         2009
}

@book{Zienkiewicz.etal:2005,
  author =       {Zienkiewicz, Olek C and Taylor, Robert L},
  publisher =    {Elsevier},
  title =        {The finite element method for solid and structural
                  mechanics},
  year =         2005
}

\end{document}